\documentclass[12pt]{article}

\usepackage[british]{babel}

\usepackage[a4paper,top=2cm,bottom=2cm,left=2.5cm,right=2.5cm,marginparwidth=1.75cm]{geometry}

\usepackage{amsmath,amssymb,amsthm}
\usepackage{multirow}
\usepackage{graphicx}
\usepackage[colorlinks=true, allcolors=blue]{hyperref}
\usepackage{hyperref}
\usepackage[title]{appendix}
\usepackage{mathrsfs}
\usepackage{amsfonts}
\usepackage{booktabs} 
\usepackage{caption}  
\usepackage{threeparttable} 
\usepackage{algorithm}
\usepackage{algorithmicx}
\usepackage{algpseudocode}
\usepackage{listings}
\usepackage{enumitem}
\usepackage{booktabs}
\usepackage{subcaption}
\usepackage[T1]{fontenc}    

\usepackage{setspace}
\usepackage{xcolor}
\usepackage{titlesec}
\titleformat{\section} 
  {\normalfont\Large\bfseries}{\thesection.}{1em}{}
  
\usepackage{float}   
\usepackage{caption} 
\makeatletter

\newtheorem{theorem}{Theorem}
\newtheorem{Lemma}{Lemma}
\renewcommand{\P}{\mathbb{P}}

\newcommand{\E}{\mathbb{E}}

\newcommand{\Var}{\mathrm{Var}}
\newcommand{\SD}{\mathrm{SD}}

\makeatother

\title{Optimal Adjustment and Combination of Independent Discrete $p$-Values}

\author{Gonzalo Contador \thanks{Department of Mathematics, Universidad T\'ecnica Federico Santa Mar\'ia,  Vicu\~na Mackenna 3300, Santiago, Chile}
\and
Zheyang Wu \thanks{ Department of Mathematical Sciences, Worcester Polytechnic Institute, 100 Institute Road, Worcester, MA, 01609, USA}}

\date{}  

\begin{document}
\maketitle
\begin{abstract}

Combining $p$-values from multiple independent tests is a fundamental task in statistical inference, but presents unique challenges when the $p$-values are discrete. We extend a recent optimal transport-based framework for combining discrete $p$-values, which constructs a continuous surrogate distribution by minimizing the Wasserstein distance between the transformed discrete null and its continuous analogue. We provide a unified approach for several classical combination methods, including Fisher’s, Pearson’s, George’s, Stouffer’s, and Edgington’s statistics. Our theoretical analysis and extensive simulations show that accurate Type I error control is achieved when the variance of the adjusted discrete statistic closely matches that of the continuous case. We further demonstrate that, when the likelihood ratio test is a monotonic function of a combination statistic, the proposed approximation achieves power comparable to the uniformly most powerful (UMP) test. The methodology is illustrated with a genetic association study of rare variants using case-control data, and is implemented in the R package \texttt{DPComb}.
\end{abstract}

\textbf{Keywords}: Meta-Analysis, Continuous Approximations, Optimal Transport, Global Hypothesis Testing, Integral Transform.  


\newpage
\section{Introduction}
\label{intro}


Combining multiple $p$-values to test a global hypothesis is a fundamental statistical technique with broad applications, including signal detection, meta-analysis, and integrative data analysis. In this framework, a set of $p$-values $P_1, \ldots, P_n$, each representing evidence from an individual hypothesis test, are aggregated into a single summary statistic. This combined statistic is then used to assess the overall evidence against the global null hypothesis. For example, in signal detection, $P_j$, $j=1, \ldots, n$, may represent the significance of the $j$th feature or marker; in meta-analysis, $P_j$ corresponds to the result from the $j$th study. By combining these $p$-values, researchers can evaluate the collective significance across multiple tests or studies, thereby increasing power and robustness \cite{fisher1925statistical,edgington1972additive}.

Specifically, consider the global null hypothesis 
\begin{equation}
	\label{eq:globalNull}
	H_0 = \bigcap_{j=1}^n H_{0j},
\end{equation}
where each $H_{0j}$ is an individual null hypothesis for $j = 1, \ldots, n$. Let $P_j$ denote the $p$-value from the $j$th test. A common approach for testing $H_0$ is to aggregate these $p$-values into a single statistic that summarizes the overall evidence against the global null. In this paper, we consider a general family of such statistics:
\begin{equation}
	\label{eq:sumstat}
	T = \sum_{j=1}^n G^{-1}(P_j) \quad \text{or} \quad T = \sum_{j=1}^n G^{-1}(1 - P_j),
\end{equation}
where $G$ is a strictly increasing cumulative distribution function (CDF). Different choices of $G$ yield various classical combination statistics, as summarized in Table \ref{table:combstat}. For example, Fisher’s and Pearson’s combinations correspond to $G^{-1}(1 - P_j)$ and $G^{-1}(P_j)$, respectively, with $G$ being the CDF of the $\chi^2_2$ distribution. George’s, Stouffer’s, and Edgington’s combinations correspond to $G^{-1}(P_j)$ with $G$ being the CDF of the Logistic$(0,1)$, $\mathcal{N}(0,1)$, and Uniform$(0,1)$ distributions, respectively. Note that in some literature \cite{george1990p,heard2017choosing,long2023cauchy}, Stouffer’s and George’s statistics are defined as $\tilde{T}_S = \sum_{j=1}^n \Phi^{-1}(1 - P_j)$ and $\tilde{T}_G = \sum_{j=1}^n \log \frac{1 - P_j}{P_j}$, respectively. Since the standard normal and logistic distributions are symmetric about zero, i.e., $G^{-1}(x) = -G^{-1}(1 - x)$, it follows that $\tilde{T}_S = -T_S$ and $\tilde{T}_G = -T_G$. Thus, all results in this paper correspond to the definitions in Table \ref{table:combstat}, but can be directly translated to these alternative forms.

To test the global null hypothesis, compare the observed value of $T$ to the quantiles of its null distribution at significance level $\alpha$. Specifically, reject $H_0$ if $T = \sum_{j=1}^n G^{-1}(P_j)$ is less than or equal to the $\alpha$ quantile of its null distribution, or if $T = \sum_{j=1}^n G^{-1}(1 - P_j)$ is greater than or equal to the $1-\alpha$ quantile. This ensures that smaller values of $T$ (for $G^{-1}(P_j)$) or larger values of $T$ (for $G^{-1}(1-P_j)$) provide stronger evidence against the global null.

\begin{table}[h!]
	\caption{Classical $p$-value combination tests and their null distributions under the assumption of continuous $p$-values.}
	\centering
	\begin{tabular}{l c c }
		\toprule
		Test & Statistic & Null distribution under continuity  \\ 
		\midrule
		Fisher \cite{fisher1925statistical}    & $T_F=-2\sum_{j=1}^n \log P_j$ & $\chi^2_{2n}$   \\ 
		Pearson  \cite{Pearson1933}    & $T_P=-2\sum_{j=1}^n \log (1-P_j)$ & $\chi^2_{2n}$  \\ 
		George  \cite{mudholkar1979logit}    & $T_G= \frac{T_P-T_F}{2}=\sum_{j=1}^n \log \frac{P_j}{1-P_j}$ & $\approx$Logistic$(0,n)$ \cite{george1983convolution,ojo2003generalized}\\ 
		Stouffer  \cite{Stouffer1949}   & $T_S=\sum_{j=1}^n \Phi^{-1}(P_j)$ & $\mathcal{N}(0,n)$  \\ 
		Edgington  \cite{edgington1972additive}   & $T_E=\sum_{j=1}^n P_j$ & $\text{Irwin-Hall}(n)$  \\ 
		\bottomrule
	\end{tabular}
	\label{table:combstat}	
\end{table}

In much of the literature, it is assumed that the underlying $p$-values are continuous. Under this assumption, the $P_j$ are independent and identically distributed as Uniform$(0, 1)$ random variables. If $G$ in \eqref{eq:sumstat} is the CDF of a continuous distribution, then both $G^{-1}(P_j)$ and $G^{-1}(1-P_j)$ follow this distribution, making it straightforward to derive the null distribution of $T$. For the statistics in Table \ref{table:combstat}, Fisher's and Pearson's combination tests correspond to the chi-squared distribution, Stouffer's combination corresponds to the normal distribution, Edgington's combination corresponds to the Irwin-Hall distribution, and George's combination test corresponds approximately to the logistic distribution.

In practice, however, the data and resulting test statistics are often discrete. For example, rare genetic mutations lead to discrete test statistics in genetic association studies, where treating them as continuous can result in poor performance \cite{Neale2011}. Discrete tests also pose unique challenges: they tend to be conservative \cite{Berry}, and the resulting $p$-values are not uniformly distributed under the null.

In this paper, we consider a general setting where the input $p$-values are discrete and may arise from left-, right-, or two-sided tests under an arbitrary discrete null distribution (see Appendix \ref{sec:p_sided} for details on the definitions and distributions of left-, right-, and two-sided $p$-values). Specifically, we assume the possible values of a discrete $p$-value $P$ form a finite or countable, increasingly ordered set $0 = F_0 < F_1 < \ldots < F_i < \ldots \leq 1$ (with $F_\infty = 1$ if needed), and
\begin{equation}
	\label{eq:disc_P_left}
	P \in \{F_i : i \in \mathbb{N}\}, \quad \text{with} \quad \P(P = F_i) = p_i = F_i - F_{i-1}.
\end{equation}
The CDF of $P$ is $F(x) = \P(P \leq x) = F_i$ for $x \in (F_{i-1}, F_i]$, and the probability mass function (PMF) $p_i$ is non-negative and sums to 1. In general, the distribution of $P$ is not uniform. Throughout this paper, the index $i$ refers to the possible values of $P$, while the index $j$ in \eqref{eq:sumstat} refers to the independent tests and their corresponding $p$-values.

The combination procedure for global hypothesis testing presents additional challenges. For discrete $P_j$'s, obtaining the exact distribution of the statistics in Table~\ref{table:combstat} requires an $n$-fold convolution, which quickly becomes computationally infeasible as $n$ increases \cite{mielke2004combining}. Resampling-based methods can be used to approximate the null distribution of $T_n$, but they are computationally intensive and may lack accuracy for controlling $p$-values at small significance levels $\alpha$ \cite{routledge1997p}.


Several strategies have been proposed to address these challenges. One approach, the randomization strategy \cite{dickhaus2013randomized}, constructs a test variable with a uniform distribution by combining the discrete $p$-value with an independent random variable. This enables the use of continuous-case testing procedures \cite{pearson1950questions, ochieng2023multiple, dickhaus2024randomized}. However, randomized $p$-values do not satisfy Birnbaum’s admissibility criterion—meaning the same observed data can lead to different decisions \cite{birnbaum1954combining}—and, more importantly, they introduce additional randomness into the combination process \cite{rubin2019meta,stevens1950fiducial}. For these reasons, we do not consider randomization strategies in this work.

Another strategy is to adjust the observed test statistic in \eqref{eq:disc_P_left} to better approximate a continuous null distribution. A classic approach is the mid-$p$-value, defined as $\tilde{P} = \frac{F_i+F_{i-1}}{2}$ when $P=F_i$ \cite{lancaster1961significance}. Barnard \cite{barnard1989alleged} argued that any reasonable significance measure for $P=F_i$ should fall between $F_{i-1}$ and $F_i$. For Fisher's combination test, Lancaster \cite{lancaster1949combination} proposed mean-value and median-value $\chi^2$ adjustments for $-2\log (P_j)$ to better match the $\chi^2_2$ null distribution. More recently, Contador and Wu \cite{contador2023minimum} extended this idea using a Wasserstein distance framework, providing an optimal adjustment for Fisher's combination and showing that using a Gamma distribution with matched moments, instead of $\chi^2_{2n}$, yields improved type I error control and statistical power.


This paper makes the following contributions. First, we generalize the optimal transport-based framework of \cite{contador2023minimum} to a broad class of additive $p$-value combination statistics as in \eqref{eq:sumstat}. The approach consists of two main steps: (1) adjusting the discrete statistic by minimizing the Wasserstein distance to its continuous analogue, which ensures the first moments are matched; and (2) constructing a surrogate null distribution by matching the second moment within an appropriate parametric family, yielding asymptotically accurate Type I error control. We provide explicit formulas and implementation details for the classical combination methods in Table \ref{table:combstat}, and note that the framework applies to any strictly increasing $G$ in \eqref{eq:sumstat} with finite first and second moments.

Second, we provide practical guidelines for comparing and selecting among various $p$-value combination tests under discreteness. While previous comparative studies under the continuous case have focused mainly on statistical power \cite{heard2017choosing}, the discrete scenario introduces additional challenges, including computational efficiency and accurate Type I error control. We address these issues in detail. Specifically, we show that the accuracy of Type I error control in our framework is closely related to the Wasserstein distance between the adjusted discrete statistic and its continuous surrogate null distribution. We propose a simple, interpretable criterion based on the ratio of variances, which can be easily computed and used to guide the choice among different combination methods. Additionally, we characterize when a combination statistic is equivalent to a likelihood ratio test, and empirically show that the proposed approximation scheme preserves statistical power relative to the UMP test.


Third, we conduct extensive simulation studies to evaluate the performance of the proposed testing framework and to compare various $p$-value combination methods under discreteness. We illustrate the methodology with a case-control genetic association study. To facilitate practical implementation, we provide the R package \texttt{DPComb}, available on CRAN, which supports hypothesis testing with input statistics from common discrete distributions (including binomial, Poisson, negative binomial, hypergeometric, noncentral hypergeometric, geometric, and any user-specified discrete CDF). The package also includes tools for computing the variance-based criteria and other guidelines for selecting among $p$-value combination methods.


The remainder of the paper is organized as follows. Section \ref{sec:wass} introduces the Wasserstein distance as a metric for quantifying the difference between discrete and continuous random variables, describes the minimum distance adjustment of discrete $p$-values, and presents a moment-matching testing procedure with theoretical results on Type I error control. Section \ref{sec:t1e} provides practical guidelines for selecting a combination strategy based on the accuracy of Type I error control. Section \ref{sec:UMPU} compares the statistical power of different methods and discusses their relationship to UMP tests. Section \ref{sec:example} demonstrates the application of the proposed testing procedures using a case-control study in the context of a gene-based genetic association analysis of rare variants. Section \ref{sec:discussion} concludes with a discussion and directions for future research. Mathematical proofs of theorems and lemmas, additional numerical examples, figures, and supplementary information are provided in the Appendix.


\section{Testing Procedure} \label{sec:wass}

We extend the optimal transport-based framework of \cite{contador2023minimum} to the general family of additive $p$-value combination statistics in \eqref{eq:sumstat}. The procedure consists of two main steps:

\begin{enumerate}[leftmargin=2em]
	\item \textbf{Minimum Distance Adjustment:} Each discrete statistic $G^{-1}(P_j)$ (or $G^{-1}(1-P_j)$) is replaced by an adjusted value $Z_j$ that minimizes the Wasserstein distance to its continuous analogue, $G^{-1}(U)$, where $U \sim \text{Uniform}(0,1)$. This ensures the first moments are matched and yields the adjusted sum
	\begin{equation}
		\label{eq:sumstat_adj}
		S = \sum_{j=1}^n Z_j.
	\end{equation}
	\item \textbf{Moment-Matching Surrogate Distribution:} To approximate the null distribution of $S$, we construct a proper continuous surrogate distribution $\tilde{S}$:
\begin{equation}
	\label{eq:sumstat_adj2}
	\tilde{S} = \sum_{j=1}^n \tilde{Y}_j,
\end{equation}
where each $\tilde{Y}_j$ is drawn from a surrogate distribution with matched first and second moments with $Z_j$ within a suitable parametric family (e.g., Gamma or Normal). This surrogate is then used to compute the global $p$-value for testing the global null hypothesis.
\end{enumerate}

This two-step procedure preserves the discrete structure of the data while providing asymptotically accurate Type I error control for the global test.

\subsection{Minimum Distance Adjustment}

The Wasserstein distance is a metric for quantifying the difference between probability distributions, particularly suitable for comparing discrete and continuous distributions as it accounts for their underlying geometry. For two probability measures $\mu$ and $\nu$ on $\mathbb{R}$, the Wasserstein distance of order $2$ is defined as \cite{villani2003topics}:
\begin{equation}
	W_2(\mu,\nu)=\left(\inf_{\gamma \in C(\mu,\nu)}\int_{\mathbb{R}^2} |x-y|^2\,d\gamma (x,y)\right)^{1/2},
	\label{eq:wassgen}
\end{equation}
where $C(\mu, \nu)$ is the set of all couplings of $\mu$ and $\nu$, i.e., all joint distributions on $\mathbb{R}^2$ with marginals $\mu$ and $\nu$. In this context, the Wasserstein distance measures the ``cost'' of transforming a continuous distribution $\nu$ into the distribution of a transformation of a discrete $p$-value, such as $G^{-1}(P)$ or $G^{-1}(1-P)$ as defined in \eqref{eq:sumstat} and \eqref{eq:disc_P_left}, using the Euclidean metric.
We use $W_2(\mu,\nu)$ and $W_2(X,Y)$ interchangeably, where $X$ and $Y$ are random variables with distributions $\mu$ and $\nu$, respectively. The index $j$ in $P_j$ is omitted when clear from context.

For the statistic adjustment step, each component $G^{-1}(P)$ or $G^{-1}(1-P)$ is replaced by an adjusted statistic $Z$ that minimizes the Wasserstein distance to the continuous distribution $G^{-1}(U)$, where $U \sim \text{Uniform}(0,1)$. Specifically, $Z$ takes value $z_i$ whenever $P=F_i$, with $z_i$ chosen so that the distribution of $Z$ is as close as possible to $G$. Theorem \ref{thm:zopti} provides the explicit formula for $Z$. The proof is given in Appendix \ref{sec:proof}.

\begin{theorem}
	Consider a discrete $p$-value $P$ with distribution given in \eqref{eq:disc_P_left} and a continuous CDF $G$ with finite second moment. On the set where $P=F_i$, the modifications of $G^{-1}(P)$ and $G^{-1}(1-P)$ that are closest to $G$ under the metric defined in \eqref{eq:wassgen} take the respective values
	$$\frac{\int_{F_{i-1}}^{F_{i}} G^{-1}(w)dw}{F_i-F_{i-1}} \quad \text{and} \quad \frac{\int_{F_{i-1}}^{F_{i}} G^{-1}(1-w)dw}{F_i-F_{i-1}} .$$
	\label{thm:zopti}
\end{theorem}

Observe that, since $G^{-1}$ is strictly increasing, the adjusted statistic $Z$ is an increasing function of $P$ when adjusting $G^{-1}(P)$, and a decreasing function of $P$ when adjusting $G^{-1}(1-P)$. Thus, evidence against the null corresponds to small values of an adjustment of $G^{-1}(P)$ or large values of an adjustment of $G^{-1}(1-P)$. The explicit forms of $Z$ for each method in Table \ref{table:combstat}, evaluated at $P=F_i$, are given in Table \ref{table:zstat}.

\begin{table}[h!]
	\caption{Adjusted $Z$ statistics for the tests in Table \ref{table:combstat} under the discrete $p$-value distribution in \eqref{eq:disc_P_left}. Notation: $\overline{F_i} \equiv 1-F_i$, $K_i \equiv (2\pi)^{-1/2} \exp[-(\Phi^{-1}(F_i))^2/2]$, where $\Phi^{-1}$ is the standard normal quantile function.}
	\centering
	\begin{tabular}{l c c }
		\toprule
		Method & $Z$ Statistic & $z_i$ value when $P=F_i$\\ 
		\midrule
		Fisher     & $Z_F$ & $2 - 2(F_i - F_{i-1})^{-1} (F_i \log F_i - F_{i-1} \log F_{i-1})$ \\
		Pearson    & $Z_P$ & $2 - 2(F_i - F_{i-1})^{-1} (\overline{F_{i-1}} \log \overline{F_{i-1}} - \overline{F_i} \log \overline{F_i})$ \\
		George     & $Z_G$ & $(Z_P - Z_F)/2$ \\
		Stouffer   & $Z_S$ & $(F_i - F_{i-1})^{-1} (K_{i-1} - K_i)$ \\
		Edgington  & $Z_E$ & $(F_i + F_{i-1})/2$ \\
		\bottomrule
	\end{tabular}
	\label{table:zstat}
\end{table}


\subsection{Moment Matching}\label{sec:appissues}

The adjusted $Z$ statistic is the minimum distance approximation to the continuous distribution of $Y=G^{-1}(U)$. By Theorem \ref{thm:zopti}, each $z_i$ is a conditional expectation, so $Z$ is unbiased: $\E(Z)=\E(Y)$. However, the variance of $Z$ is always less than that of $Y$, as stated in Lemma \ref{lem:lowervariance}.

\begin{Lemma}\label{lem:lowervariance}
	Consider $Y$ and $Z$ as in Theorem \ref{thm:zopti}. Then $$\Var(Y)=\Var(Z)+W_2^2(Z,Y)>\Var(Z).$$
\end{Lemma}

This variance reduction explains why using the distribution of $Y$ directly in a testing procedure can be conservative. As discussed in Section \ref{sec:t1e}, the extent of this discrepancy depends on both the choice of $G$ and the CDF $F$ of the discrete $p$-value.

For Fisher's combination of discrete $p$-values, we proposed a further adjustment of the null distribution, from  $Y$ (i.e., the $\chi^2_2$ distribution) to a gamma distributed $\tilde{Y}$ with parameters chosen to minimize the Wasserstein distance to $Z$ within the family of gamma distributions (c.f. \cite{contador2023minimum} Theorem 2). The choice of gamma distribution is because it is a more general distribution covering the chi-squared distribution, and is closed under sum of i.i.d. random variables when the scale parameter is fixed. This Wasserstein-optimal adjustment garantees matching the first two moments. We extend this strategy to general scenario -- for a given $Z$, we find $\tilde{Y}$ under proper distribution with the first two moments matched: $\E (\Tilde{Y})=\E(Z)$ and $\Var(\Tilde{Y})=\Var(Z)$. 

For computational convenience and accuracy, we propose to choose $\tilde{Y}$ from a distributional family that contains the original null distribution $Y$ and is preferably closed under convolution. For Fisher's and Pearson's combinations, the gamma distribution is used. For Stouffer's combination, the normal distribution is exact. For George's and Edgington's combinations, the continuous counterparts are the logistic and uniform distributions, respectively. However, neither the logistic nor uniform families are closed under convolution, but the sum of independent random variables from these distributions is well approximated by a normal distribution, even for moderate $n$ \cite{hoyt1968teacher, marengo2017geometric, epps2005tests}. Therefore, we use the normal approximation for these cases: $\tilde{Y} \sim \mathcal{N}(0, \nu_G)$ for George's method and $\tilde{Y} \sim \mathcal{N}(0.5, \nu_E)$ for Edgington's method. The choices of $\tilde{Y}$ are summarized in Table \ref{table:variances}.

\begin{table}[h!]
	\caption{Variance and surrogate testing distribution for each discrete statistic. Notation: $\overline{F_i} \equiv 1-F_i$, $O_i = F_i/\overline{F_i}$, $K_i = (2\pi)^{-1/2} \exp[-(\Phi^{-1}(F_i))^2/2]$, where $\Phi^{-1}$ is the standard normal quantile function.}
	\centering
	\begin{tabular}{l c c}
		\toprule
		Method & Variance of $Z$ & Surrogate $\tilde{Y}$ \\ 
		\midrule
		Fisher  &  $\nu_F = 4\sum \frac{\left( F_i\log F_i - F_{i-1}\log F_{i-1} \right)^2}{F_i-F_{i-1}}$ & $\text{Gamma}(4/\nu_F, \nu_F/2)$ \\
		Pearson &  $\nu_P = 4\sum \frac{\left( \overline{F_{i-1}}\log \overline{F_{i-1}} - \overline{F_i}\log \overline{F_i} \right)^2}{F_i-F_{i-1}}$ &  $\text{Gamma}(4/\nu_P, \nu_P/2)$ \\
		George  & $\nu_G = \sum \frac{\left[ F_i\log O_i - F_{i-1}\log O_{i-1} + \log(\overline{F_{i-1}}^{-1} \overline{F_i}) \right]^2}{F_i-F_{i-1}}$ & $\approx \mathcal{N}(0, \nu_G)$ \\ 
		Stouffer & $\nu_S = \sum \frac{[K_i - K_{i-1}]^2}{F_i - F_{i-1}}$ & $\mathcal{N}(0, \nu_S)$ \\ 
		Edgington &  $\nu_E = \sum \frac{F_i F_{i-1} (F_i - F_{i-1})}{4}$ &  $\approx \mathcal{N}(0.5, \nu_E)$ \\ 
		\bottomrule
	\end{tabular}
	\label{table:variances}	
\end{table}

Accordingly, instead of using $\sum_{j=1}^n Y_j$, we approximate the distribution of $S = \sum_{j=1}^n Z_j$ in \eqref{eq:sumstat_adj} by $\tilde{S} = \sum_{j=1}^n \tilde{Y}_j$, where $\tilde{Y}_j$ is drawn from the surrogate distribution with matched first and second moments. This moment-matching approximation provides asymptotically correct Type I error control as $n$ increases, as stated in Lemma \ref{lem:asymptoticpvalue}.

\begin{Lemma} \label{lem:asymptoticpvalue}
	Let $\{P_j\}_{j \in \mathbb{N}}$ be a sequence of i.i.d. discrete random variables with distribution in \eqref{eq:disc_P_left}. Let $G$ be a continuous CDF in a scale family $\mathcal{G}$ such that $Y_j \overset{i.i.d.}{\sim} G$ have moments of order two. Define $Z_j$ as the minimum distance approximations of $G^{-1}(P_j)$ (or $G^{-1}(1-P_j)$) to $Y_j$ given by Theorem \ref{thm:zopti}, and define $S_{n,G}=\sum_{j=1}^n Z_j$. Suppose that there exists an i.i.d. sequence with distribution in $\mathcal{G}$, $\tilde{Y_j}$, such that $\E (\Tilde{Y_j})=\E(Z_j)=\int_{0}^{1}G^{-1}(w)dw$ and $\Var(\tilde{Y}_j)=\Var(Z_j)$, and denote by $q_{p,n,\tilde{Y}}$ the $p$ quantile of the distribution of $\tilde{S}_n = \sum_{j=1}^n \Tilde{Y}_j$. For any $p \in (0,1)$, 
	$$
	\lim_{n \rightarrow \infty} \P(S_{n,G} < q_{p,n,\tilde{Y}})=p.
	$$
\end{Lemma}

Observing that all surrogate distributions in Table \ref{table:variances} have second-order moments matched to their corresponding discrete statistics, we summarize the continuous approximations for the combination statistics in Table \ref{table:zstat}. Under the global null \eqref{eq:globalNull}, these approximations provide asymptotically accurate Type I error control:
\begin{subequations}
	\label{eq:combinations}
	\begin{align}
		\label{eq:sumF}
		S_{F,n} = \sum_{j=1}^n Z_{F,j} & \approx \text{Gamma}(4n/\nu_F, \nu_F/2); \\
		\label{eq:sumP}
		S_{P,n} = \sum_{j=1}^n Z_{P,j} & \approx \text{Gamma}(4n/\nu_P, \nu_P/2); \\
		\label{eq:sumS}
		S_{S,n} = \sum_{j=1}^n Z_{S,j} & \approx \mathcal{N}(0, n\nu_S); \\
		\label{eq:sumG}
		S_{G,n} = \sum_{j=1}^n Z_{G,j} & \approx \mathcal{N}(0, n\nu_G); \\
		\label{eq:sumE}
		S_{E,n} = \sum_{j=1}^n Z_{E,j} & \approx \mathcal{N}(n/2, n\nu_E).
	\end{align}
\end{subequations}

The distributions in \eqref{eq:sumF}, \eqref{eq:sumP}, and \eqref{eq:sumS} correspond to sums of independent Gamma or Gaussian variables, while \eqref{eq:sumG} and \eqref{eq:sumE} use normal approximations as described above. At a global significance level $\alpha \in (0,1)$, evidence against the global null $H_0$ in \eqref{eq:globalNull} (i.e., small input $p$-values $P_1, \ldots, P_n$) results in Fisher’s statistic in \eqref{eq:sumF} exceeding the $1-\alpha$ quantile of its surrogate Gamma distribution, and the statistics in \eqref{eq:sumP}, \eqref{eq:sumS}, \eqref{eq:sumG}, and \eqref{eq:sumE} falling below the $\alpha$ quantile of their respective surrogate distributions. Thus, to control the significance level at $\alpha$, we use the $1-\alpha$ quantile for Fisher’s combination test and the $\alpha$ quantile for the others.


\subsection{Combining Non-Identically Distributed $p$-Values}\label{sec:nonid}

We now address the case where the $p$-values are independent but not identically distributed. Suppose $\{P_j\}_{j=1}^n$ are independent discrete $p$-values, each with its own distribution $F^{(j)}$ as in \eqref{eq:disc_P_left}. For a strictly increasing CDF $G$ with finite second moments, let $Z_{G,j}$ denote the minimum distance adjustment of $G^{-1}(P_j)$ (or $G^{-1}(1-P_j)$) as in Theorem \ref{thm:zopti}, and let $\nu_{G,j} = \Var(Z_{G,j})$. 

To construct the surrogate null distribution for the sum statistic, we match the first two moments: the mean is $n\E(G^{-1}(U))$ and the variance is $\sum_{j=1}^n \nu_{G,j}$. The surrogate distribution is chosen from the same family as in the i.i.d. case, but with parameters reflecting the average or total variance across all $p$-values. Lemma \ref{thm:asymptoticpvaluenoniid} below extends the asymptotic Type I error control result to this non-identically distributed setting under mild regularity conditions.

\begin{Lemma} \label{thm:asymptoticpvaluenoniid}
	Let $\{P_j\}_{j \in \mathbb{N}}$ be a sequence of independent, discrete $p$-values, each with distribution $F^{(j)}$. Let $G$ be a continuous CDF in a scale family $\mathcal{G}$ such that $Y_j \overset{i.i.d.}{\sim} G$ have moments of order two. Define $Z_j$ as the minimum distance approximation of $G^{-1}(P_j)$ (or $G^{-1}(1-P_j)$) to $Y_j$ given by Theorem \ref{thm:zopti}, and define $T_{n,G}=\sum_{j=1}^n Z_j$. Suppose that there exists an independent sequence with distribution in $\mathcal{G}$, $\tilde{Y}_j$, such that $\E (\tilde{Y}_j)=\E(Z_j)=\int_{0}^{1}G^{-1}(w)dw$ and $\Var(\tilde{Y}_j)=\Var(Z_j)$, and denote $q_{p,n,G}$ as the $p$ quantile of the distribution of $S_n = \sum_{j=1}^n \tilde{Y}_j$. Suppose further that the parameter sequence $\nu_{j}=\Var(Z_{j})$ satisfies the Lyapunov condition, i.e., 
	$$
	\lim_{n \to \infty} \frac{\sum_{j=1}^n\E(|Z_j-\E(Y_j)|^{2+\delta})}{(\sum_{j=1}^n \nu_j )^{1+\delta/2}}=0, \quad \text{ for some }\delta>0.
	$$
	Then, for any $p \in (0,1)$,
	$$
	\lim_{n \rightarrow \infty} \P(S_n < q_{p, n,G})=p.
	$$
\end{Lemma}

In accordance with Lemma \ref{thm:asymptoticpvaluenoniid}, for each combination method, we propose using the same surrogate distributions as in \eqref{eq:combinations}, but with the variance parameter replaced by the average (or sum) of the method-specific variances across the distributions $F^{(j)}$, $j = 1, \ldots, n$.

Lemmas \ref{lem:asymptoticpvalue} and \ref{thm:asymptoticpvaluenoniid} ensure that comparing the values of the adjusted statistics with the quantiles of their proposed surrogate distributions will produce asymptotically accurate type error control as $n$ increases. For practical applications with small $n$, however, it is important to assess and compare the accuracy of Type I error control. We address this in the next section.

\section{Type I Error Control}\label{sec:t1e}

Lemma \ref{lem:asymptoticpvalue} establishes that, under the null hypothesis $H_0$ in \eqref{eq:globalNull}, all combination methods in \eqref{eq:combinations} provide asymptotically correct Type I error control as the number of combined $p$-values $n$ increases. However, for finite $n$, the accuracy of Type I error control becomes a key factor in selecting among different testing procedures. This accuracy is influenced by both the distribution of the discrete $p$-values and the properties of the chosen combination statistic. In this section, we introduce practical and interpretable metrics to guide this choice, discuss the relative performance of different methods under various $p$-value distributions, and present examples and simulation results to demonstrate the effectiveness of the proposed procedures and the validity of the selection criteria.

The testing procedures use the distribution of $\tilde{S} = \sum_{j=1}^n \tilde{Y}_j$ in \eqref{eq:sumstat_adj2} as an approximation to the null distribution of $S = \sum_{j=1}^n Z_j$ in \eqref{eq:sumstat_adj}. Thus, the accuracy of Type I error control depends on how closely the adjusted discrete statistic $Z$ (from Theorem \ref{thm:zopti}) matches the continuous surrogate $\tilde{Y}$ in distribution. We recommend two practical metrics for quantifying this accuracy: the scaled Wasserstein distance $W_2(Z,\tilde{Y})/\SD(Y)$ and the variance ratio $\Var(Z)/\Var(Y)$. For the non-i.i.d. case, we suggest using the average variance ratio, $\sum_{j=1}^n \Var(Z_j)/(n\Var(Y))$.

A testing procedure with a lower value of $W_2(Z,\tilde{Y})/\SD(Y)$ is preferable, indicating higher accuracy of Type I error control. Since the Wasserstein distance is not scale-invariant, we scale it by the standard deviation. Specifically, $W_2(aZ, a\tilde{Y}) = |a| W_2(Z, \tilde{Y})$ for any constant $a \in \mathbb{R}$. We use $\SD(Y)$ for scaling because it is easy to compute and interpret as the standard deviation under the continuous case, providing a meaningful reference for comparison across different statistics.



	To further understand the accuracy of the testing procedures, we study a scaled lower bound for the Wasserstein distance between the adjusted statistic $Z$ and its continuous surrogate distribution $\tilde{Y}$ (see Appendix Section~\ref{sec:proof} for its proof): 
	\begin{equation}
		\label{eq:lowerbound}
		W_2(Z,\tilde{Y})/\SD(Y) \geq \SD(Y)^{-1}\sqrt{\max_{i \in \mathbb{N}}\int_{\tilde{G}^{-1}(\P(Z<z_i))}^{\tilde{G}^{-1}(\P(Z\leq z_i))} |z_i-y|^2\tilde{g}(y)dy }.
	\end{equation} 

The lower bound reveals how the discrepancy depends on the domain and the integrand of the integral, which are related to the $\tilde{G}$ distribution and its corresponding density $\tilde{g}$, the $z_i$ values, and their respective probabilities $\P(Z=z_i) = \P(Z\leq z_i)-\P(Z<z_i)$, corresponding to the probability masses of the $p$-value. In particular, the integral is larger when $z_i$ is larger with a higher probability at the long tail of $\tilde{G}$ (so the domain of the integral is larger). This provides insights regarding the advantages and disadvantages of different combination methods under different distributions of the discrete $p$-values. For instance, for Fisher's combination test, since $\tilde{G}$ is a Gamma distribution (always skewed to the right), the integral is large when the discrete $p$-value $P$ in \eqref{eq:disc_P_left} has a large mass at small values (note the $z_i$ value here corresponds to $G^{-1}(1-P)$ with $G(x)=1-e^{-x/2}$, so that small $P$ correspond to larger $z_i$). In this case, Fisher's combination test by \eqref{eq:sumF} will have a larger lower bound of the ratio, indicating a poor finite-sample fit and thus poor Type I error control. On the contrary, with $\tilde{G}$ also being a Gamma distribution, Pearson's combination test, which uses $G^{-1}(P)$ to get $z_i$, will have a large lower bound when $P$ has a large mass at large values (i.e., close to one), leading to a poor finite-sample fit and Type I error control. George's, Stouffer's, and Edgington's proposed surrogates, based on the normal distribution, have a density $\tilde{g}(y)=O(e^{-y^2/2})$ based on the normal distribution with faster exponential decay, making the integrands at both tails of the $P$ distribution smaller than their gamma counterparts. The numerical results below evidence such patterns of performance.

	While $W_2(Z,\tilde{Y})$ can be computed or approximated, a particularly simple and practical metric for selecting the best combination method is the variance ratio $\Var(Z)/\Var(Y)$. A higher variance ratio indicates that $Z$ and $\tilde{Y}$ are closer in distribution, leading to more accurate Type I error control in finite samples. This metric is easy to compute, always lies between 0 and 1 (since $\Var(Z) \leq \Var(Y)$ by Lemma \ref{lem:lowervariance}), and directly reflects the variance reduction of the adjusted statistic. Moreover, Lemma \ref{lem:vardifference} shows that, for all testing methods proposed in \eqref{eq:combinations} $W_2(Z,\tilde{Y})$ is bounded above by a constant times $\sqrt{\Var(Y)-\Var(Z)}$ (a result whose proof can easily be extended to $\tilde{Y}$ in a location-scale family), so maximizing the variance ratio is equivalent to minimizing this upper bound. Thus, both a higher variance ratio and a smaller $W_2(Z,\tilde{Y})$ indicate that the distributions of $Z$ and $\tilde{Y}$ are closer, resulting in better Type I error control.

	\begin{Lemma}
		\label{lem:vardifference}
		Let $Z$ be any of the statistics in Table \ref{table:zstat}, $Y$ its distribution under continuity and $\tilde{Y}$ its surrogate distribution described in \eqref{eq:combinations}. It holds that, for some $C>0$ not depending on the distribution of $P$, 
		
		$$	W_2(Z,\tilde{Y})\leq C \sqrt{\Var(Y)-\Var(Z)}.$$
	\end{Lemma}

We illustrate the practical use of these metrics with four representative discrete $p$-value distributions, each with a large probability mass concentrated at a different location:

\begin{align*}
	P_L: \quad& F_1-F_0=0.4;\quad F_i-F_{i-1}=0.01 \text{ for } i=2,\ldots,61;\\
	P_R: \quad& F_{61}-F_{60}=0.4;\quad F_i-F_{i-1}=0.01 \text{ for } i=1,\ldots,60;\\
	P_C: \quad& F_{31}-F_{30}=0.4;\quad F_i-F_{i-1}=0.01 \text{ for } i=1,\ldots,61,\ i\neq 31;\\
	P_S: \quad& F_1-F_0=F_{42}-F_{41}=0.3;\quad F_i-F_{i-1}=0.01 \text{ for } i=2,\ldots,41.
\end{align*}

\begin{figure}[h!]
	\centering
	\includegraphics[width=\textwidth]{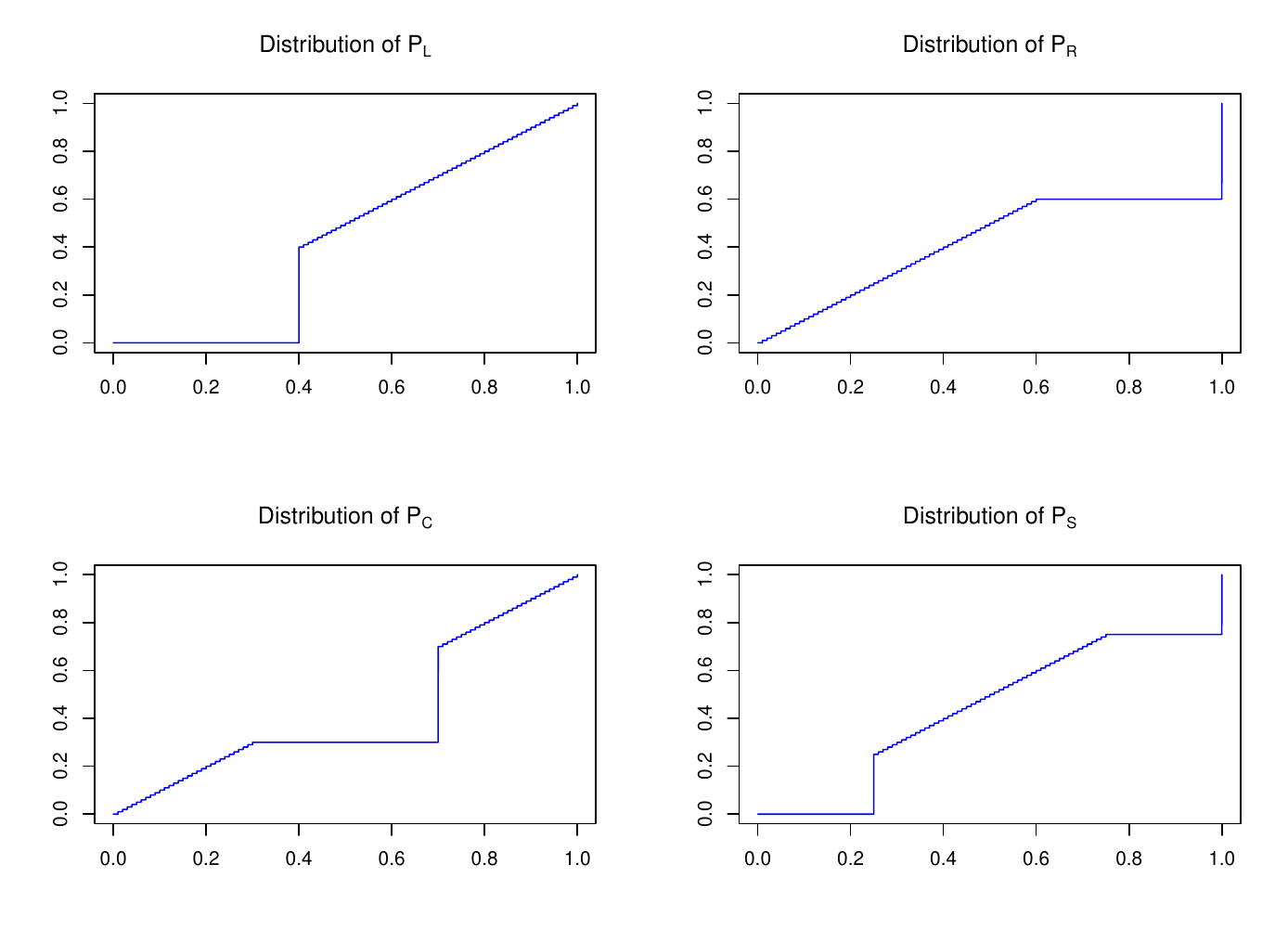}
	\caption{Cumulative distribution functions of $P_L$ (large mass at $0.4$, top left), $P_R$ (large mass at $1$, top right), $P_C$ (large mass at $0.7$, bottom left), and $P_S$ (large masses at both ends, bottom right).}
	\label{fig:qoaexample}
\end{figure}

Here, $P_L$ has a large mass ($0.4$) at the leftmost point, $P_R$ at the rightmost point, $P_C$ at the center, and $P_S$ has two large masses ($0.3$ each) at both extremes. Figure~\ref{fig:qoaexample} shows the cumulative distribution functions for these four cases.

	Table \ref{tab:exqoa} summarizes the variances of $Z$, scaled Wasserstein distances $W_2(Z,\tilde{Y})/\SD(Y)$, and variance ratios $\Var(Z)/\Var(Y)$ for the combination methods in \eqref{eq:combinations} across the four representative $p$-value distributions. As expected from the lower bound discussion, Fisher's statistic performs best (highest ratio, lowest distance) for $P_R$ and worst for $P_L$, while Pearson's statistic shows the opposite pattern. Edgington's variance ratios are identical for $P_L$, $P_R$, and $P_C$, but the actual distances differ, illustrating that the variance ratio, while easy to compute, may not fully capture distributional differences.
	
	\begin{table}[!h]
		\caption{Variance, normalized distance, and variance ratio of minimum distance statistics for each approximation in Table \ref{table:zstat} and each of the $p$-value distributions $P_L$, $P_R$, $P_C$, and $P_S$. For each $p$-value, the highest ratio and lowest normalized distance are bolded; the lowest ratio and highest normalized distance are italicized.
		} 
		\centering
		\begin{tabular}{l c c c c c c}
			\toprule
			$p$-value & Metric & Fisher & Pearson & Stouffer & Edgington & George \\ 
			\midrule
			\multirow{3}{*}{$P_L$} & $\Var(Z)$ & 2.4   & 3.922 & 0.874 & 0.077 & 2.771 \\
			& $W_2(Z,\tilde{Y})/\SD(Y)$ &  \textit{0.469}   & \textbf{0.139}  & 0.337 & 0.379 & 0.337\\
			& $\Var(Z)/\Var(Y)$ & \textit{0.6}   & \textbf{0.98}  & 0.874 & 0.936 & 0.842  \\  
			\midrule
			\multirow{3}{*}{$P_R$} & $\Var(Z)$ & 3.922 & 2.4   & 0.874 & 0.077 & 2.771 \\
			& $W_2(Z,\tilde{Y})/\SD(Y)$ & \textbf{0.139}    & \textit{0.469}  & 0.337 & 0.379 & 0.337 \\
			& $\Var(Z)/\Var(Y)$ & \textbf{0.98}   & \textit{0.6} & 0.874 & 0.936 & 0.842 \\ 
			\midrule
			\multirow{3}{*}{$P_C$} & $\Var(Z)$ & 3.864 & 3.864 & 0.962 & 0.077 & 3.178 \\
			& $W_2(Z,\tilde{Y})/\SD(Y)$ & \textbf{0.182} & \textbf{0.182}  & 0.191 & \textit{0.27} & 0.207 \\
			& $\Var(Z)/\Var(Y)$ & \textbf{0.966} & \textbf{0.966} & 0.962 & \textit{0.936} & \textbf{0.966} \\ 
			\midrule
			\multirow{3}{*}{$P_S$} & $\Var(Z)$ & 2.787  & 2.787  & 0.841& 0.078  & 2.578 \\
			& $W_2(Z,\tilde{Y})/\SD(Y)$ & \textit{0.446}   & \textit{0.446}   & \textbf{0.36} & 0.399 & 0.373 \\
			& $\Var(Z)/\Var(Y)$ & \textit{0.696} & \textit{0.696} & 0.841 & \textbf{0.945} & 0.784 \\
			\bottomrule
		\end{tabular}
		\label{tab:exqoa}
	\end{table}

	\begin{center}
		\begin{figure}[h!] 
			\includegraphics[width=0.5\textwidth]{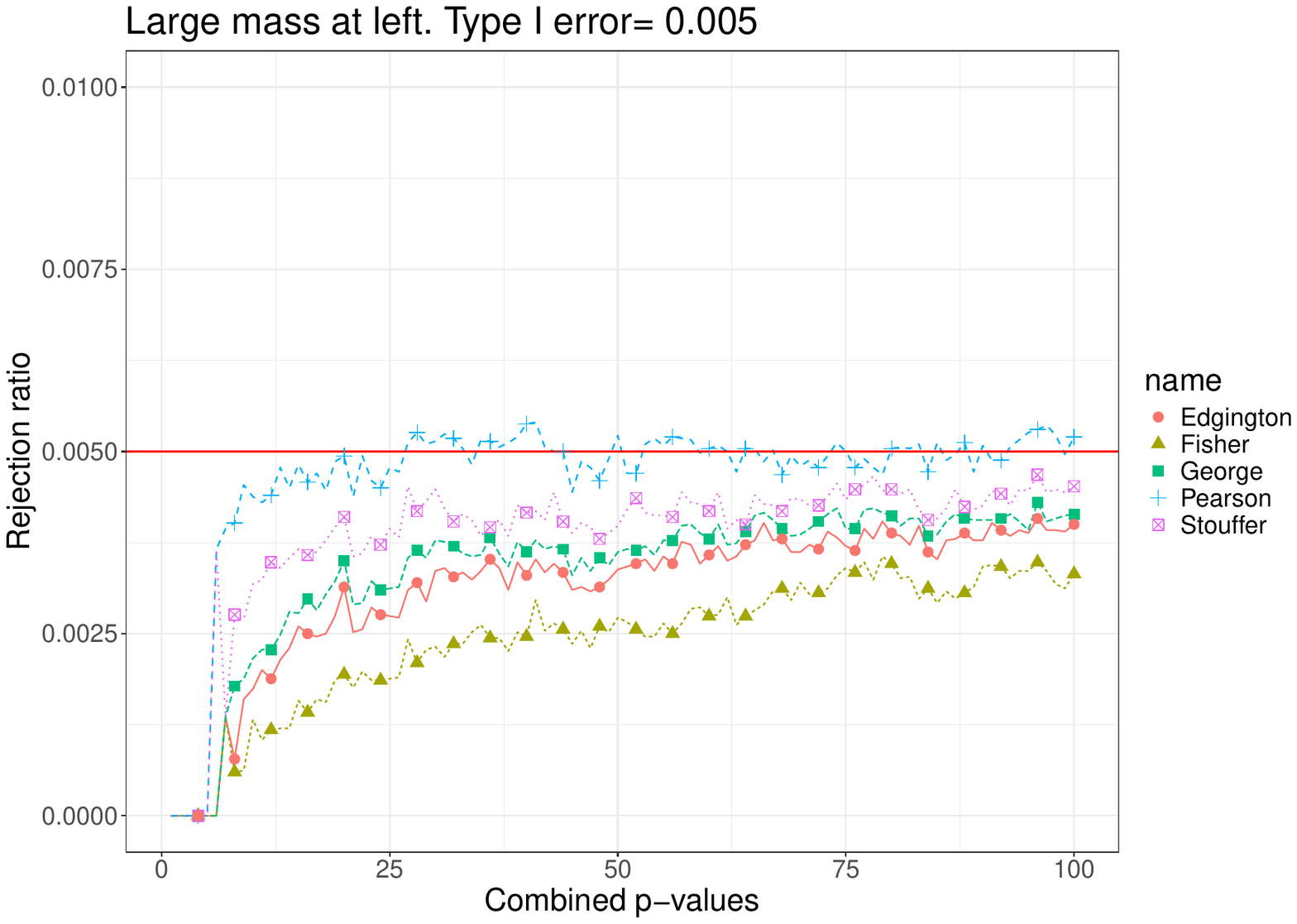} 
			\includegraphics[width=0.5\textwidth]{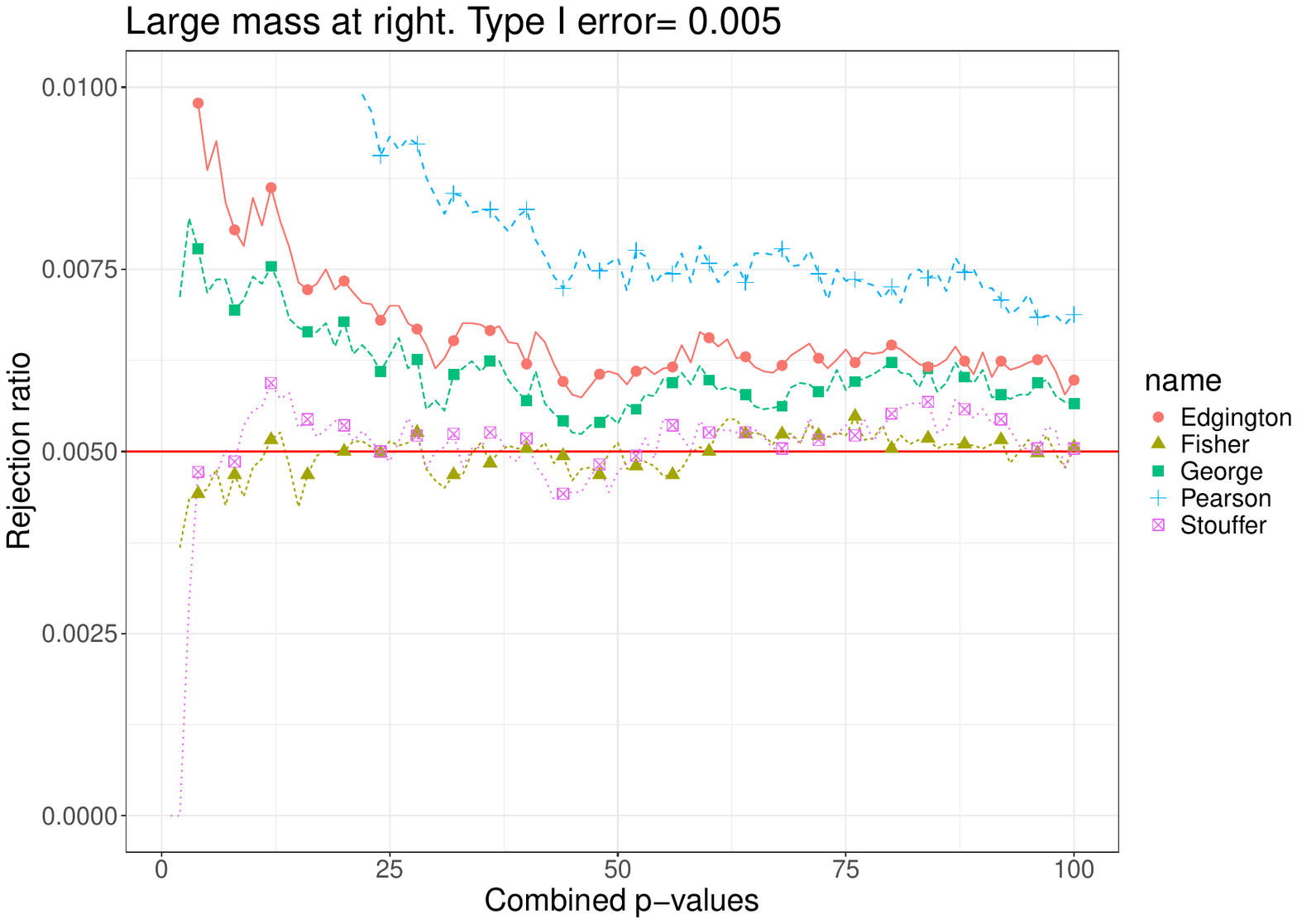} 
			\includegraphics[width=0.5\textwidth]{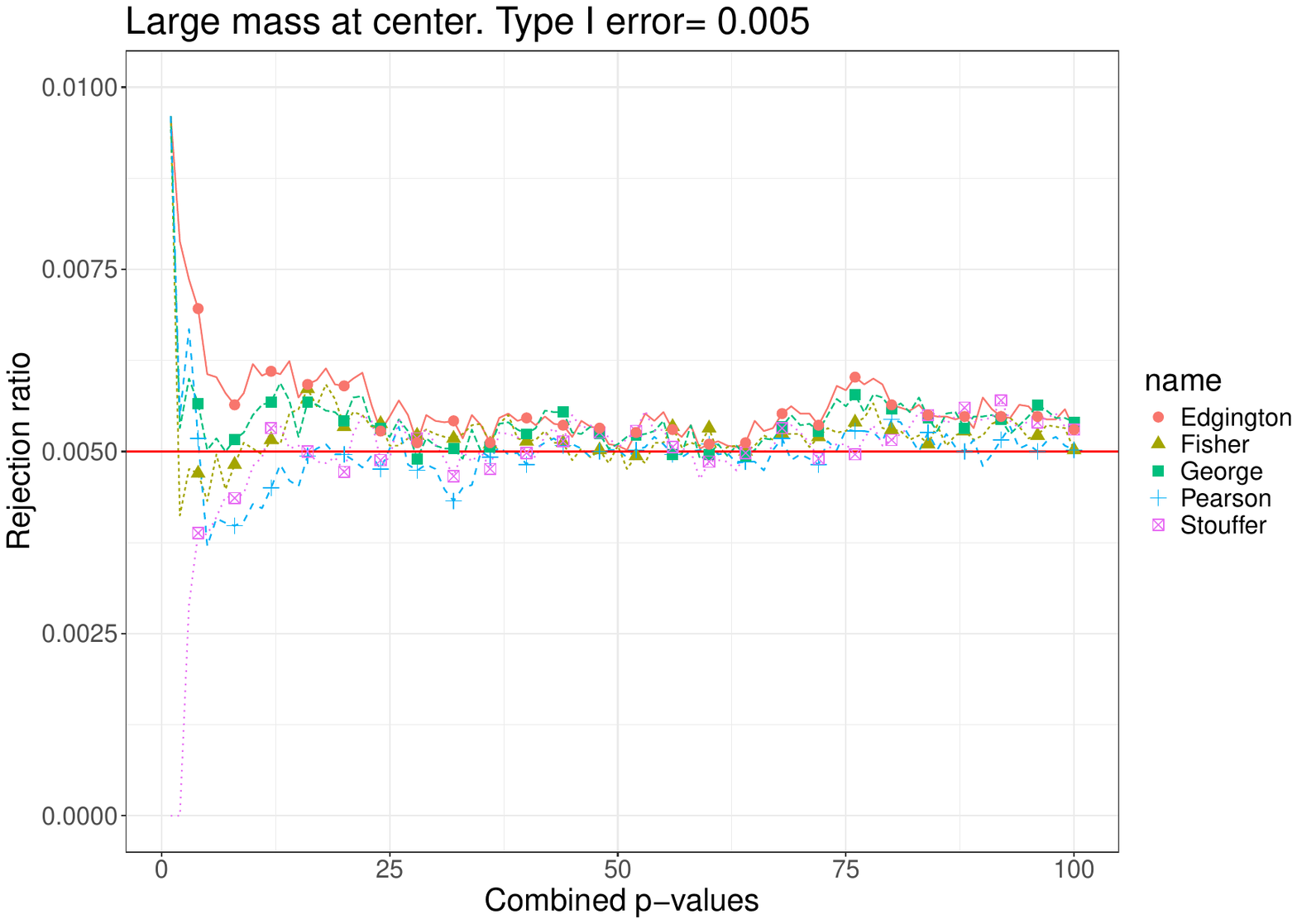} 
			\includegraphics[width=0.5\textwidth]{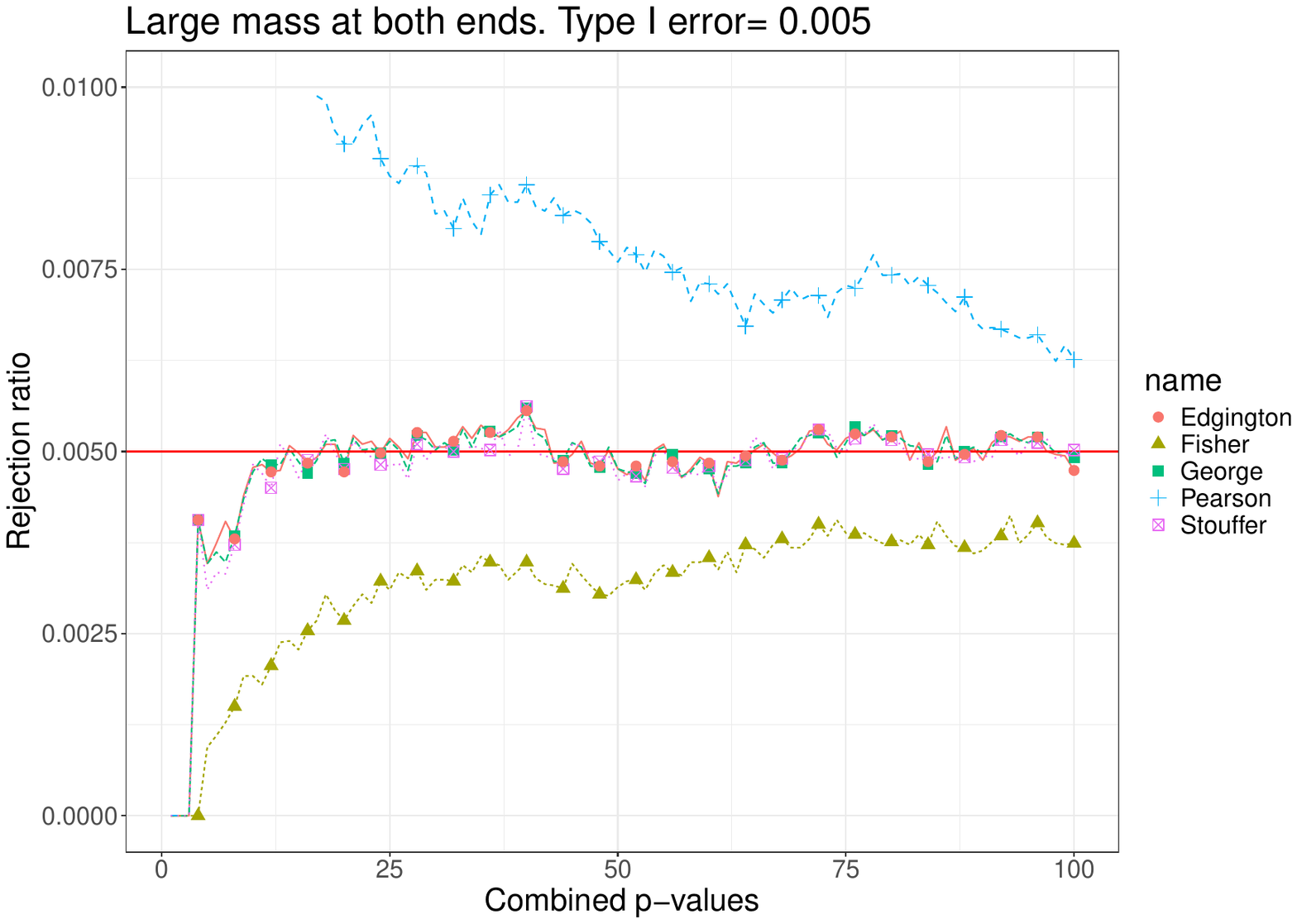} \\
			\caption{Empirical Type I error rates (proportion of rejections) at nominal $\alpha = 0.005$ as a function of the number of combined $p$-values $n$ (x-axis), for $P_L$, $P_R$, $P_C$, and $P_S$ (corresponding to large probability mass at the left, right, center, and both ends of the $p$-value distribution, respectively). Results are based on 50,000 simulation replicates.}
			\label{fig:exqoa005}
		\end{figure}
	\end{center}

We conduct a simulation study to demonstrate the accuracy of Type I error control for the testing procedures in \eqref{eq:combinations}. We combine up to $n=100$ i.i.d. $p$-values with null distributions given by $P_L$, $P_R$, $P_C$, or $P_S$ under the global null hypothesis. At significance level $\alpha = 0.005$, Figure \ref{fig:exqoa005} shows the empirical Type I error rates. Several observations follow. First, as expected and consistent with Lemma \ref{lem:asymptoticpvalue}, all combination methods provide asymptotically consistent Type I error control, with the empirical error rate converging to the nominal $\alpha$ as $n$ increases. Second, the finite-sample accuracy of Type I error control aligns with the metrics in Table \ref{tab:exqoa}. For example, for $P_L$ (large mass at small $p$-values), Pearson's statistic achieves the most accurate Type I error control, while Fisher's is the least accurate, matching their respective lowest scaled Wasserstein distance and highest variance ratio (and vice versa for Fisher). For $P_R$ (large mass at large $p$-values), Fisher performs best and Pearson worst, again consistent with the metrics. For $P_C$ (large mass at medium $p$-values), all methods perform similarly, with Pearson and Fisher slightly better and Edgington slightly worse, as indicated by the metrics. For $P_S$ (large masses at both ends), Fisher and Pearson perform worst, while the other three statistics perform better, again matching the variance ratio and distance metrics. Similar results are observed for $\alpha=0.001$, as shown in Figure \ref{fig:exqoa001} in the Appendix.


\section{The UMP test and Power Comparisons} \label{sec:UMPU}

Another important question is which $p$-value combination method yields higher statistical power for a given dataset. For continuous $p$-values, it is known \cite{heard2017choosing} that, under certain alternatives for the $p$-value distribution, some combination statistics are equivalent to the likelihood ratio test (LRT) and thus, by the Neyman-Pearson Lemma (Theorem 8.3.17 in \cite{casella2002statistical}), are uniformly most powerful (UMP): no other method with the same type I error control can achieve higher power.

Here, we extend this reasoning to the combination of discrete $p$-values. The discrete setting presents additional challenges: the exact null distributions of combination statistics are often unavailable or computationally infeasible, and our proposed procedures rely on surrogate distribution approximations. We therefore consider two scenarios: (1) cases where the LRT distribution is known (e.g., geometric data), and (2) cases where the LRT distribution is unknown but can be well approximated (e.g., circular data). We use simulation studies to compare empirical power, illustrating how the surrogate-based procedures closely match the UMP test when the combination statistic is monotonic in the LRT.

\subsection{Known LRT Distribution: Geometric Data}\label{sec:geom}
We first consider the scenario where the null distribution of the LRT, which is UMP, is explicitly available—using the geometric distribution as an example. In this setting, we derive the UMP $p$-value combination tests and compare their power with all surrogate distribution-based testing procedures defined in \eqref{eq:combinations}.  ss

Specifically, let $X$ be a geometric random variable with probability mass function $p(1-p)^{x-1}$, $x \in \mathbb{N}$, and consider the hypotheses
\begin{equation}
	\label{eq:H0A_geometric}
	H_0: p=p_0 \quad \text{and} \quad H_A: p=p_1 \neq p_0.
\end{equation}
The likelihood ratio for this test is
$$
\mathcal{L} =\left(\frac{p_0(1-p_1)}{p_1(1-p_0)}\right)\left(\frac{1-p_0}{1-p_1}\right)^X,
$$ 
which is a monotonic function of $X$ (increasing if and only if $p_0 < p_1$), and thus the test is UMP.

The geometric CDF under the null is $F_0(x)= 1-(1-p_0)^x$, $x\in \mathbb{N}$. The right-sided $p$-value associated with $X$ is $P=1-F_0(X-1)=(1-p_0)^{X-1}$. Rearranging gives $X=1+\log(P)/\log(1-p_0)$, so the likelihood ratio is a monotonic function of Fisher's statistic, making Fisher's combination the most powerful in this case. Conversely, the left-sided $p$-value is $P=F_0(X)=1-(1-p_0)^{X}$, which rearranges to $X=\log(1-P)/\log(1-p_0)$, so the likelihood ratio is a monotonic function of Pearson's statistic, making Pearson's combination most powerful for the left alternative.

We conduct simulation studies to compare the performance of the surrogate distribution procedures for each of the proposed combination methods in \eqref{eq:combinations} under both i.i.d. and non-i.i.d. settings.

\subsubsection{I.I.D. Geometric Data}\label{sec:iidgeom}

We consider $n=1000$ independent geometric random variables with the same null parameter $p_0=0.5$ and alternative values $p_1 \in (0.4, 0.6)$. In this setting, the rejection region for the UMP test is based on the quantiles of the negative binomial distribution, while the rejection regions for the other statistics are based on the quantiles of their surrogate distributions, depending on whether $p_1 < p_0$ or $p_1 > p_0$. See Appendix \ref{sec:more} for the specific rejection regions.

\begin{center}
		\begin{figure}[h!] 
			\includegraphics[width=1\textwidth]{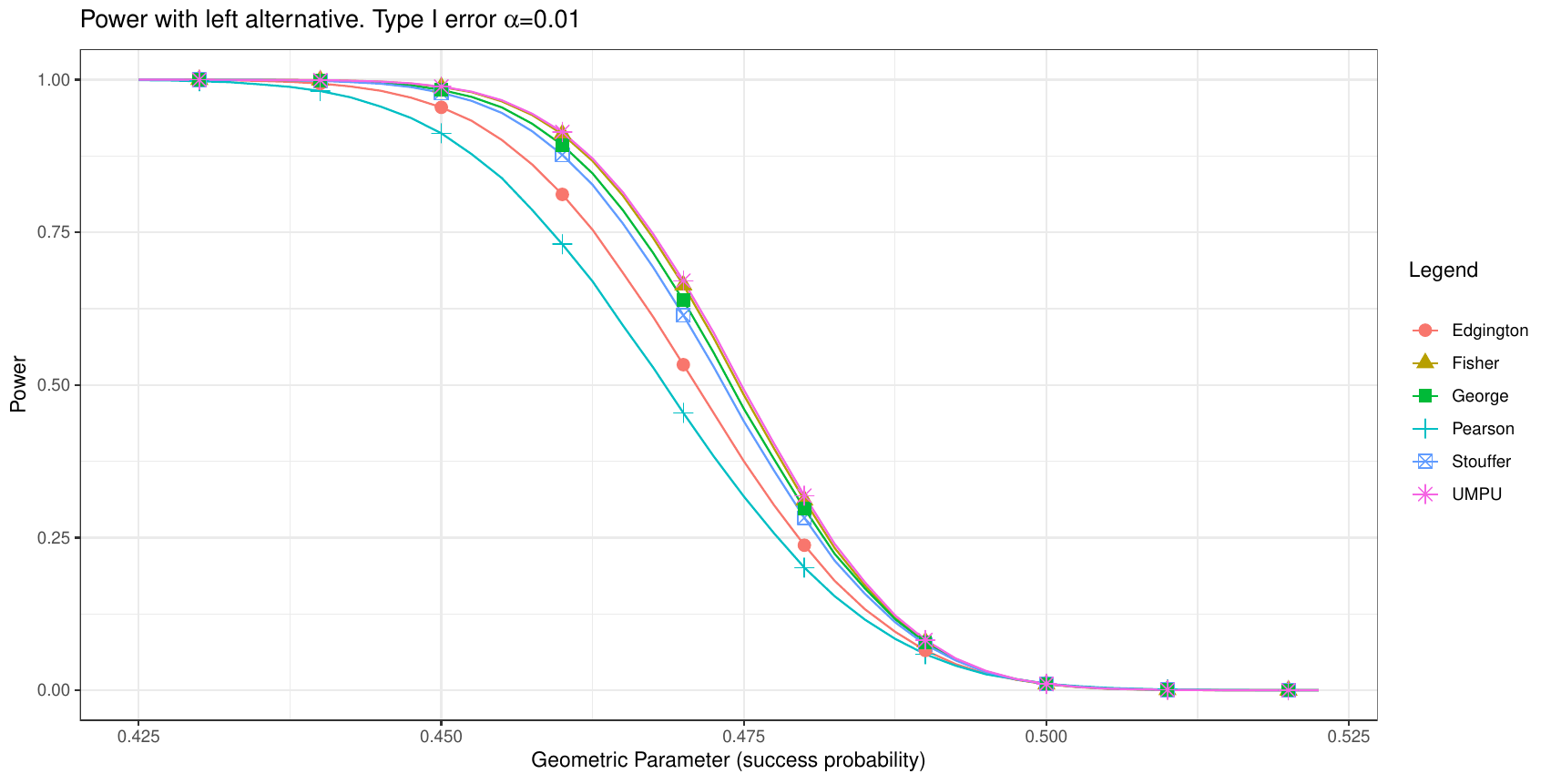} 
			\includegraphics[width=1\textwidth]{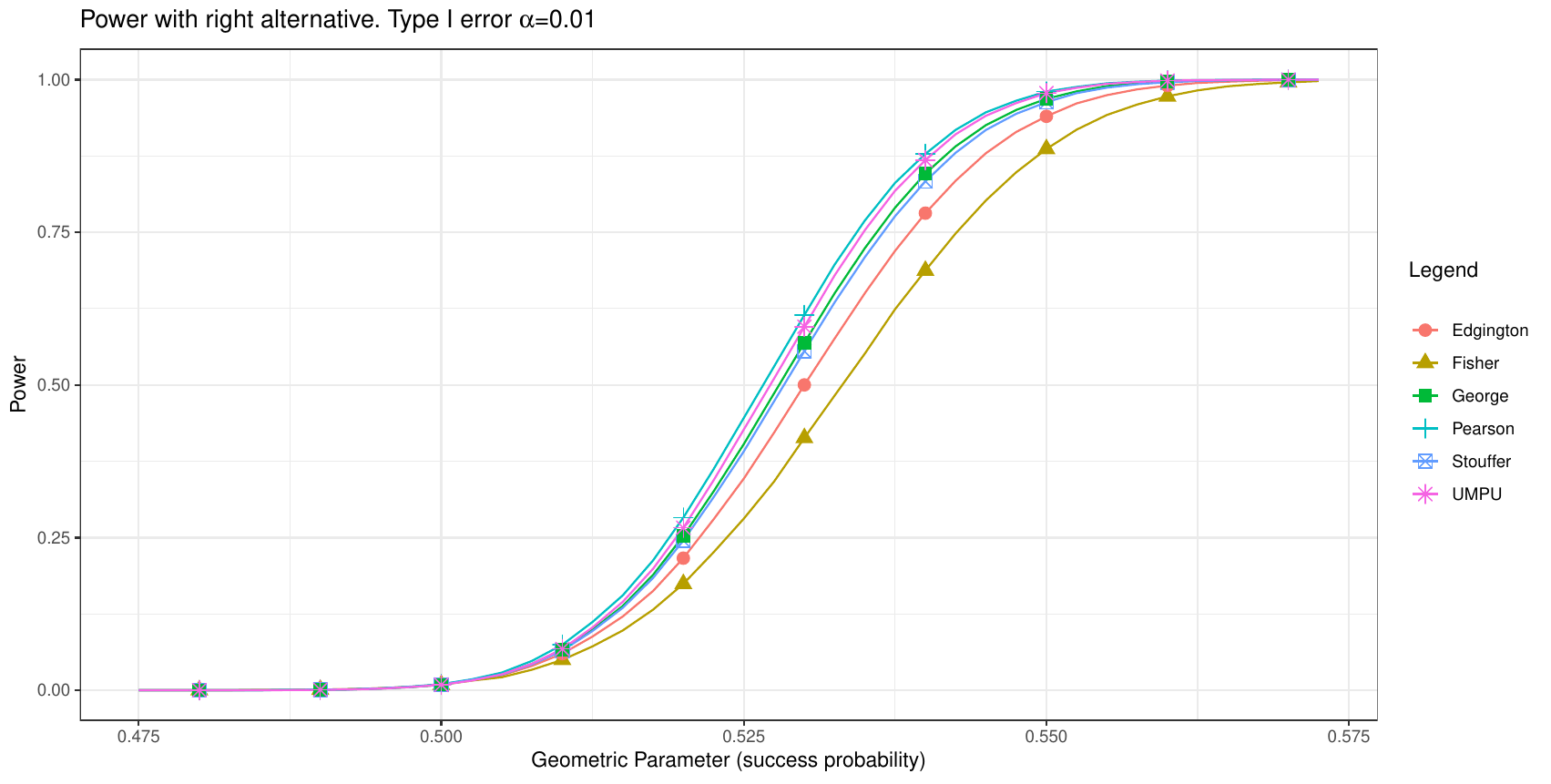} 
			\caption{Power curves for combining $n=1000$ geometric $p$-values under left-sided (upper panel) and right-sided alternatives. Each curve shows the proportion of rejections (empirical power) over $N=100{,}000$ simulated datasets, as a function of the geometric probability parameter $p$. Results are shown for nominal Type I error rates $\alpha=0.01$.
			}
			\label{fig:powergeom_alpha001}
		\end{figure}
	\end{center}

Figure~\ref{fig:powergeom_alpha001} shows the power curves at significance level $\alpha=0.01$. All methods control the Type I error at the nominal rate. Fisher's and Pearson's tests (using the adjusted statistics $S_{F,n}$ and $S_{P,n}$ with surrogate Gamma distributions from \eqref{eq:combinations}) achieve the highest power, matching the UMP LRT in the cases $p_1 < p_0$ and $p_1 > p_0$, respectively. The other combination methods show slightly lower power. These results confirm that Fisher's and Pearson's tests are equivalent to the LRT for the sided alternatives in this setting, and that the proposed adjustment and surrogate distribution procedure preserves the high power of the likelihood ratio test.

Notably, in the right-sided case ($p_1 < p_0$), Pearson's power is slightly higher than that of the LRT. This occurs because the LRT rejection region is based on discrete negative binomial quantiles, which in this setting do not achieve exact $\alpha=0.01$ control; a conservative adjustment is applied to ensure proper Type I error control, resulting in marginally reduced power. Such conservativeness is a known property of discrete tests \cite{contador2023minimum}. Power comparisons at $\alpha=0.05$ are similar and shown in Figure~\ref{fig:powergeom_alpha005} in the Appendix.

	
\subsubsection{Non-I.I.D. Geometric Data}\label{sec:noniidgeom}	
 
We consider combining $p$-values from independent but non-identically distributed geometric random variables, where each null parameter $p_{0j}$ is chosen independently and with equal probability from $\{0.2, 0.5, 0.8\}$, and the true parameter is $p_{1j} \in (p_{0j}-0.1,\, p_{0j}+0.1)$ for $j = 1, \ldots, n$. In this non-i.i.d. setting, the exact LRT distribution is generally unavailable. Therefore, we focus on the adjusted statistics in \eqref{eq:combinations}, using surrogate null distributions constructed by componentwise moment matching as described in Section~\ref{sec:nonid} (i.e., replacing $\nu$ in \eqref{eq:combinations} by the average $\bar{\nu} = \sum_{j=1}^n \nu_j / n$, where each $\nu_j$ is computed from the corresponding $p_{0j}$).
 
\begin{center}
	\begin{figure}[h!] 
		\includegraphics[width=0.5\textwidth]{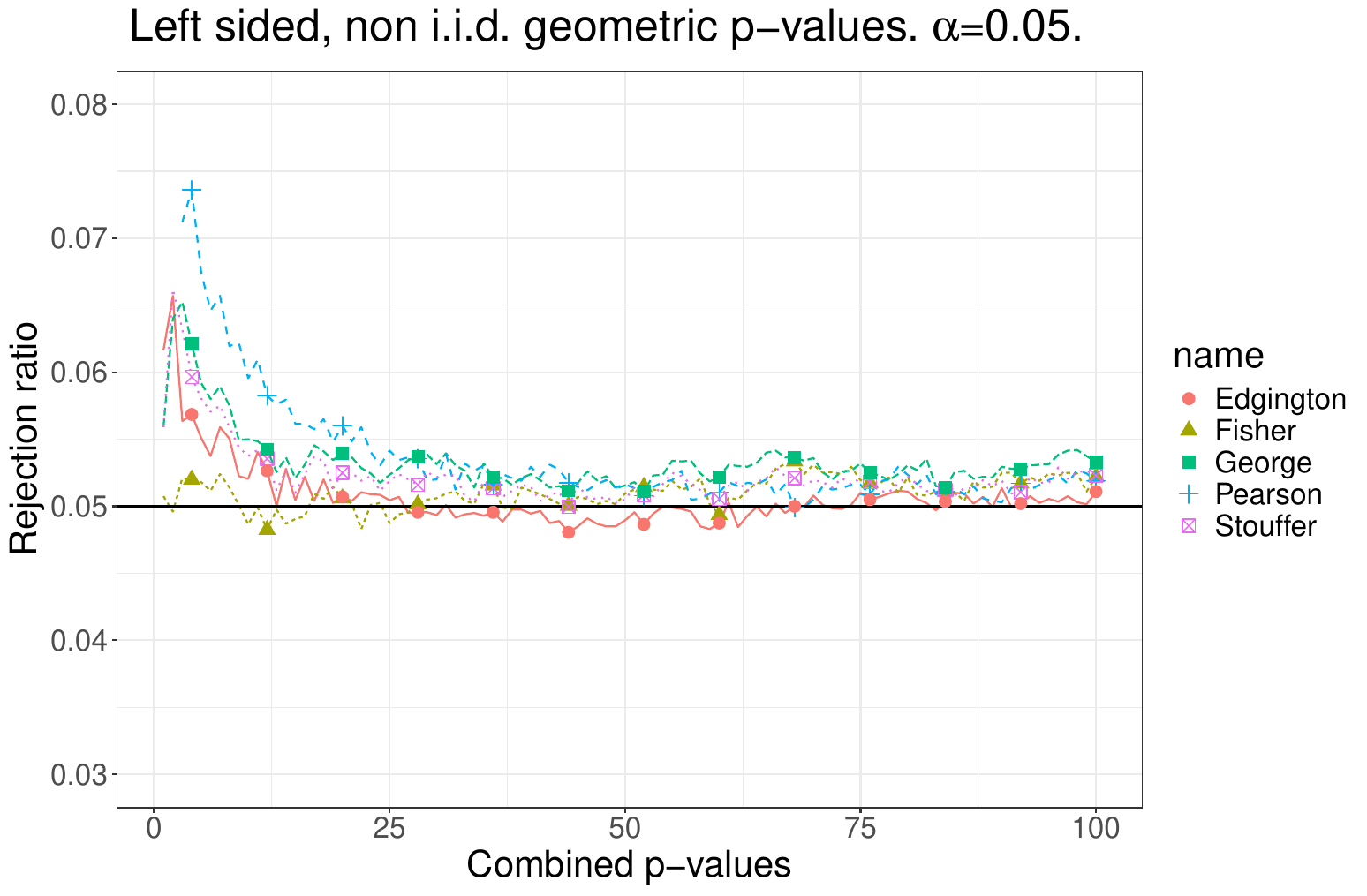} 
		\includegraphics[width=0.5\textwidth]{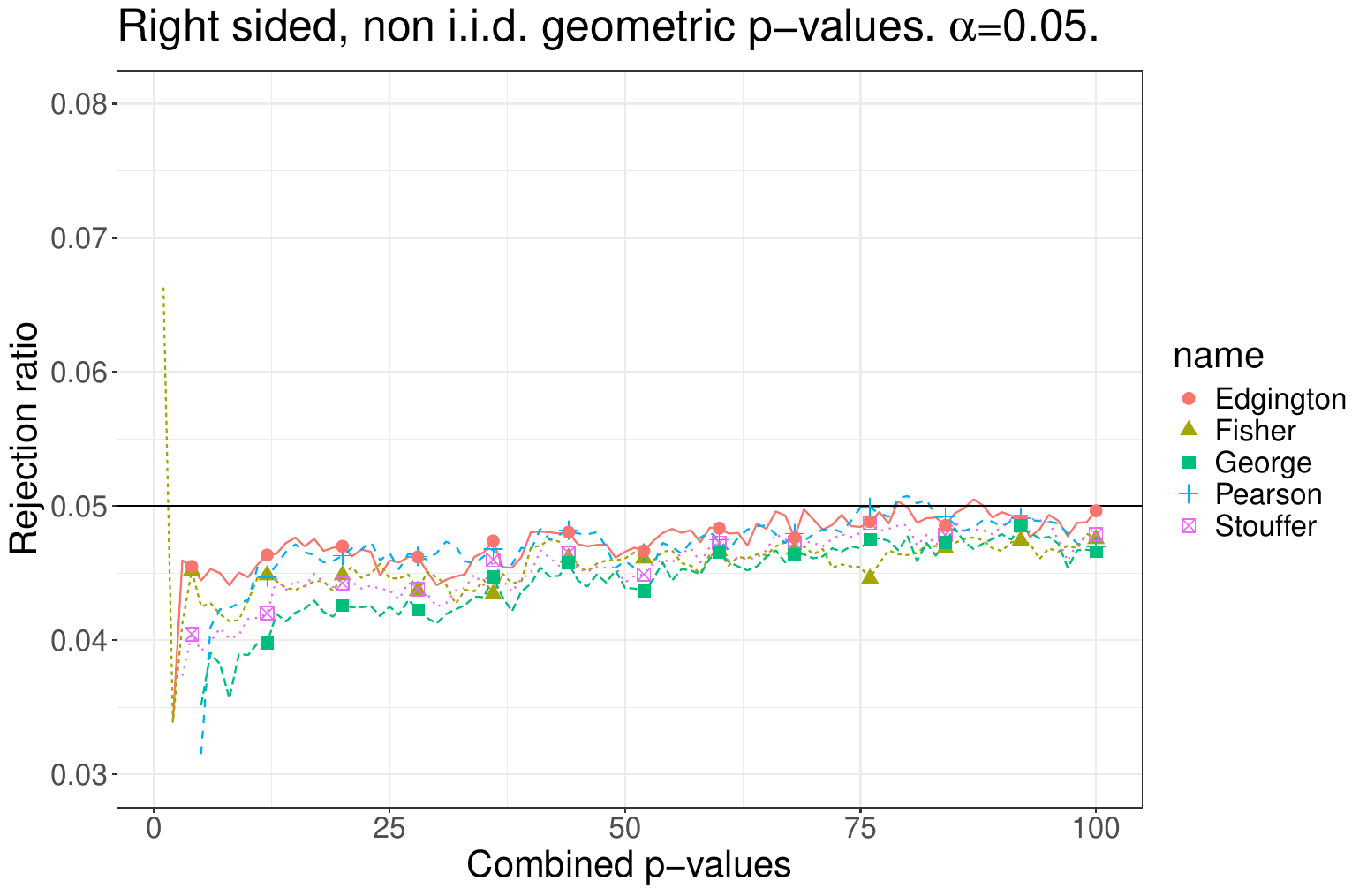} \\
		\includegraphics[width=0.5\textwidth]{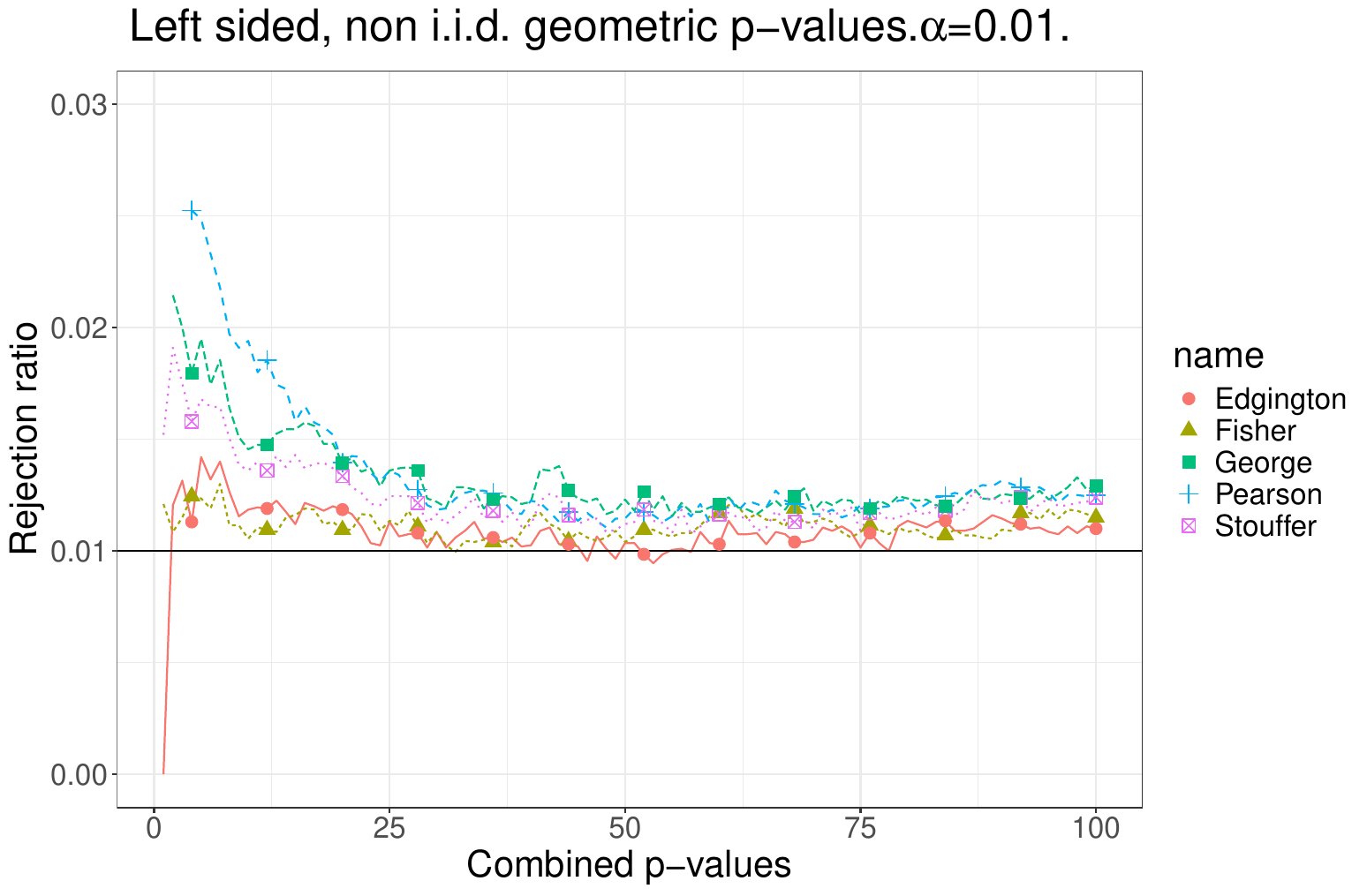} 
		\includegraphics[width=0.5\textwidth]{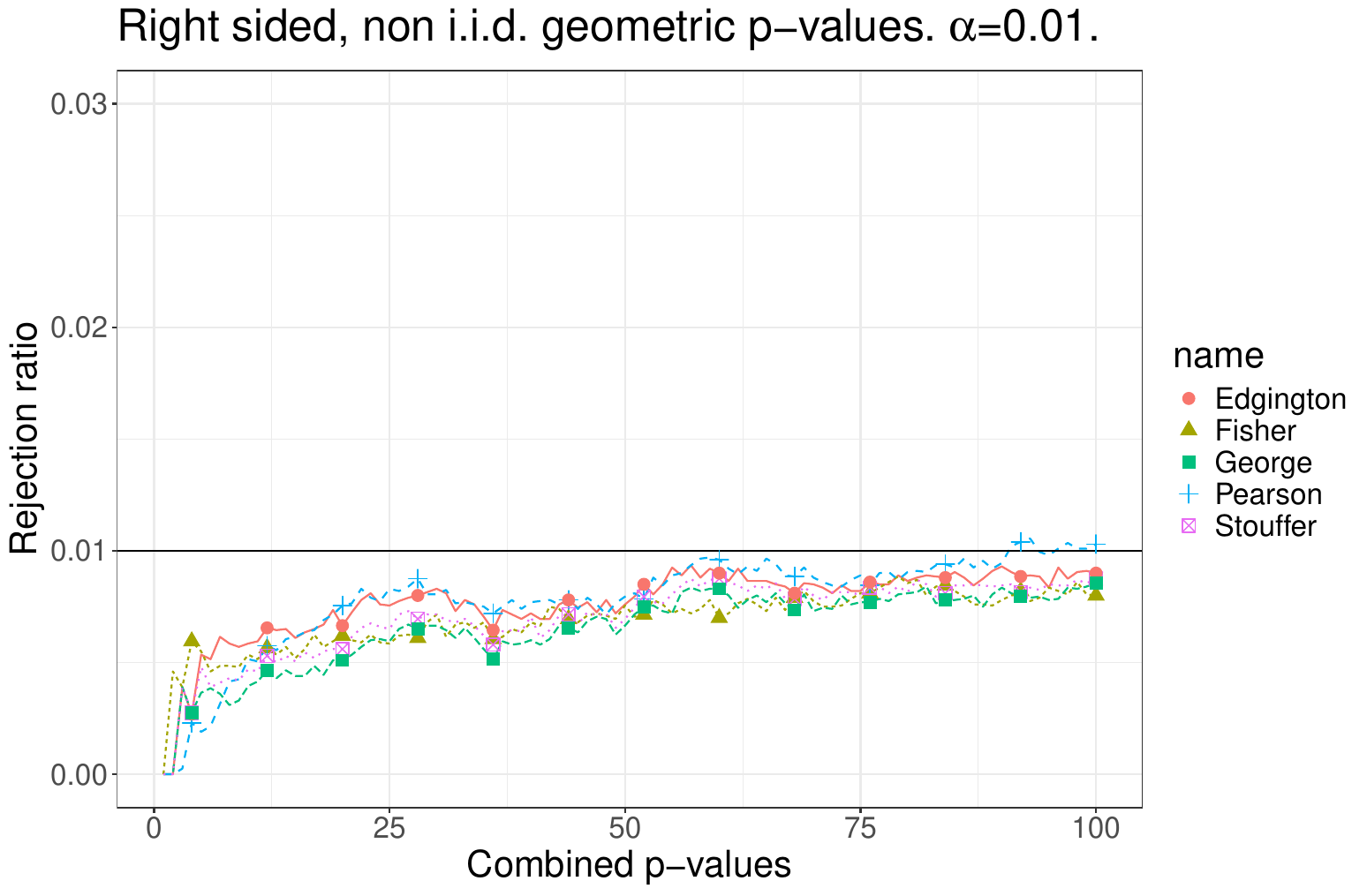} \\
		\caption{Empirical Type I error rates (proportion of rejections) as a function of the number of combined $p$-values $n$, for left-sided (left column) and right-sided (right column) geometric $p$-values. Results are shown for nominal Type I error rates $\alpha=0.05$ (top row) and $\alpha=0.01$ (bottom row), based on 20,000 simulation replicates.
		}
		\label{fig:t1geomn}
	\end{figure}
\end{center}

We first assess Type I error control by combining up to $n=100$ $p$-values. Figure~\ref{fig:t1geomn} displays the empirical Type I error rates at nominal levels $\alpha=0.05$ and $\alpha=0.01$. For small $n$, the procedures tend to be slightly conservative when testing the right-sided alternative ($p_1 < p_0$), or slightly liberal when testing the left-sided alternative ($p_1 > p_0$). As $n$ increases beyond 10, the empirical Type I error rates approach the nominal $\alpha$ levels, indicating that the proposed methods provide accurate Type I error control even when combining a moderate number of $p$-values. Among the combination methods, Fisher's test achieves the most accurate Type I error control for left-sided alternatives, while Pearson's test performs best for right-sided alternatives. This pattern matches the i.i.d. case and is well predicted by the variance ratio metric ($\Var(Z)/\Var(Y)$), as shown in Appendix Tables \ref{table:distnoniidl} and \ref{table:distnoniidr}.

\begin{center}
	\begin{figure}[h!] 
		\includegraphics[width=0.9\textwidth]{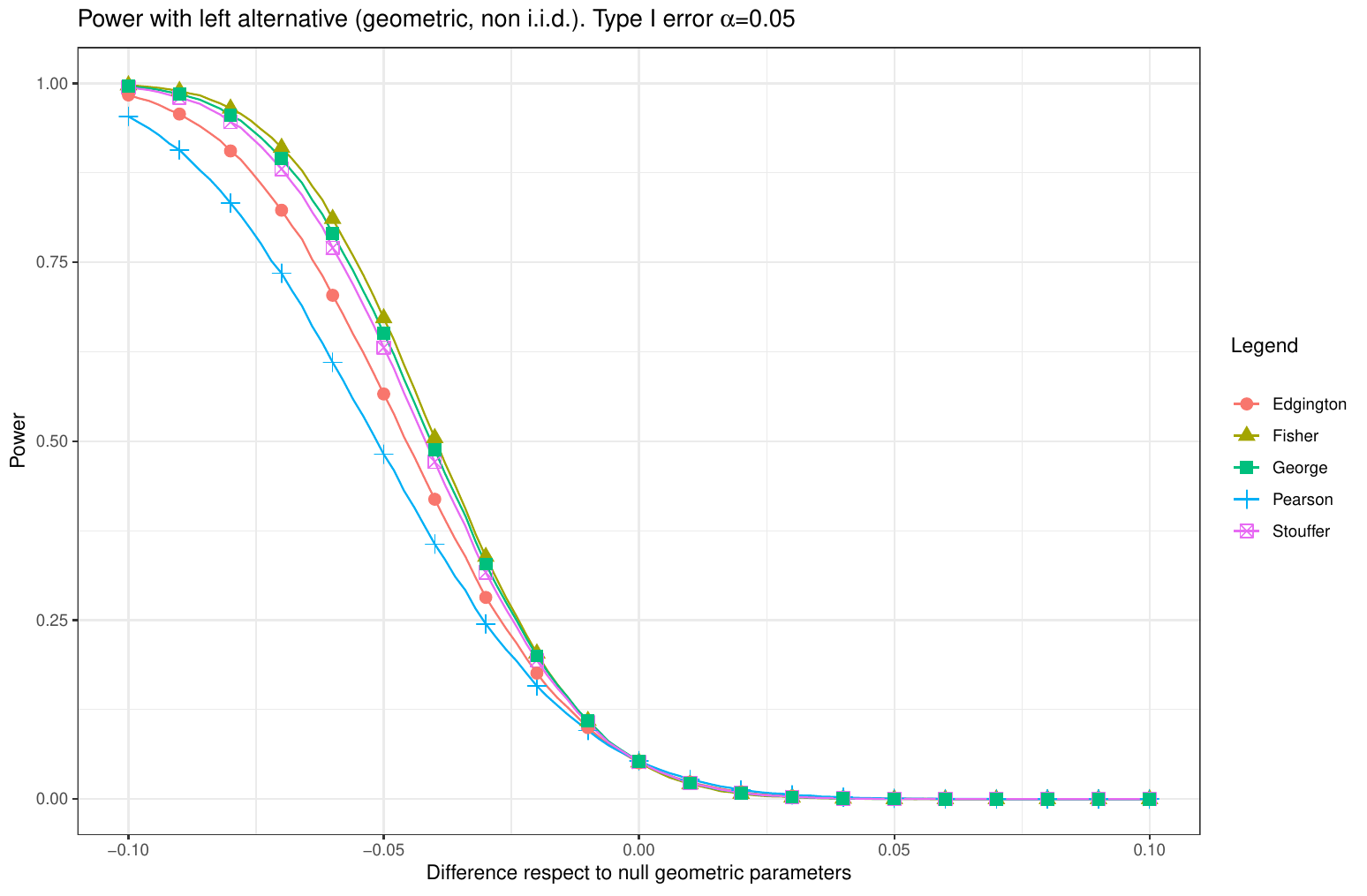} \\
		\includegraphics[width=0.9\textwidth]{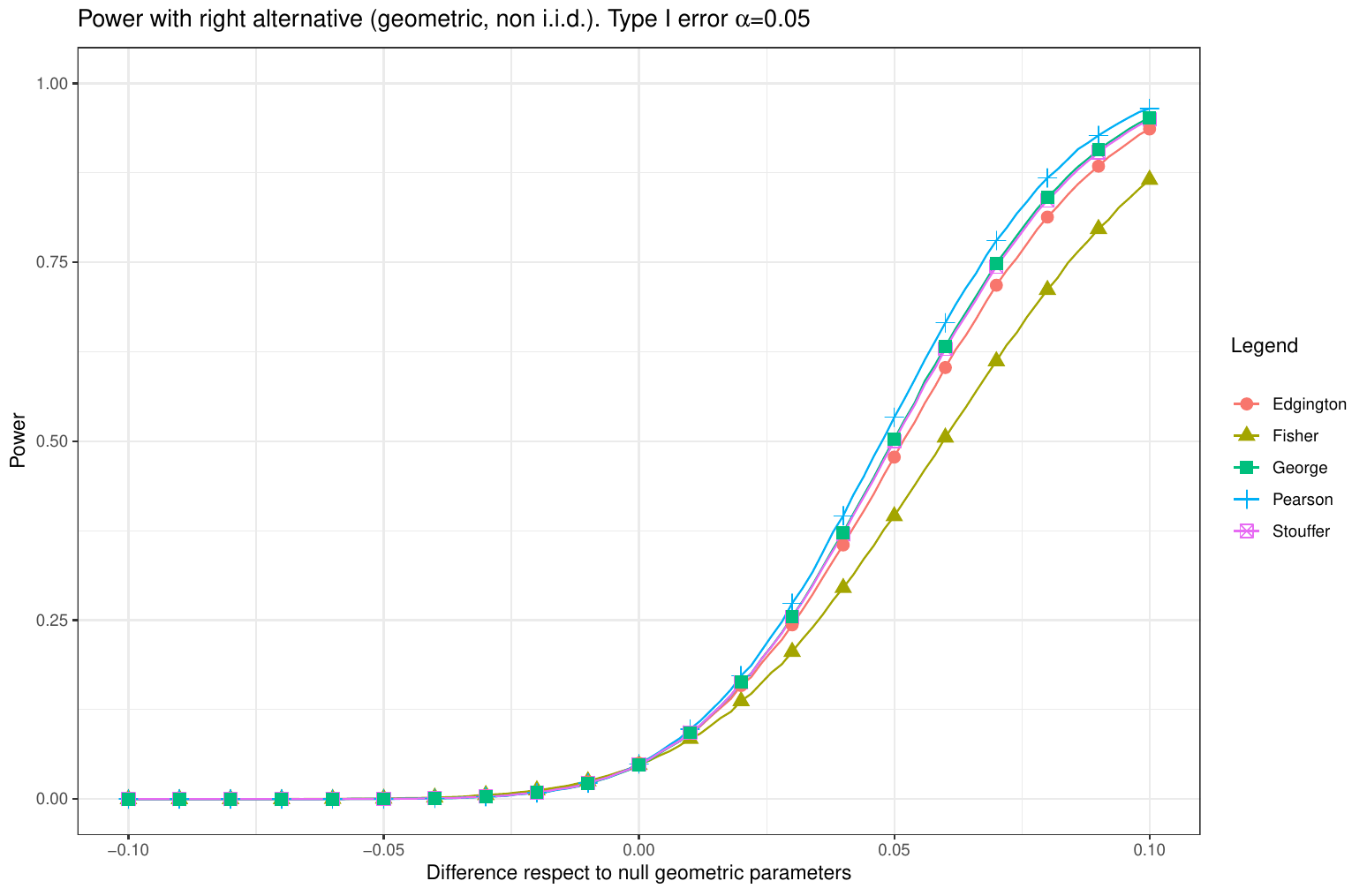} 
		\caption{Power curves for combining $n=100$ non-i.i.d. geometric $p$-values under left-sided (top) and right-sided (bottom) alternatives. Each curve shows the proportion of rejections (empirical power) over $N=100{,}000$ simulated datasets, as a function of the geometric probability parameter $p$. Results are shown for nominal Type I error rate $\alpha=0.05$. 
		}
		\label{fig:powergeomni}
	\end{figure}
\end{center}

To assess the power of each combination and approximation method, we fix $n=100$ $p$-values to combine and estimate power as the proportion of rejections under the alternative. Figure \ref{fig:powergeomni} shows that our proposed Fisher's and Pearson's procedures achieve the highest power for left- and right-sided alternatives, respectively, for nominal type I error rate $\alpha=0.05$ (see figure ~\ref{fig:powergeomni} for $\alpha=0.01$ in the Appendix). This matches the i.i.d. case, where these tests are equivalent to the UMP LRT. Their strong performance extends to the non-i.i.d. setting, even though the exact LRT distribution is not straightforward. The other combination methods (Stouffer's, Edgington's, and George's) provide robust power across both alternatives, with performance close to the best, which is advantageous in practice.
 
\subsection{Unknown LRT Distribution: Circular Data}

We now consider a scenario where the likelihood ratio test (LRT) statistic is a monotonic function of one of the $p$-value combination statistics, but the exact null distribution of the LRT is not readily available. This situation commonly arises in the analysis of circular data, which occur in contexts where measurements are periodic, such as angles, time of day, or compass directions (see, for example, astronomical data or time-to-event modulo a period \cite{bell2024wrapped,fisher1995statistical}). Statistical models for such data often employ wrapped distributions \cite{jammalamadaka2004new}.

	Consider the problem of sampling with replacement from $N$ equispaced points on the unit circle; that is, for integer $i \in \{1, \ldots, N\}$, the corresponding point is $y_i = (\cos(2\pi i / N), \sin(2\pi i / N))$. Let $X$ be a random integer in $\{1, \ldots, N\}$, with the following hypotheses:
	\begin{equation*}
		\label{eq:H0A_circular}
		H_0: \P(X = i) = N^{-1} \quad \text{vs.} \quad H_A: \P(X = i) \propto \exp(-\lambda \min\{i, N - i\}),
	\end{equation*}
	where $\lambda > 0$ is a parameter that increases the likelihood of sampling points closer to $x_N = (1, 0)$. The null hypothesis corresponds to $\lambda = 0$.
	
	Define $T_j = \min\{X_j, N - X_j\}$. The $p$-value associated with observation $X_j$ is the probability, under $H_0$, of observing a point with an $x$-coordinate greater than or equal to that of $X_j$, i.e., $P_j = \frac{2T_j + 1}{N}$ (assuming $N$ is odd for simplicity). The likelihood ratio for this test, based on an i.i.d. sample $X_1, \ldots, X_n$, is 
	\[
	L(X_1,\ldots,X_n)=N^n c_\lambda^{-n} \exp\left[-\lambda \sum_{j=1}^nT_j\right],
	\]
	where $c_\lambda = 1 + 2\sum_{k=1}^{(N-1)/2} e^{-\lambda k}$ is a constant independent of the sample values. Thus, the likelihood ratio is a monotonic function of $\sum_{j=1}^n T_j$, which can be written as a monotonic function of $\sum_{j=1}^n P_j$. Therefore, the original Edgington's statistic in Table \ref{table:combstat} is equivalent to the LRT, and thus UMP, in this setting.
	
	Obtaining the exact distribution of $\sum_{j=1}^n T_j$ or $\sum_{j=1}^n P_j$ is computationally intensive and impractical for moderate or large $n$. Therefore, we focus on comparing the power of the proposed testing procedures in \eqref{eq:combinations}, and show that the adjusted Edgington's test in \eqref{eq:sumE} achieves the highest power among these methods. To provide a comprehensive evaluation, we consider two scenarios: a sparse grid with $N=11$ points (highly discrete data), and a fine grid with $N=199$ points (data closer to the continuous case).

\begin{center}
		\begin{figure}[H] 
			\includegraphics[width=0.85\textwidth]{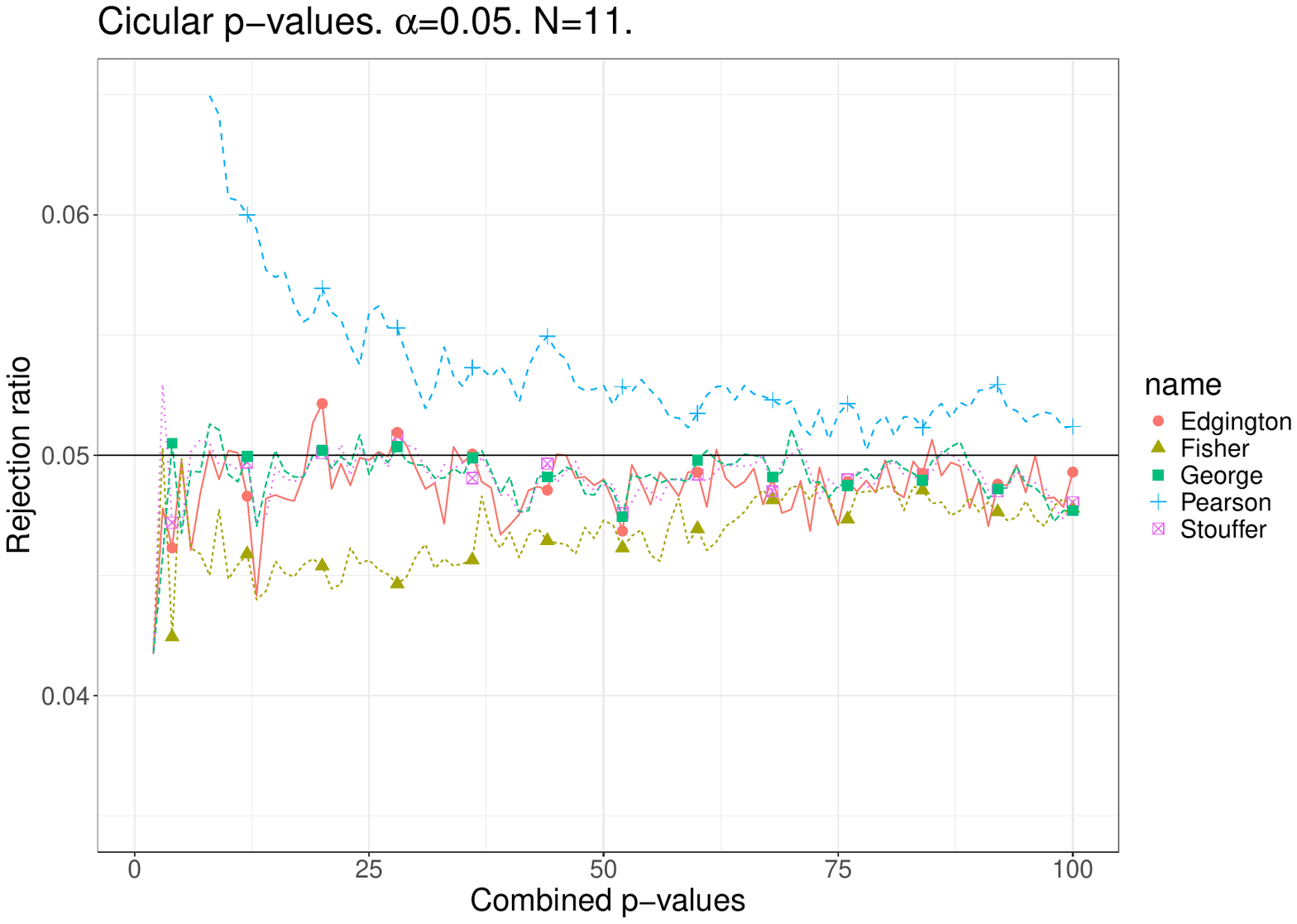}  \\
			\includegraphics[width=0.85\textwidth]{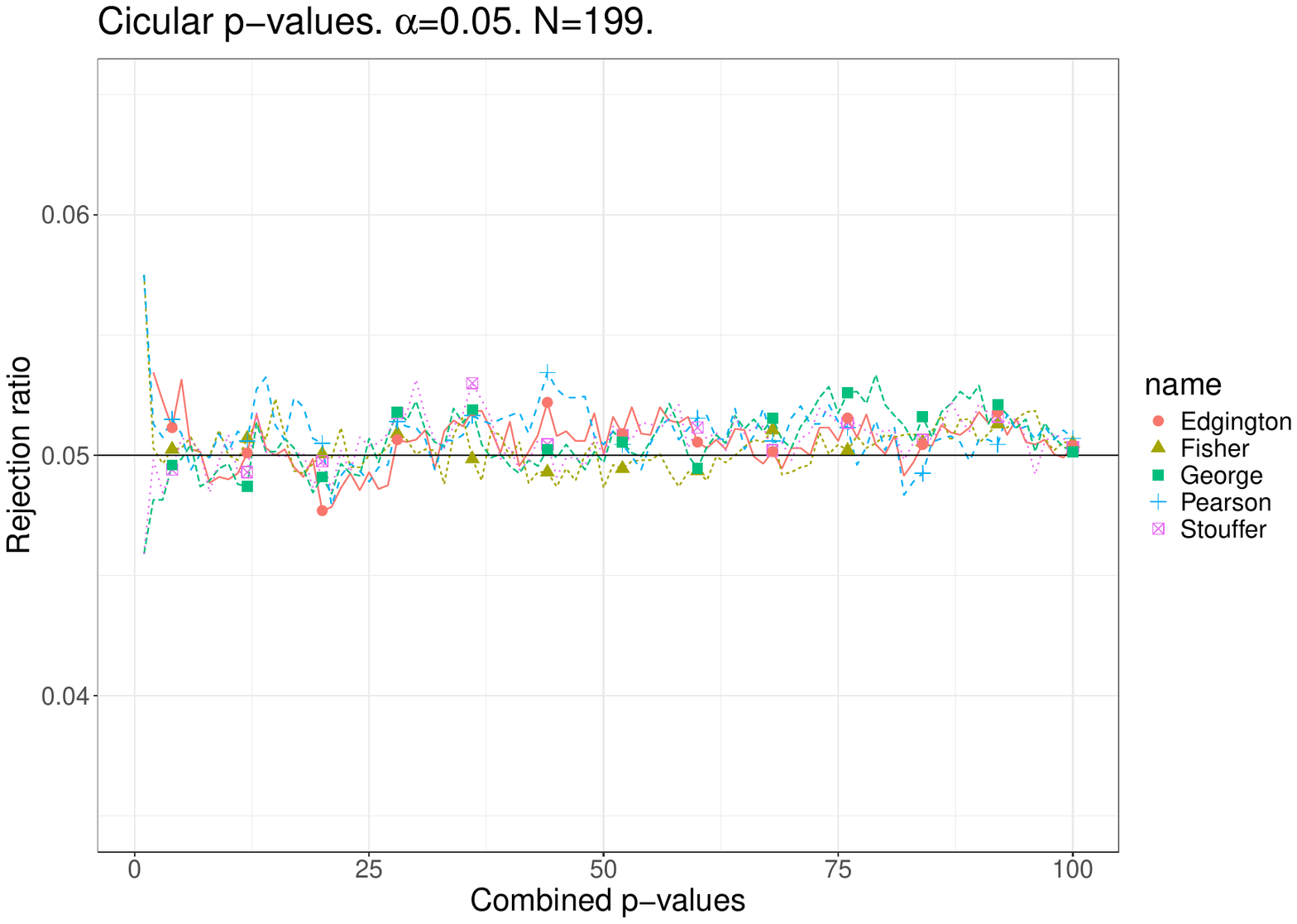}  \\
			\caption{Empirical Type I error rates (proportion of rejections) as a function of the number of combined $p$-values $n$, for circular data with $N=11$ (top) and $N=199$ (bottom) points. Nominal Type I error rate $\alpha \in \{0.05, 0.01, 0.005, 0.001\}$. Results are based on 20,000 simulation replicates.}
			\label{fig:pvcontcirchere}
		\end{figure}
	\end{center}

	To ensure rigorous assessment, we first evaluate Type I error control. Figure~\ref{fig:pvcontcirchere} displays the empirical Type I error rates when combining up to $n=100$ i.i.d. $p$-values at nominal level $0.05$ (see Figure~\ref{fig:pvcontcirc} in Appendix Section \ref{sec:more} for $\alpha \in \{0.05, 0.01, 0.005, 0.001\}$). As expected, the empirical Type I error rates converge to the nominal $\alpha$ as $n$ increases, with faster convergence for the denser grid ($N=199$) compared to the sparse grid ($N=11$). The accuracy of Type I error control is also influenced by the variance ratio $\Var(Z)/\Var(Y)$, as discussed in Section~\ref{sec:t1e}. In particular, Edgington's, George's, and Stouffer's tests have higher variance ratios (see Table~\ref{table:circ}) and exhibit more accurate Type I error control than Fisher's and Pearson's tests, especially in the highly discrete $N=11$ case.
	
		\begin{table}[H]
		\caption{Variance $\Var(Z)$ and variance ratio $\Var(Z)/\Var(Y)$ for circular data $p$-values with $N=11$ and $N=199$ grid points. Edgington's method, which matches the UMP test in this setting, achieves the highest variance ratio.}
		\centering
		\begin{tabular}{l c c c c}
			\toprule
			Statistic  & Variance ($N=11$)  & Ratio ($N=11$)  &  Variance ($N=199$)  & Ratio ($N=199$) \\ 
			\midrule
			Fisher         & 3.5388 & 0.8846                 & 3.9740               &      0.9935       \\ 
			Pearson     & 3.2261   & 0.8065                       & 3.9670             &    0.9891        \\ 
			Stouffer     & 0.9276  & 0.9276                 & 0.9982             &       0.9982       \\ 
			Edgington    & 0.0808 & 0.9691               & 0.0833       &     0.9998        \\ 
			George      & 2.9380  & 0.8930                    & 3.2722         &   0.9946      \\ 
			\bottomrule
		\end{tabular}
		\label{table:circ}	
	\end{table}

	
	We conduct a power study for the different statistics in \eqref{eq:combinations} at various nominal Type I error levels, combining $n=100$ $p$-values and varying the true parameter $\lambda > 0$ under the alternative. Empirical power curves are shown in Figure~\ref{fig:powercirc} for $N=11$ and Figure~\ref{fig:powercirc2} for $N=199$, which can be found in Appendix section \ref{sec:more}. In all settings, Edgington's statistic achieves the highest power, consistent with its equivalence to the UMP LRT in this scenario. Fisher's and Pearson's statistics consistently yield the lowest power. The advantage of Edgington's combination becomes more pronounced as the grid becomes finer ($N=199$) and at more stringent significance levels (smaller $\alpha$).

	
\section{Application: Gene-Based Association Analysis in a Case-Control Study}\label{sec:example}

Our proposed testing procedures have a wide range of potential applications. For instance, Appendix Section \ref{sec:bin} presents simulations and power analyses for data following a binomial distribution. To further demonstrate their use in real data analysis, this section applies the procedures to a case-control study dataset, combining $p$-values from Fisher’s exact tests in the context of a gene-based genetic association study. The data and analysis procedure are provided in the \texttt{DPComb} R package.

In genetic association studies, researchers investigate the relationship between genetic variants (such as single nucleotide polymorphisms, or SNPs) and specific traits (e.g., disease status). The goal is to determine whether certain genetic factors are associated with the trait by testing whether specific SNPs are more prevalent in individuals with the disease (cases) compared to those without it (controls). This approach helps to uncover the genetic basis of diseases and may identify potential targets for treatment or prevention.

A challenging aspect of these studies is handling rare variants, where the number of mutations is small, making standard logistic regression or chi-squared tests inappropriate \cite{Neale2011}. In such cases, it is important to account for the discrete nature of the mutation data. Furthermore, small numbers of mutations often result in weak signals in association studies. A common strategy to address this is to combine $p$-values from multiple SNPs—such as those within gene regions in gene-based analyses—to assess the overall association of a gene with disease status. While individual SNPs may not show significant associations on their own, they may collectively indicate a strong association with the disease.

When a SNP is not associated with disease status, its mutations are randomly distributed between cases and controls. In this situation, the number of mutations observed in the cases follows a hypergeometric distribution, given the total number of mutations, which naturally leads to the use of Fisher's exact test. In gene-based association studies, the goal is to test the null hypothesis that none of the SNPs within a gene are associated with the disease. This is done by combining the SNP-level $p$-values obtained from Fisher’s exact tests. The methods we propose can be effectively applied to combine these $p$-values, taking into account both the discrete nature of the data and the specific hypotheses being tested.


We simulated a genotype–phenotype dataset consisting of 2,000 study subjects, including 1,000 cases and 1,000 controls. The genotype data were generated to mimic typical real-world data and include 15 SNPs across two genes: SNPs 1–5 belong to Gene 1, and SNPs 6–15 belong to Gene 2. The number of mutations for each SNP was randomly assigned between 5 and 20. SNPs 1–3 in Gene 1 and SNPs 6–10 in Gene 2 are associated with disease status, with mutations occurring preferentially in cases (with a 70\% probability), while the remaining SNPs are not associated, and their mutations are randomly distributed between cases and controls. The distribution of mutations for each SNP in the resulting dataset is summarized in Appendix Table \ref{table:casecontrol}.

The objective is to assess whether each gene is associated with disease status by combining SNP-level $p$-values calculated using Fisher’s exact test. In practice, the true direction of the effects for disease-associated SNPs is unknown. As a result, SNP $p$-values may come from right-sided tests (for detecting deleterious mutations more frequent in cases), left-sided tests (for detecting protective mutations more frequent in controls), or two-sided tests (for detecting both types). 

The discrete SNP $p$-values are assumed to be independent, which is reasonable as rare variants often exhibit weak correlation in practice. Therefore, they can be combined using each of the combination tests described in \eqref{eq:combinations}. The resulting gene-level $p$-value reflects the strength of evidence for an association between the gene and disease status. The combination statistics and corresponding $p$-values (in parentheses) are summarized in Table \ref{table:gene1gene2}.

The results indicate that Gene 1 shows significant evidence of association with disease status under both right-sided and two-sided alternatives, while Gene 2 shows significant evidence only under the right-sided alternative. The left-sided tests for both genes yield very weak evidence against the null hypothesis, suggesting no negative association between mutation counts and disease status. These findings are consistent with the simulation setup, in which disease-associated SNPs have a higher prevalence of mutations in cases. Furthermore, for Gene 1, Pearson’s statistic provides the strongest evidence of association, while for Gene 2, Edgington’s statistic yields the most significant result. This analysis illustrates the practical application of the proposed testing procedures in a realistic context, demonstrating their effectiveness in genetic association studies.

\begin{table}[!h]
	\caption{Sum statistics and $p$-values for Gene 1 and Gene 2 under the global null of no association with disease status. Rows correspond to the side of the alternative hypothesis, and columns correspond to the combination method employed. Each cell shows the sum statistic and the corresponding $p$-value in parentheses, with those below $0.05$ in bold.}
	\centering
	\begin{tabular}{ll c c c c c}
		\toprule
		Gene & Alternative & Fisher & Pearson & Edgington & Stouffer & George \\ 
		\midrule
		\multirow{3}{*}{Gene 1} 
		& Two   & \begin{tabular}[c]{@{}c@{}}\textbf{19.00} \\ (\textbf{0.0370})\end{tabular} & \begin{tabular}[c]{@{}c@{}}\textbf{1.77} \\ (\textbf{0.0003})\end{tabular} & \begin{tabular}[c]{@{}c@{}}\textbf{0.8} \\ (\textbf{0.0030})\end{tabular} & \begin{tabular}[c]{@{}c@{}}\textbf{-5.11} \\ (\textbf{0.0075})\end{tabular} & \begin{tabular}[c]{@{}c@{}}\textbf{-8.61} \\ (\textbf{0.0111})\end{tabular} \\ 
		& Right & \begin{tabular}[c]{@{}c@{}}\textbf{25.93} \\ (\textbf{0.0034})\end{tabular} & \begin{tabular}[c]{@{}c@{}}\textbf{0.84} \\ (\textbf{0.0001})\end{tabular} & \begin{tabular}[c]{@{}c@{}}\textbf{0.4} \\ (\textbf{0.0005})\end{tabular} & \begin{tabular}[c]{@{}c@{}}\textbf{-7.16} \\ (\textbf{0.0006})\end{tabular} & \begin{tabular}[c]{@{}c@{}}\textbf{-12.54} \\ (\textbf{0.0009})\end{tabular} \\ 
		& Left  & \begin{tabular}[c]{@{}c@{}}0.84 \\ (0.9999)\end{tabular} & \begin{tabular}[c]{@{}c@{}}25.93 \\ (0.9966)\end{tabular} & \begin{tabular}[c]{@{}c@{}}4.6 \\ (0.9995)\end{tabular} & \begin{tabular}[c]{@{}c@{}}7.16 \\ (0.9994)\end{tabular} & \begin{tabular}[c]{@{}c@{}}12.54 \\ (0.9991)\end{tabular} \\ 
		\midrule
		\multirow{3}{*}{Gene 2} 
		& Two   & \begin{tabular}[c]{@{}c@{}}22.26 \\ (0.3232)\end{tabular} & \begin{tabular}[c]{@{}c@{}}13.96 \\ (0.1079)\end{tabular} & \begin{tabular}[c]{@{}c@{}}4.05 \\ (0.1347)\end{tabular} & \begin{tabular}[c]{@{}c@{}}-2.57 \\ (0.1899)\end{tabular} & \begin{tabular}[c]{@{}c@{}}-4.15 \\ (0.2145)\end{tabular} \\ 
		& Right & \begin{tabular}[c]{@{}c@{}}\textbf{31.20} \\ (\textbf{0.0496})\end{tabular} & \begin{tabular}[c]{@{}c@{}}\textbf{9.72} \\ (\textbf{0.0244})\end{tabular} & \begin{tabular}[c]{@{}c@{}}\textbf{3.08} \\ (\textbf{0.0160})\end{tabular} & \begin{tabular}[c]{@{}c@{}}\textbf{-6.23} \\ (\textbf{0.0227})\end{tabular} & \begin{tabular}[c]{@{}c@{}}\textbf{-10.74} \\ (\textbf{0.0284})\end{tabular} \\ 
		& Left  & \begin{tabular}[c]{@{}c@{}}9.72 \\ (0.9756)\end{tabular} & \begin{tabular}[c]{@{}c@{}}31.20 \\ (0.9504)\end{tabular} & \begin{tabular}[c]{@{}c@{}}6.92 \\ (0.9840)\end{tabular} & \begin{tabular}[c]{@{}c@{}}6.23 \\ (0.9773)\end{tabular} & \begin{tabular}[c]{@{}c@{}}10.74 \\ (0.9716)\end{tabular} \\ 
		\bottomrule
	\end{tabular}
	\label{table:gene1gene2}
\end{table}

\section{Discussion}\label{sec:discussion}

This paper introduced a general framework for combining independent discrete $p$-values, using optimal transport (Wasserstein distance) to adjust discrete statistics and match them to continuous surrogate distributions by aligning the first two moments. We generalized previous work to a broad class of classical combination methods, providing explicit formulas for the adjusted statistics, their surrogate distributions, and practical implementation in the R package \texttt{DPComb}. Theoretical results show that the proposed procedures achieve asymptotically accurate Type I error control.

To guide the choice of combination method for accurate Type I error control, we proposed two practical metrics: the scaled Wasserstein distance between the adjusted and surrogate distributions, and the variance ratio between the adjusted statistic and its continuous analogue. These metrics are easy to compute and interpret, and are supported by theoretical lower and upper bounds. The lower bound, in particular, clarifies how the location of large probability masses in the $p$-value distribution and the shape of the surrogate distribution affect the accuracy of Type I error control. Simulation studies confirm that these metrics effectively predict the finite-sample performance of different combination methods.

We also investigated statistical power from the UMP perspective. When the likelihood ratio test (LRT) statistic and its distribution are available, the corresponding surrogate-distribution-based procedure achieves power nearly identical to the UMP test. When the LRT distribution is unknown, the related combination tests remain computationally efficient and retain high power. For example, Fisher's and Pearson's tests are UMP in the geometric case (depending on the alternative), and Edgington's test is UMP for circular data. In both cases, the surrogate-based procedures closely match the power of the UMP test.

The current framework assumes independence and the existence of second moments for the chosen quantile function. Future work should address relaxing these assumptions, especially for dependent $p$-values or for combination statistics (such as the Cauchy Combination Statistic \cite{Lui2018Cauchy}) that lack finite second moments. Extending the methodology to handle arbitrary dependency structures and heavy-tailed statistics remains an important direction for further research.

\section*{Funding}
Gonzalo Contador's work was supported by the Chilean Agencia Nacional de Investigaci\'on y Desarrollo (Grant 13220097). Zheyang Wu's work was partially supported by the US NSF grant DMS-2113570.





\renewcommand\theequation{\Alph{section}\arabic{equation}} 
\counterwithin*{equation}{section} 
\renewcommand\thefigure{\Alph{section}\arabic{figure}} 
\counterwithin*{figure}{section} 
\renewcommand\thetable{\Alph{section}\arabic{table}} 
\counterwithin*{table}{section} 

\begin{appendices}

\section{Proofs of Mathematical Results}\label{sec:proof}
\subsection*{Proof of Theorem 	\ref{thm:zopti}}
\begin{proof}
	Observe that $G^{-1}$ is a strictly increasing function.

	We consider first the case of $X=G^{-1}(P)$, which takes the values $G^{-1}(0)< \ldots< G^{-1}(F_{i-1}) < G^{-1}(F_i) <\ldots \leq G^{-1}(1)$. According to Theorem 1 in \cite{contador2023minimum}, when $X=G^{-1}(F_i)$ (which is equivalent to $P=F_i$), the closest random variable takes the value $\frac{\int_{F_{i-1}}^{F_{i}} G^{-1}(w)dw}{F_i-F_{i-1}}.$
	
	For the case $X=G^{-1}(1-P)$, which takes the values $G^{-1}(1-1)\leq \ldots< G^{-1}(1-F_i) < G^{-1}(1-F_{i-1}) <\ldots < G^{-1}(1-0)$. Observe here that the $i$-th \textit{largest} value of $X$, i.e. $G^{-1}(1-F_i)$ happens when $P=F_{i}$, and invoking again Theorem 1 in \cite{contador2023minimum}, the closest random variable takes the value $\frac{\int_{F_{i}}^{F_{i-1}} G^{-1}(z)dz}{F_i-F_{i-1}}$, with the result following after the change of variable $w=1-z$.
\end{proof}

\subsection*{Proof of Lemma \ref{lem:lowervariance}}
\begin{proof}
	We have
	\begin{align*}
		W_2^2 (Z,Y)&=\text{Var}(Z)+\text{Var}(Y)-2\text{Cov}(Z,Y)  \\
		&= \text{Var}(Z)+\text{Var}(Y)-2\left(\sum_{i \in \mathbb{N}} z_i\int_{G^{-1}(F_{i-1})}^{G^{-1}(F_i)} yg(y) dy-E(Y)E(Z)\right)\\
		&= \text{Var}(Z)+\text{Var}(Y)-2\left(\sum_{i \in \mathbb{N}} z_i[z_iP(Z=z_i)]-E(Z)^2\right)\\
		&= \text{Var}(Z)+\text{Var}(Y)-2\left(E(Z^2)-E(Z)^2\right)\\
		&= \text{Var}(Y)-\text{Var}(Z).
	\end{align*} 
	
	Rearranging terms, we obtain $\Var(Y)=\Var(Z)+W_2^2(Z,Y)$, and the inequality follows from $W_2$ being a metric and $Y\neq Z$.
\end{proof}

\subsection*{Proof of Lemmas \ref{lem:asymptoticpvalue} and \ref{thm:asymptoticpvaluenoniid}}

We prove Lemma \ref{thm:asymptoticpvaluenoniid}, with Lemma \ref{lem:asymptoticpvalue} following immediately from noting that any i.i.d. condition verifies the Lyapunov condition with $\delta=1$. 
\begin{proof}
Let $\epsilon>0$ be arbitrary.
	If we denote by $\Phi$ the CDF of a standard normal, the Lyapunov conditions guarantee (see page 362 on \cite{billingsley2013convergence}) that, as $n \to \infty$, for any $p \in (0,1)$
	$$\left|P(T_{n,G} < q_{p,n,G})-\Phi\left(\frac{\E(T_{n,G} )-q_{p,n,G}}{\sqrt{\Var(T_{n,G} )}}\right) \right|\to 0,$$
which means that one can find $n_0$ such that for any $n\geq n_0$, 
$$\left|P(T_{n,G} < q_{p,n,G})-\Phi\left(\frac{\E(T_{n,G} )-q_{p,n,G}}{\sqrt{\Var(T_{n,G} )}}\right) \right|\leq \frac{\epsilon}{2}.$$

As the sequence $\tilde{Y}_j$ shares the first two moments with the sequence $Z_j$, it will also verify the Lyapunov condition, and since $p=\P(\sum_{j=1}^n\Tilde{Y}_j<q_{p; n, G,Z}) $, one can find $n_1$ such that for $n\geq n_1$,
$$\left|P(S_n < q_{p,n,G})-\Phi\left(\frac{\E(T_{n,G} )-q_{p,n,G}}{\sqrt{\Var(T_{n,G} )}}\right) \right|=\left|p-\Phi\left(\frac{\E(T_{n,G} )-q_{p,n,G}}{\sqrt{\Var(T_{n,G} )}}\right) \right|\leq \frac{\epsilon}{2}.$$

Combining the two preceding inequalities and the triangle inequality, we conclude that for $n>\max \{n_0, n_1\}$, 
$$\left|\P(T_{n,G} < q_{p,n,G})-p\right|\leq \epsilon.$$

As $\epsilon$ is arbitrary, the result follows.
\end{proof}

%

\subsection*{Proof of the lower bound \eqref{eq:lowerbound} in Section \ref{sec:t1e}}


\begin{proof}
	The one dimensional formulation of the Wasserstein distance available in \cite{panaretos2019wasserstein} for $Z\sim F$ and $\tilde{Y}\sim G$ is
	$$W_2^2(Z,Y)=\int_0^1 |F^{-1}(w)-\tilde{G}^{-1}(w)|^2dw.$$
	Observe that $F^{-1}(w)=z_i$ for $\P(Z< z_i)= F(z_{i-1})\leq w\leq F(z_i)=\P(Z\leq z_i)$. This leads to
	\begin{align*}
		W_2^2(Z,\tilde{Y})
		&=\sum_{i \in \mathbb{N}}\int_{\P(Z< z_i)}^{\P(Z\leq z_i)} |z_i-\tilde{G}^{-1}(w)|^2dw \\
		&=\sum_{i \in \mathbb{N}}\int_{\tilde{G}^{-1}(\P(Z< z_i))}^{\tilde{G}^{-1}(\P(Z\leq z_i))} |z_i-y|^2\tilde{g}(y)dw\\
		&>\max_{i \in \mathbb{N}}\int_{G_1(x_{i-1})}^{G_1(x_i)} |x_i-G_2^{-1}(w)|^2dw.\\
	\end{align*}
	The equality in the second to last line is due to the change of variable $w=\tilde{G}(y)$ and the inequality in the last line spans from all the summands being strictly positive. The result follows after taking square root.
	
	We observe that the lower bound will be largest as 1) the domain of integration $\P(Z=Z_i)=F(z_i)-F(z_{i-1})$ gets larger and/or 2) the continuous quantile function $\tilde{G}^{-1}$ varies too much on this domain.  
\end{proof} 
 
\subsection*{Proof of Lemma \ref{lem:vardifference}}

\begin{proof}
For any discrete statistic, continuous distribution and surrogate distribution triplet $(Z,Y,\tilde{Y})$ as in Theorem \ref{thm:zopti}, a straightforward application of the triangle inequality and the equality $W_2(Z,Y)=\sqrt{\Var(Y)-\Var(Z)}$, which follows from Lemma \ref{lem:lowervariance}, shows that
$$W_2(Z,\tilde{Y}) \leq W_2(Z,Y) + W_2(Y,\tilde{Y})=  \sqrt{\Var(Y)-\Var(Z)} +W_2(Y,\tilde{Y}).$$

The result follows if we can show that $W_2(Y,\tilde{Y})$ is bounded by a linear function of $\sqrt{\Var(Y)-\Var(Z)}$ for each of the distributions in our work. 

When $Y$ and $\tilde{Y}$ are in the same location-scale family, one can find constants $t,s$ such that their CDFs $G$ and $\tilde{G}$ verify 

$$\tilde{G}^{-1}(w)=t+sG^{-1}(w); \quad w \in [0,1].$$

Using the above equality in the one dimensional formulation of the Wasserstein distance available in \cite{panaretos2019wasserstein}, we get

\begin{align*}
	W_2^2(\tilde{Y},Y)&=\int_0^1 |\tilde{G}^{-1}(w)-G^{-1}(w)|^2dw \\
	&=\int_0^1 |t+(s-1)G^{-1}(w)|^2dw\\
	&=t^2 +2t(s-1)\int_0^1 G^{-1}(w)dw + (s-1)^2\int_0^1 |G^{-1}(w)|^2dw \\
	&=t^2 +2t(s-1)\E(Y)+(s-1)^2 \E(Y^2).
\end{align*}

For Stouffer's statistic, we have $\E(Y)=0$, $\E(Y^2)=1$ and $\tilde{G}^{-1}(w)=t+sG^{-1}(w)$ with $t=0$ and $s=\sqrt{\nu_S}$, which yields $W_2(\tilde{Y},Y)=1-\sqrt{\nu_S}=\sqrt{\Var(Y)}-\sqrt{\Var(Z)}$. Similarly, or George's statistic, we have $\E(Y)=0$, $\E(Y^2)=\pi^2/3$ and $\tilde{G}^{-1}(w)=t+sG^{-1}(w)$ with $t=0$ and $s=3\sqrt{\nu_S}/\pi^2$, which yields $W_2(\tilde{Y},Y)= \sqrt{(1-\nu_G\sqrt{3}/\pi)^2\pi^2/3}=\pi/\sqrt{3}-\sqrt{\nu_G}=\sqrt{\Var(Y)}-\sqrt{\Var(Z)}$. Finally, for Edgington's statistic we have $\E(Y)=0.5$, $\E(Y^2)=1/3$ and $\tilde{G}^{-1}(w)=t+sG^{-1}(w)$ with $t=0.5-\sqrt{3\nu_E}$ and $s=2\sqrt{3\nu_S}$, leading to $W_2(\tilde{Y},Y)=(0.5-\sqrt{3\nu_E})/\sqrt{3}=1/\sqrt{12}-\sqrt{\nu_E}=\sqrt{\Var(Y)}-\sqrt{\Var(Z)}$.

The preceeding paragraph shows that $W_2^2(\tilde{Y},Y)=\sqrt{\Var(Y)}-\sqrt{\Var(Z)}$ for Stouffer, George and Edgington's statistic. We observe that when $0<\sqrt{u}<\sqrt{v}$, $-u\geq-\sqrt{uv}$ and then

$$(\sqrt{v}-\sqrt{u})^2=u+v-2\sqrt{uv}\leq v-u \implies \sqrt{v}-\sqrt{u}\leq \sqrt{v-u},$$ so taking $v=\Var(Y)$ and $u=\Var(Z)$, we conclude that for all three statistics $W_2(\tilde{Y},Y)$ is bounded by a linear function of $\Var(Y)-\Var(Z)$, as desired.

The remaining statistics, Fisher's and Pearson's, have surrogates in the Gamma family, for which our previous results do not apply. However, in \cite{chhachhi20231}, the authors show that for Gamma distributed random variables $Y$ and $\tilde{Y}$ having the same mean (as is the case of the distributions obtained with moment matching), $W_2^2(Y,\tilde{Y})\propto |\Var(Y)-\Var(\tilde{Y})|$, which completes the proof.


\end{proof}

\section{On Sided Discrete $p$-Values}\label{sec:p_sided}

This section links the distribution $F$ of a discrete $p$-value in \eqref{eq:disc_P_left} with that of a test statistic in a sided test. The distribution of $P$ comes from a discrete test statistic $X$ supported in $\{x_i\}_{i \in \mathbb{N}}$ with monotone likelihood ratio \cite{NeymanPearson1933,lehmann1993fisher}, and the sequence $F_i$ is related to both the side of the alternative and the CDF of $X$: 

\begin{itemize}
	\item For left-tailed $p$-values, $F_i=\P(X\leq x_i)$ with the sequence $\{x_i\}_{i \in \mathbb{N}}$ in increasing order, thus $F_i$ would correspond to the $i$-th largest value of the CDF of $X$.
	\item For right-tailed $p$-values $F_i=\P(X\geq x_i)=1-\P(X\leq x_{i+1})$ with the sequence $\{x_i\}_{i \in \mathbb{N}}$ in decreasing order, thus $F_i$ would correspond to one minus the $i+1$-th smallest value of the CDF of $X$, with $F_\infty=1-\lim_{i \to \infty}P(X\leq x_{i+1})=1$ if needed.
	\item The case of two sided $p$-values requires a little more care, particularly when the distribution of $X$ is not symmetric. A two sided $p$-value is generally understood as the probability of observing any value in the support less likely than the observed one \cite{kulinskaya2008two,santner2012statistical,freeman1951note}, this is, the $p$-value of observation $x$ is
	
	$$\sum_{j:\P(X=x_j)\leq \P(X=x)}\P(X=x_j)$$
	
	Thus, in order to allow for the possibility of ties, the $p$-value distribution is defined iteratively according to the following algorithm
	\begin{algorithm}
		\caption{Calculation of two sided $p$-value distribution}
		\begin{algorithmic}[1]
			\State \textbf{Input:} $\{x_j\}_{j\in J}$= Support of $X$, $\P$=probability distribution of $X$, $i=1$
			\State \textbf{Output:} $\{F_i\}_{i\in \mathbb{N}}$ increasing list of numerical values for two sided $p$-value $P$
			\While{$J$ is nonempty}
			\State Calculate $p=\arg\min_{j\in J} \{\P(X=x_j)\}$;
			\State Define $\tilde{J}=\{j:\P(X=x_j)=p\}$; $F_i=F_{i-1}+p|\tilde{J}|$; 
			\State Remove all indices in $\tilde{J}$ from $J$
			\EndWhile
			\State Return $F_0, F_1, \ldots$
		\end{algorithmic}
	\end{algorithm}
	
	Consider, for instance, $X\sim \text{Binomial}(n,\theta)$ and the hypothesis contrast $H_0:\theta=0.5$ against $H_A:\theta \neq 0.5$ with even $n$, so that under $H_0$ one has $\P(X=k)=\P(X=n-k)$. The corresponding $p$-value distribution only allows for $n/2 +1$ different numerical values, with the smallest one $F_1=\P(X=0)+\P(X=n)$, the second smallest $F_2=F_1+\P(X=1)+\P(X=n-1)$, and so on.
\end{itemize}

Although an explicit formula for $F$ based on the CDF of  $X$ is not available for the two sided case, the developed software facilitates the efficient calculation of the $p$-value distribution and numerical values for all three sided alternatives.

\section{Supplementary Material for Sections \ref{sec:t1e} and \ref{sec:UMPU}}\label{sec:more}

\subsection*{Additional Type I Error Control Results for $p$-Values in Section \ref{sec:t1e}}
\begin{center}
	
	Figure \ref{fig:exqoa001} contains estimated Type I error rate for all four $p$-value distributions in Section \ref{sec:t1e} (as proportion of rejections) nominal $\alpha=0.01$ as a function of number of combined $p$-values $n$, for all surrogate distribution based testing procedures in \eqref{eq:combinations}.
	\begin{figure}[h!] 
		\includegraphics[width=0.5\textwidth]{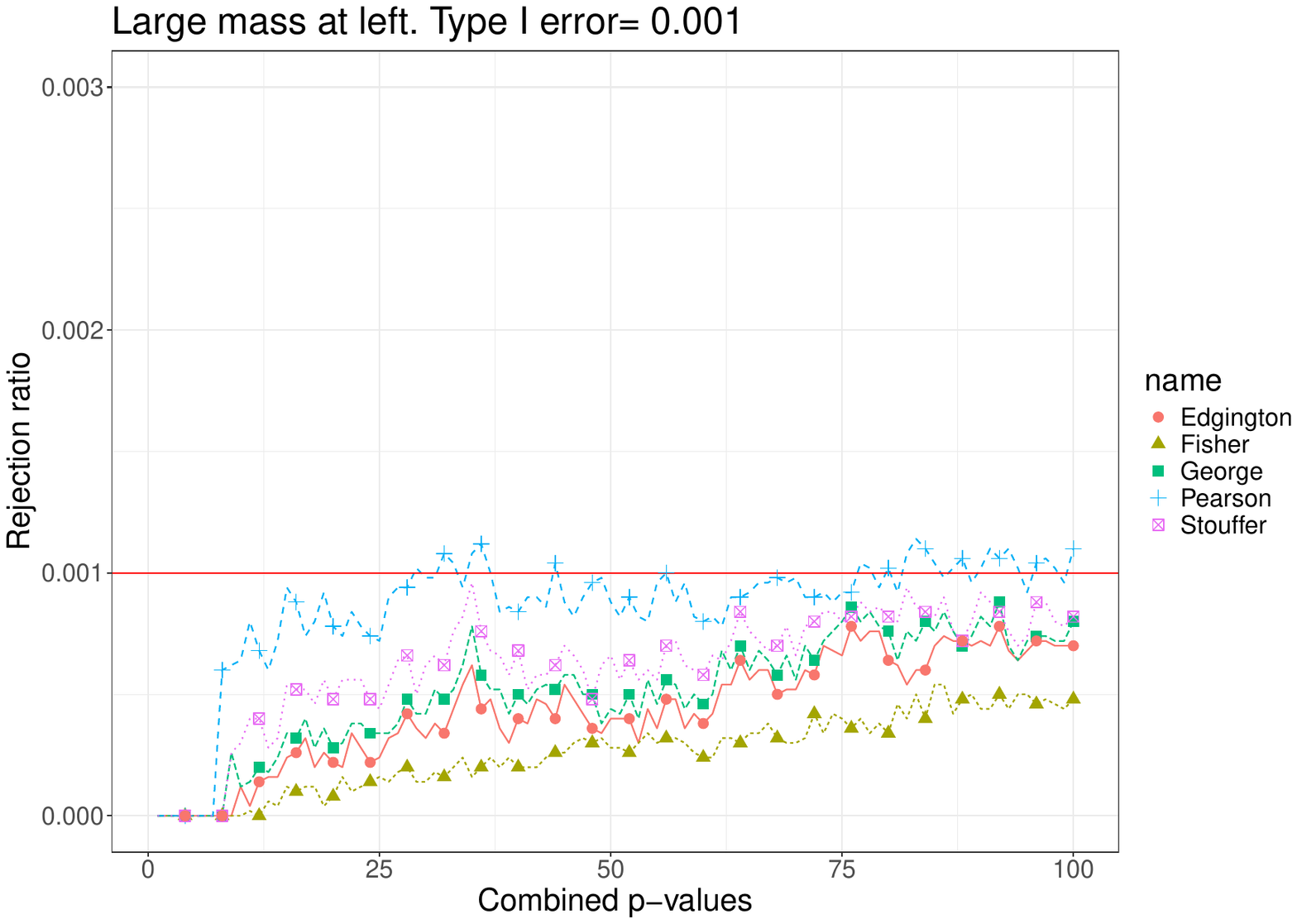} 
		\includegraphics[width=0.5\textwidth]{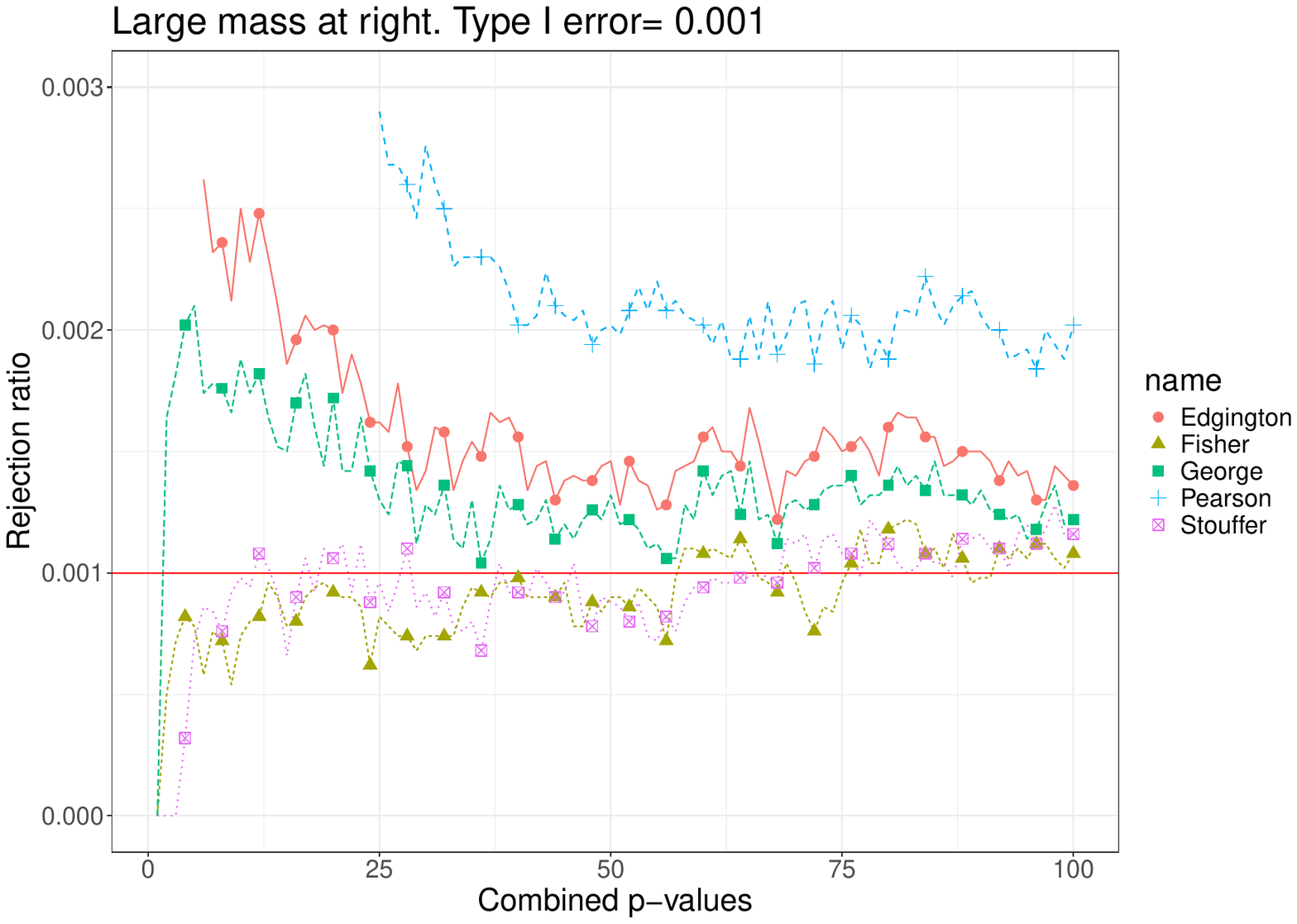} 
		\includegraphics[width=0.5\textwidth]{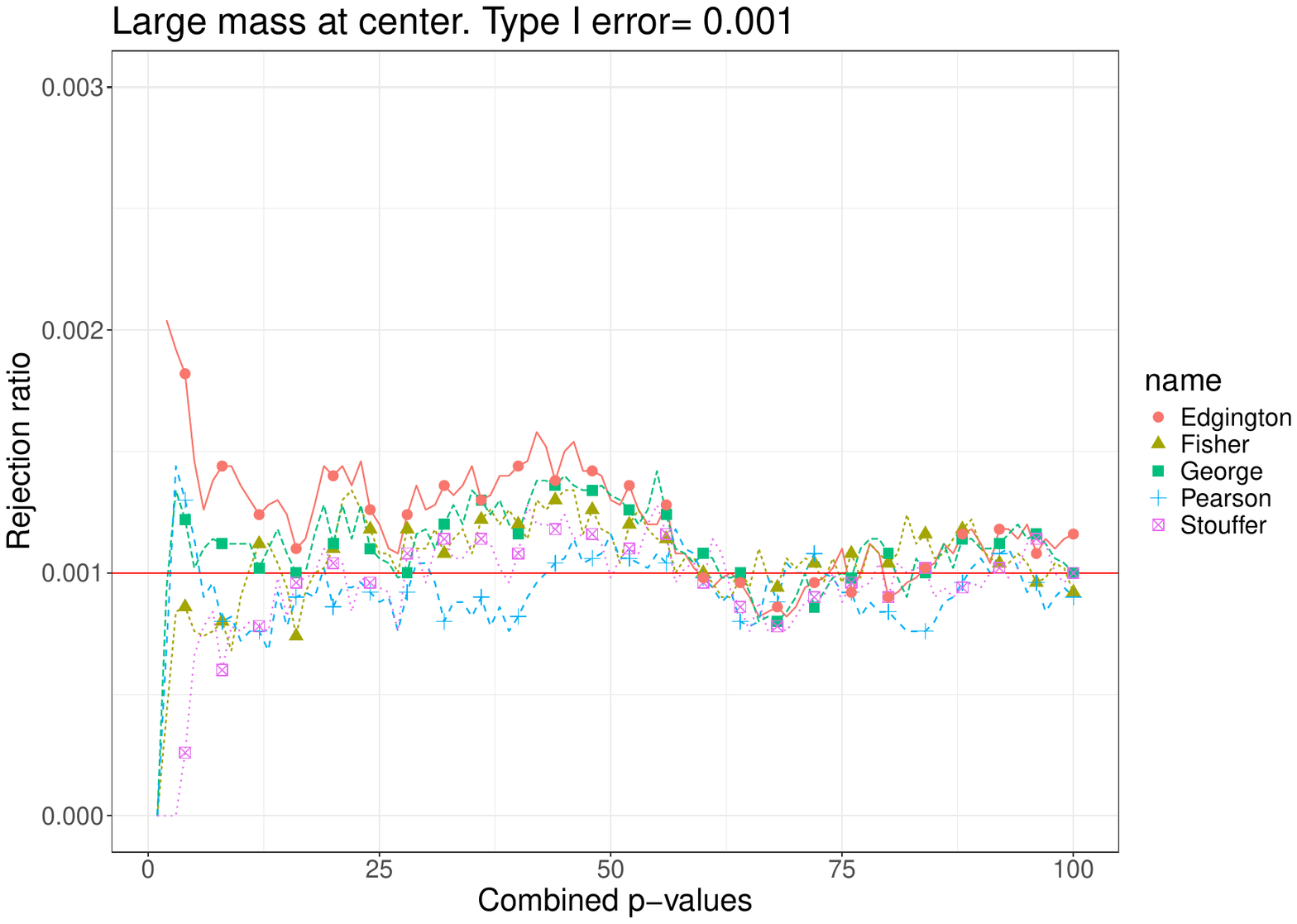} 
		\includegraphics[width=0.5\textwidth]{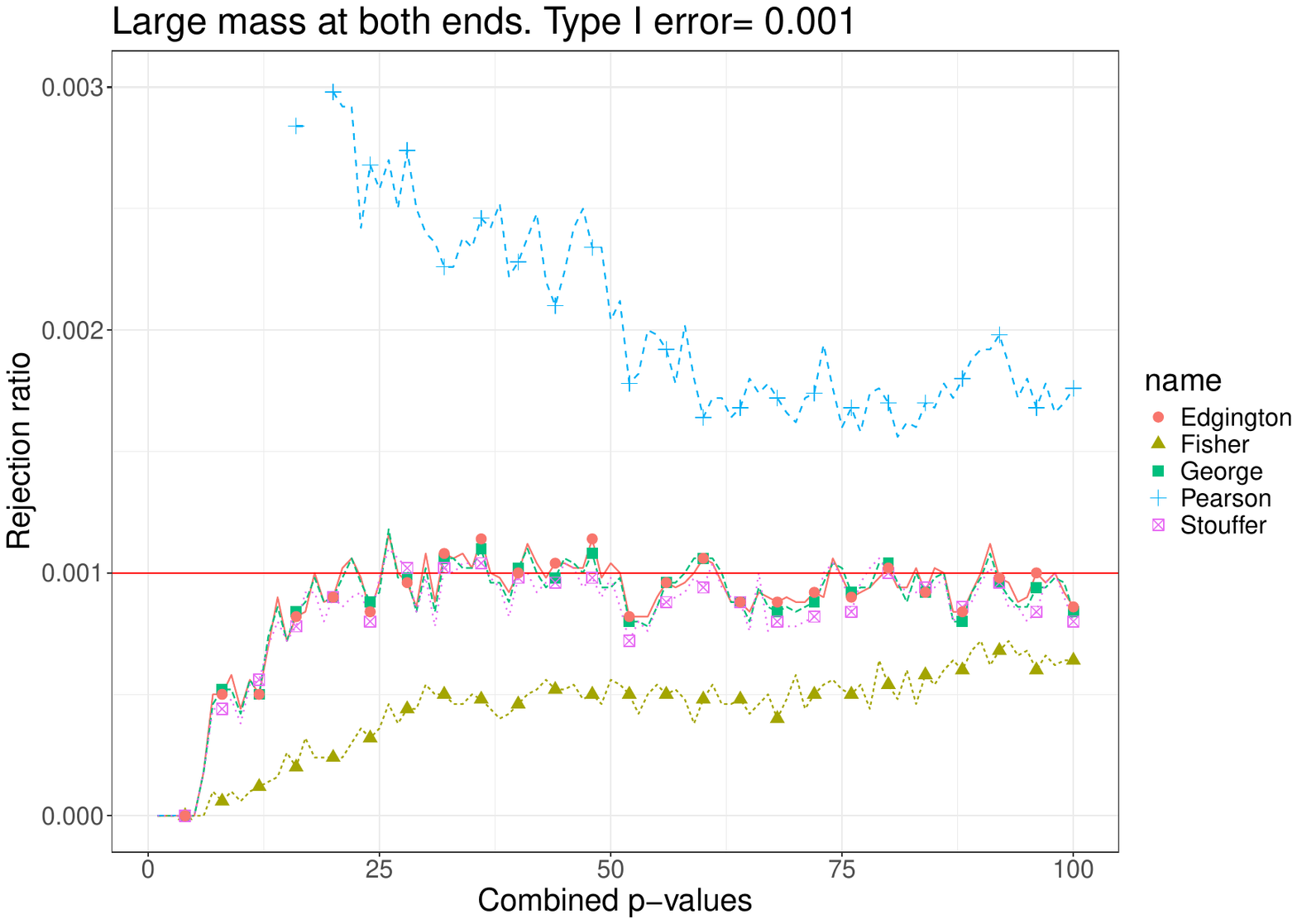} \\
		\caption{Empirical Type I error rates (proportion of rejections) at nominal $\alpha = 0.001$ as a function of the number of combined $p$-values $n$, for $P_L$ (large mass at $0.4$, top left), $P_R$ (large mass at $1$, top right), $P_C$ (large mass at $0.7$, bottom left), and $P_S$ (large masses at both ends, bottom right). Results are based on 50,000 simulation replicates. The horizontal line indicates the nominal $\alpha$ level.}
		\label{fig:exqoa001}
	\end{figure}
\end{center}

\newpage

\subsection*{Geometric Data Simulation Study in Section \ref{sec:geom}}

The rejection regions for the UMP and surrogate distribution-based testing procedures in Section~\ref{sec:iidgeom} are summarized in Table~\ref{table:distgeo2} for $\alpha = 0.05$ and Table~\ref{table:distgeo} for $\alpha = 0.01$. 
Statistical power is estimated as the proportion of times the global null is rejected across $N = 100{,}000$ simulated datasets.

\begin{table}[h!]
	\caption{Test statistics, distributions and rejection regions for geometric data with sided hypotheses. Rejection region corresponding to $\alpha=0.05$ . }
	\centering
	\begin{tabular}{l c c c c c c}
		\toprule
		Statistic  & \multicolumn{2}{c}{$H_A: p=p_1$ with $p_1<p_0$} & \multicolumn{2}{c}{$H_A: p=p_1$ with $p_1>p_0$} \\ 
		\cmidrule(lr){2-3} \cmidrule(lr){4-5}
		& Null & Rejection Region & Null & Rejection Region \\ 
		\midrule
		$T$  & NB($1000$, $0.5$) & $T \geq 2074$ & NB($1000$, $0.5$) & $T \leq 1927$  \\ 
		$S_{F,n}$ & $\Gamma$($1040.7$, $1.9$)  & $S_{F,n} \geq 2103.05$ & $\Gamma$($2015$, $0.99$) &  $S_{F,n} \geq 2073.85$  \\ 
		$S_{P,n}$ & $\Gamma$($2015$, $0.99$)  & $S_{P,n} \leq 1927.27$ ($1897.8$) & $\Gamma$($1040.7$, $1.9$) &  $S_{P,n} \leq 1899.12$  \\ 
		$S_{G,n}$ & $\mathcal{N}$($0$, $50.48$) & $S_{G,n} \leq -83.35$  & $\mathcal{N}$($0$, $50.48$) &  $S_{G,n} \leq -83.35$  \\ 
		$S_{S,n}$ & $\mathcal{N}$($0$, $28.38$) & $S_{S,n} \leq -46.68$  & $\mathcal{N}$($0$, $28.38$) &  $S_{S,n} \leq -46.68$  \\ 
		$S_{E,n}$ & $\mathcal{N}$($500$, $8.45$) & $S_{E,n} \leq 486.1$  & $\mathcal{N}$($500$, $8.45$) &  $S_{E,n} \leq 486.1$ \\ 
		\bottomrule
	\end{tabular}
	\label{table:distgeo2}
\end{table}

\begin{table}[h!]
	\caption{Test statistics, distributions and rejection regions for geometric data with sided hypotheses. Rejection region corresponding to $\alpha=0.01$. }
	\centering
	\begin{tabular}{l c c c c c c}
		\toprule
		Statistic  & \multicolumn{2}{c}{$H_A: p=p_1$ with $p_1<p_0$} & \multicolumn{2}{c}{$H_A: p=p_1$ with $p_1>p_0$} \\ 
		\cmidrule(lr){2-3} \cmidrule(lr){4-5}
		& Null & Rejection Region & Null & Rejection Region \\ 
		\midrule
		$T$  & NB($1000$, $0.5$) & $T \geq 2106$ & NB($1000$, $0.5$) & $T \leq 1897$ \\ 
		$S_{F,n}$ & $\Gamma$($1040.7$, $1.9$)  & $S_{F,n} \geq 2147.05$ & $\Gamma$($2015$, $0.99$) &  $S_{F,n} \geq 2105.11$ \\ 
		$S_{P,n}$ & $\Gamma$($2015$, $0.99$)  & $S_{P,n} \leq 1897.8$ & $\Gamma$($1040.7$, $1.9$) &  $S_{P,n} \leq 1858.6$ \\ 
		$S_{G,n}$ & $\mathcal{N}$($0$, $50.48$) & $S_{G,n} \leq -117.9$ & $\mathcal{N}$($0$, $50.48$) &  $S_{G,n} \leq -117.9$ \\ 
		$S_{S,n}$ & $\mathcal{N}$($0$, $28.38$) & $S_{S,n} \leq -66.02$ & $\mathcal{N}$($0$, $28.38$) &  $S_{S,n} \leq -66.02$ \\ 
		$S_{E,n}$ & $\mathcal{N}$($500$, $8.45$) & $S_{E,n} \leq 480.33$ & $\mathcal{N}$($500$, $8.45$) &  $S_{E,n} \leq 480.33$ \\ 
		\bottomrule
	\end{tabular}
	\label{table:distgeo}
\end{table}

	\begin{center}
		\begin{figure}[h!] 
			\includegraphics[width=.5\textwidth]{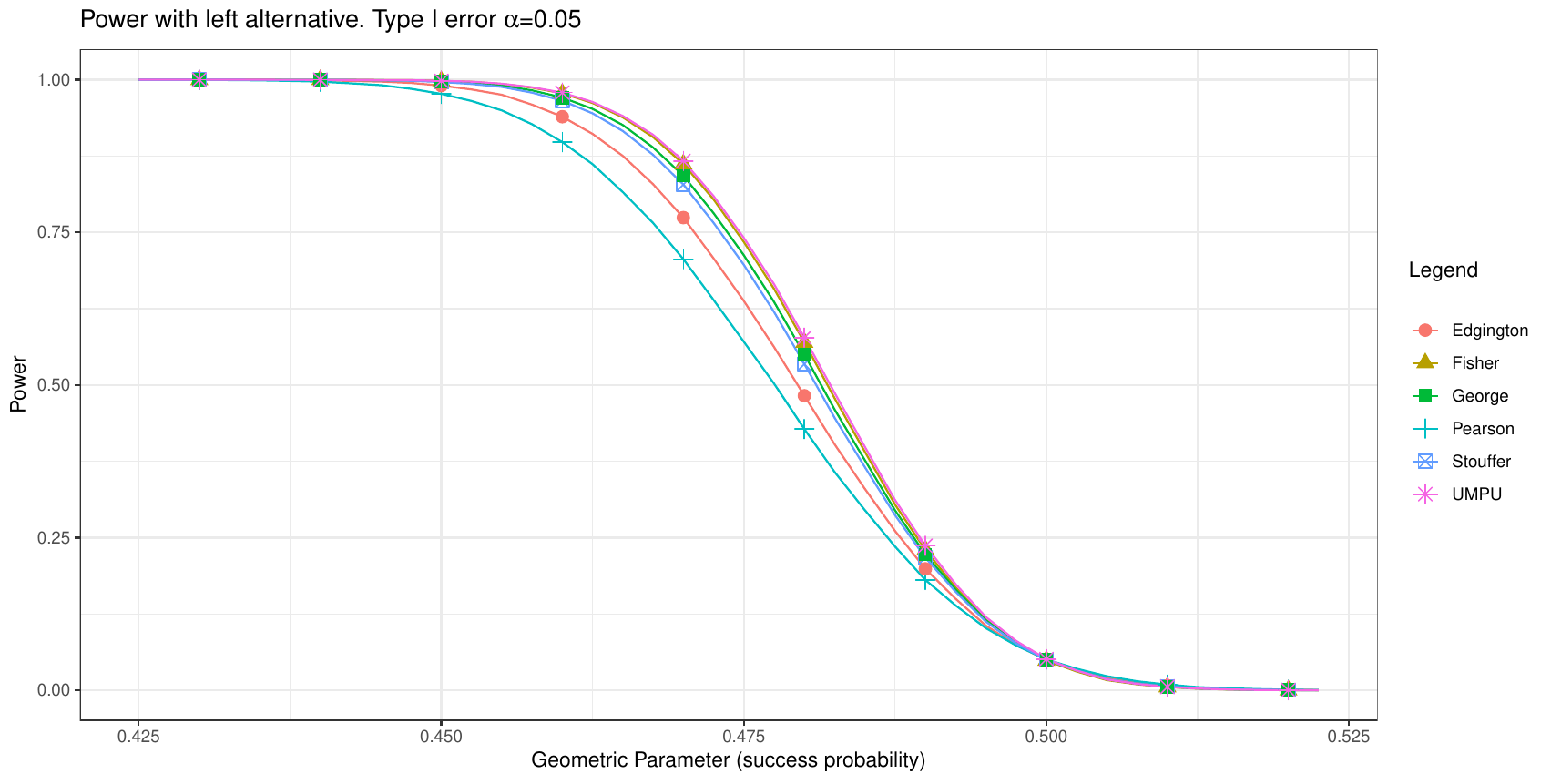} 
			\includegraphics[width=.5\textwidth]{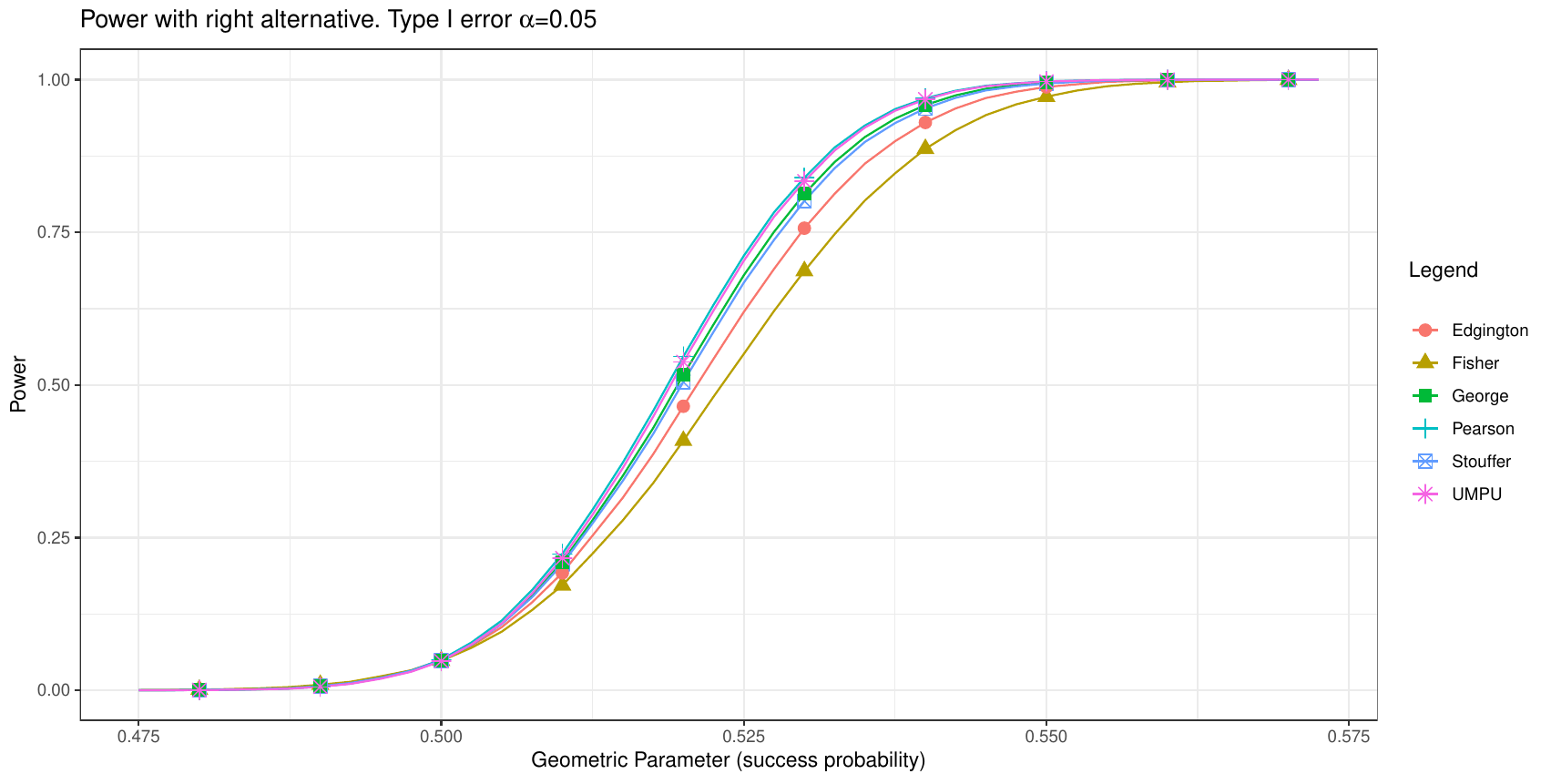} 
			\caption{Power curves for combining $n=1000$ geometric $p$-values under left-sided (left panel) and right-sided alternatives. Each curve shows the proportion of rejections (empirical power) over $N=100{,}000$ simulated datasets, as a function of the geometric probability parameter $p$. Results are shown for nominal Type I error rates $\alpha=0.05$.}
			\label{fig:powergeom_alpha005}
		\end{figure}
	\end{center}

Variances ($\Var(Z)$)and variance ratios ($\Var(Z)/\Var(Y)$) for the non-iid simulation study conducted in section \ref{sec:noniidgeom} are given in Table \ref{table:distnoniidl} for a left alternative and Table \ref{table:distnoniidr} for a right alternative.

\begin{table}[H]
	\caption{Variances (with corresponding variance ratios) for each surrogate distribution and each null parameter choice, with the rightmost column having the variance of the non i.i.d. surrogate. Left alternative. Fisher's statistic has the highest variance ratio in all cases.}
	\centering
	\begin{tabular}{l c c c c c c}
		\toprule
		& \multicolumn{4}{c}{ Variance (ratio) for $H_A: p=p_1$ with $p_1<p_0$}   \\ 
		\cmidrule(lr){2-4} \cmidrule(lr){5-5}
		Statistic 	& $p_0=0.2$ & $p_0=0.5$ & $p_0=0.8$ & Average \\ 
		\midrule
		Fisher & $3.9834 (0.9958)$  & $3.8436 (0.9609)$ & $3.2378 (0.8094)$ &  $3.6883 (0.9220)$ \\ 
		Pearson &$3.1759 (0.7939)$  & $1.9853 (0.4963)$ & $0.7982 (0.1995)$ &  $1.9865 (0.4966)$ \\ 
		George & $3.0511 (0.9274)$  & $2.5684 (0.7807)$ & $1.7419 (0.5294)$ &  $2.4538 (0.7458)$ \\ 
		Stouffer & $0.9505 (0.9505)$  & $0.8055 (0.8055)$ & $0.5223 (0.5223)$ &  $0.7594 (0.7594)$ \\ 
		Edgington & $0.0819 (0.9836)$  & $0.0714 (0.8571)$ & $0.0403 (0.4838)$ &  $0.0646 (0.7748)$ \\ 
		\bottomrule
	\end{tabular}
	\label{table:distnoniidl}
\end{table}

\begin{table}[H]
	\caption{Variances (with corresponding variance ratios) for each surrogate distribution and each null parameter choice, with the rightmost column having the variance of the non i.i.d. surrogate. Right alternative. Pearson's statistic has the highest variance ratio in all cases.}
	\centering
	\begin{tabular}{l c c c c c c}
		\toprule
		& \multicolumn{4}{c}{ Variance (ratio) for $H_A: p=p_1$ with $p_1>p_0$}   \\ 
		\cmidrule(lr){2-4} \cmidrule(lr){5-5}
		Statistic 	& $p_0=0.2$ & $p_0=0.5$ & $p_0=0.8$ & Average \\ 
		\midrule
		Fisher &$3.1759 (0.7939)$  & $1.9853 (0.4963)$ & $0.7982 (0.1995)$ &  $1.9865 (0.4966)$ \\  
		Pearson &$3.9834 (0.9958)$  & $3.8436 (0.9609)$ & $3.2378 (0.8094)$ &  $3.6883 (0.9220)$ \\ 
		George & $3.0511 (0.9274)$  & $2.5684 (0.7807)$ & $1.7419 (0.5294)$ &  $2.4538 (0.7458)$ \\ 
		Stouffer & $0.9505 (0.9505)$  & $0.8055 (0.8055)$ & $0.5223 (0.5223)$ &  $0.7594 (0.7594)$ \\ 
		Edgington & $0.0819 (0.9836)$  & $0.0714 (0.8571)$ & $0.0403 (0.4838)$ &  $0.0646 (0.7748)$ \\ 
		\bottomrule
	\end{tabular}
	\label{table:distnoniidr}
\end{table}

	\begin{center}
	\begin{figure}[h!] 
		\includegraphics[width=0.9\textwidth]{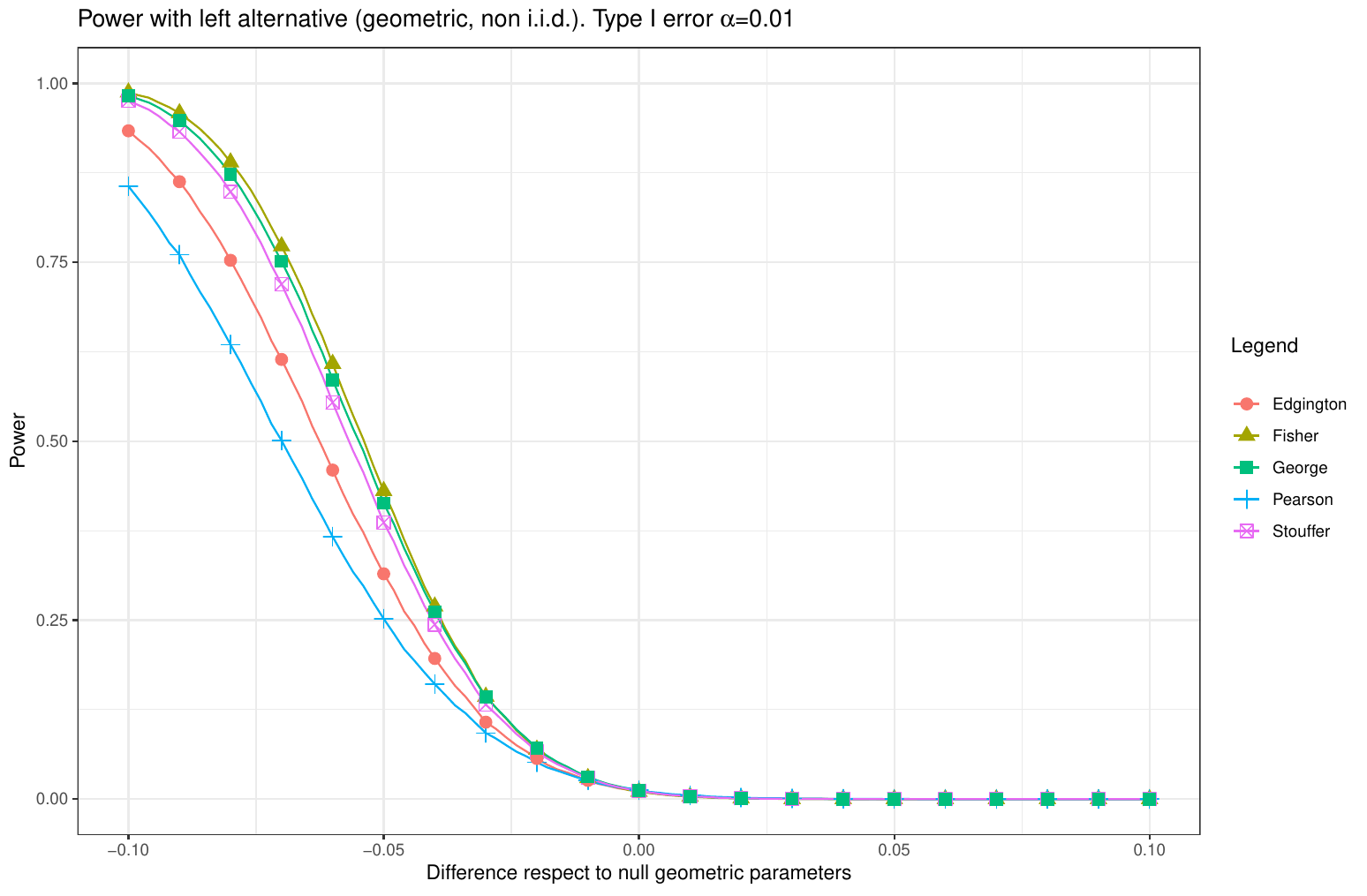} \\
		\includegraphics[width=0.9\textwidth]{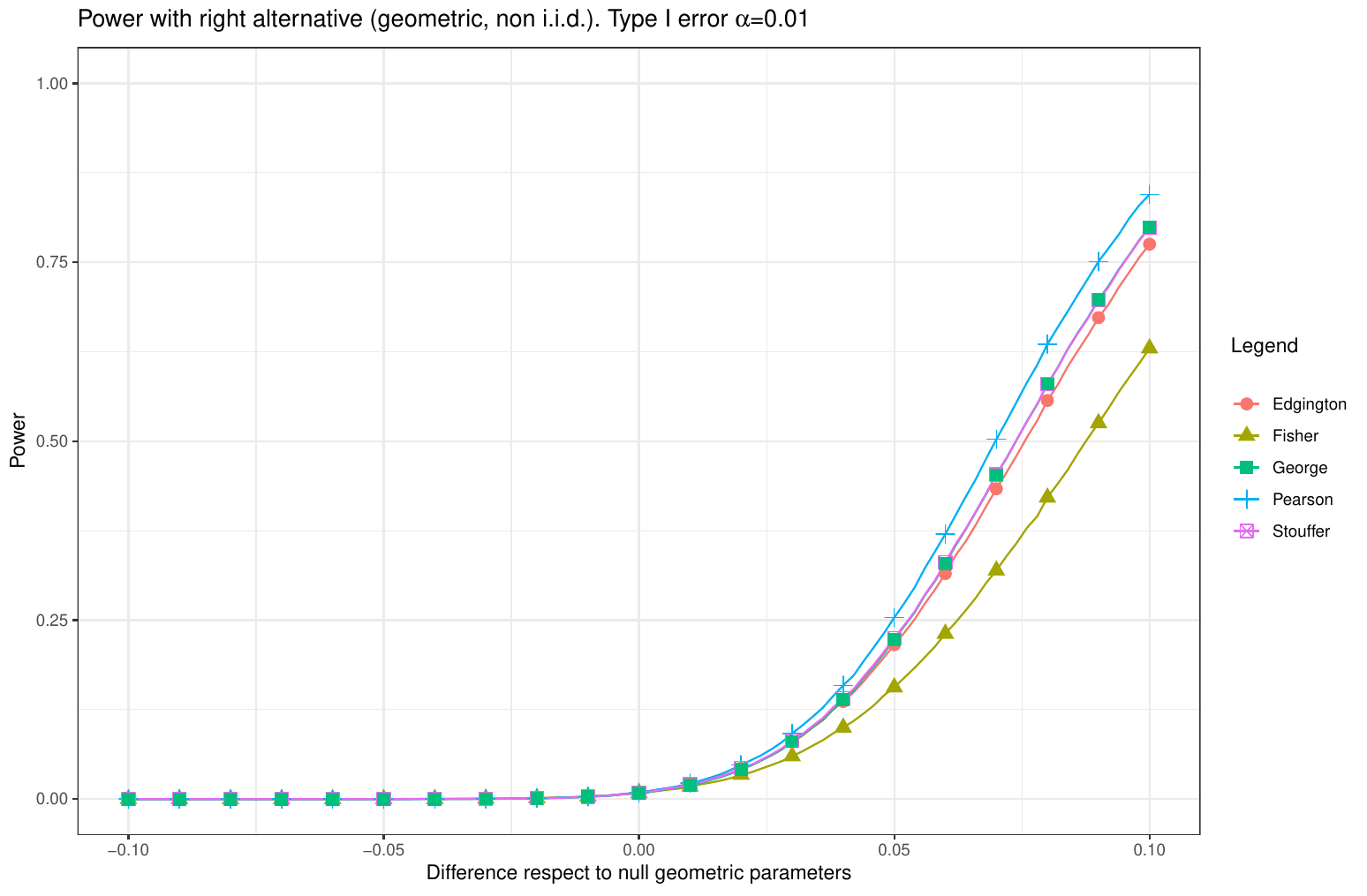} 
		\caption{Power curves for combining $n=100$ non-i.i.d. geometric $p$-values under left-sided (top) and right-sided (bottom) alternatives. Each curve shows the proportion of rejections (empirical power) over $N=100{,}000$ simulated datasets, as a function of the geometric probability parameter $p$. Results are shown for nominal Type I error rate $\alpha=0.01$. 
		}
		\label{fig:powergeomnisup}
	\end{figure}
\end{center}

\subsection*{Plots for Circular Data in Section \ref{sec:UMPU}}

	\begin{center}
	\begin{figure}[H] 
		\includegraphics[width=0.5\textwidth]{pvcontcirculara5n11} 
		\includegraphics[width=0.5\textwidth]{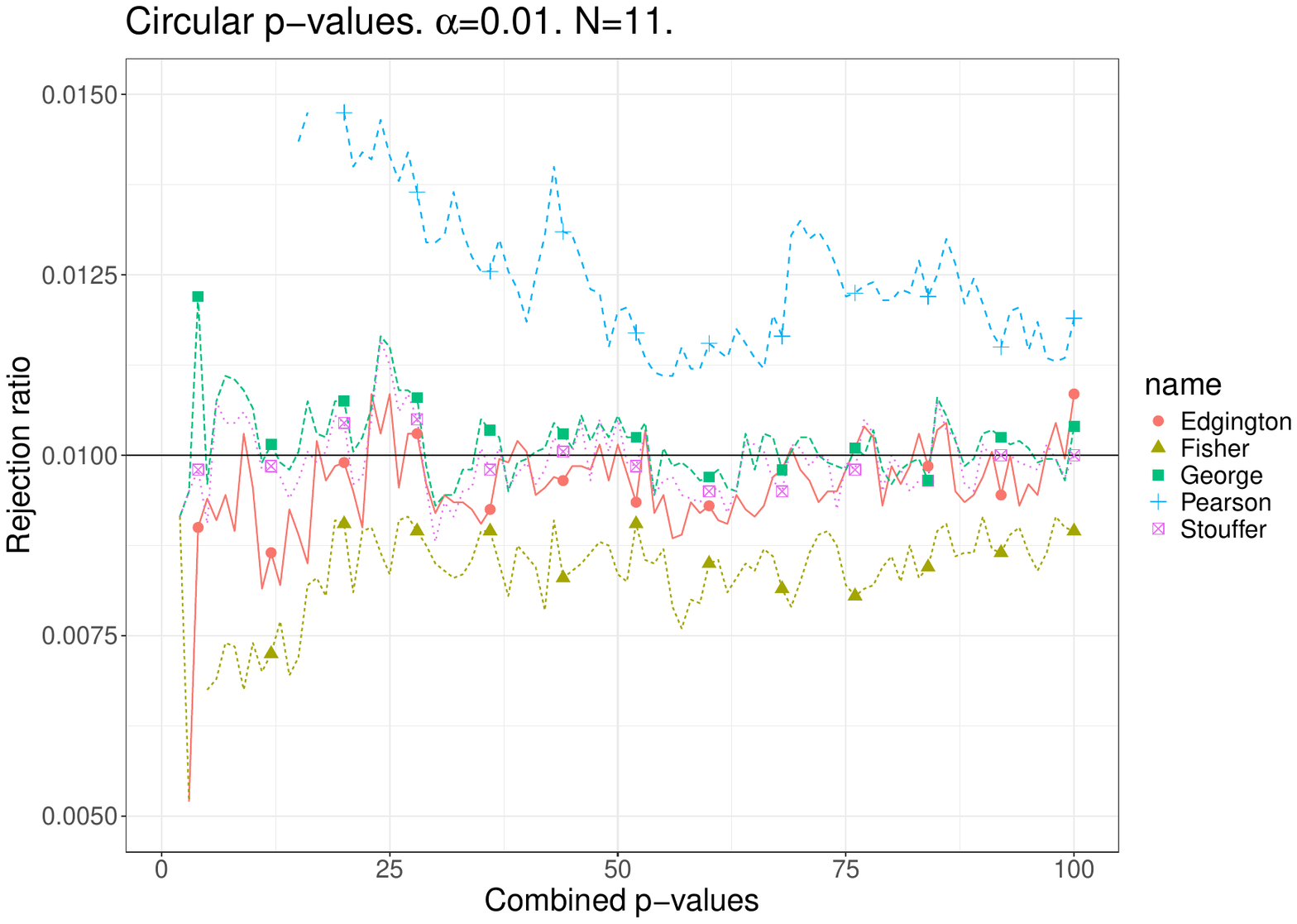} 
		\includegraphics[width=0.5\textwidth]{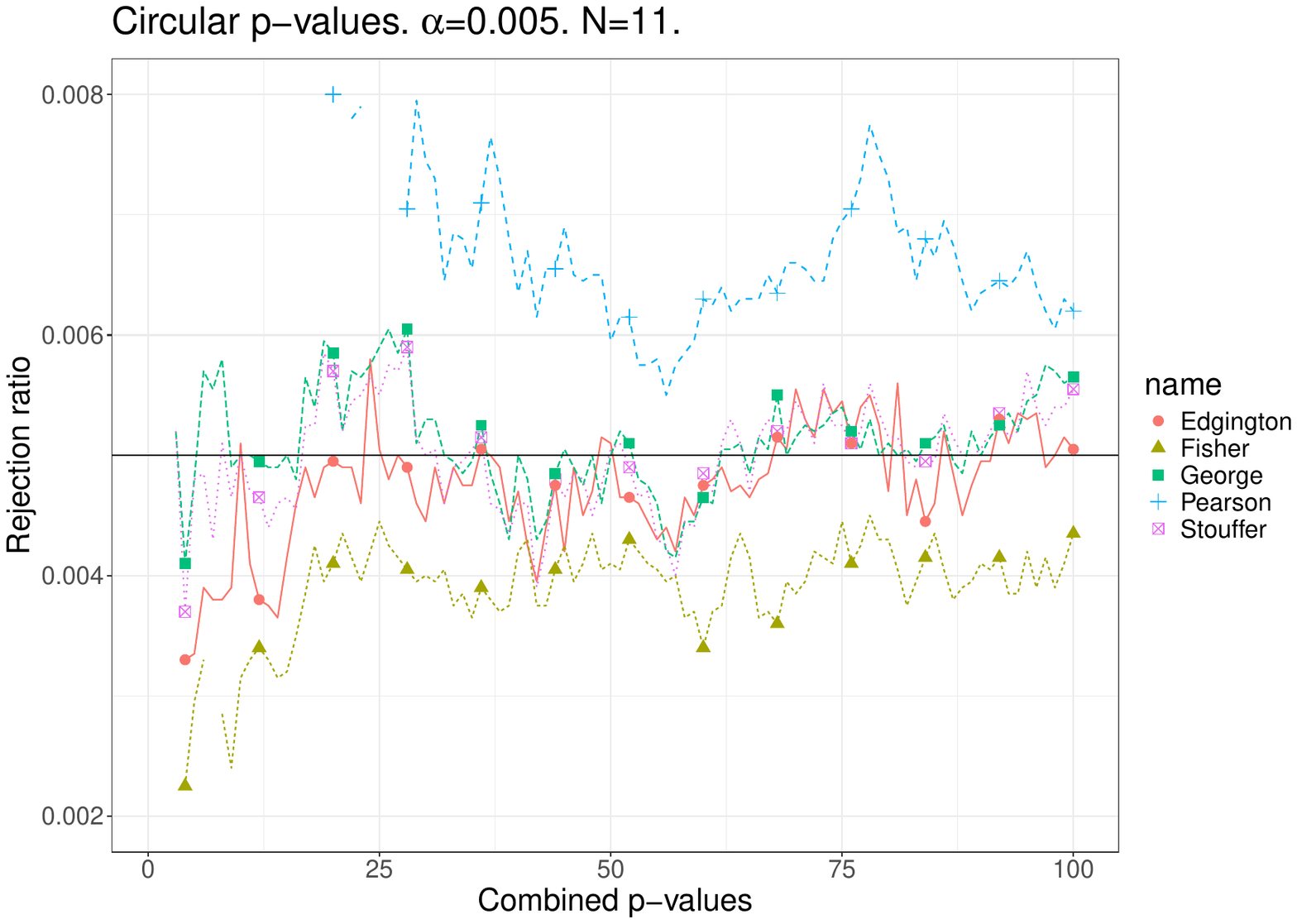} 
		\includegraphics[width=0.5\textwidth]{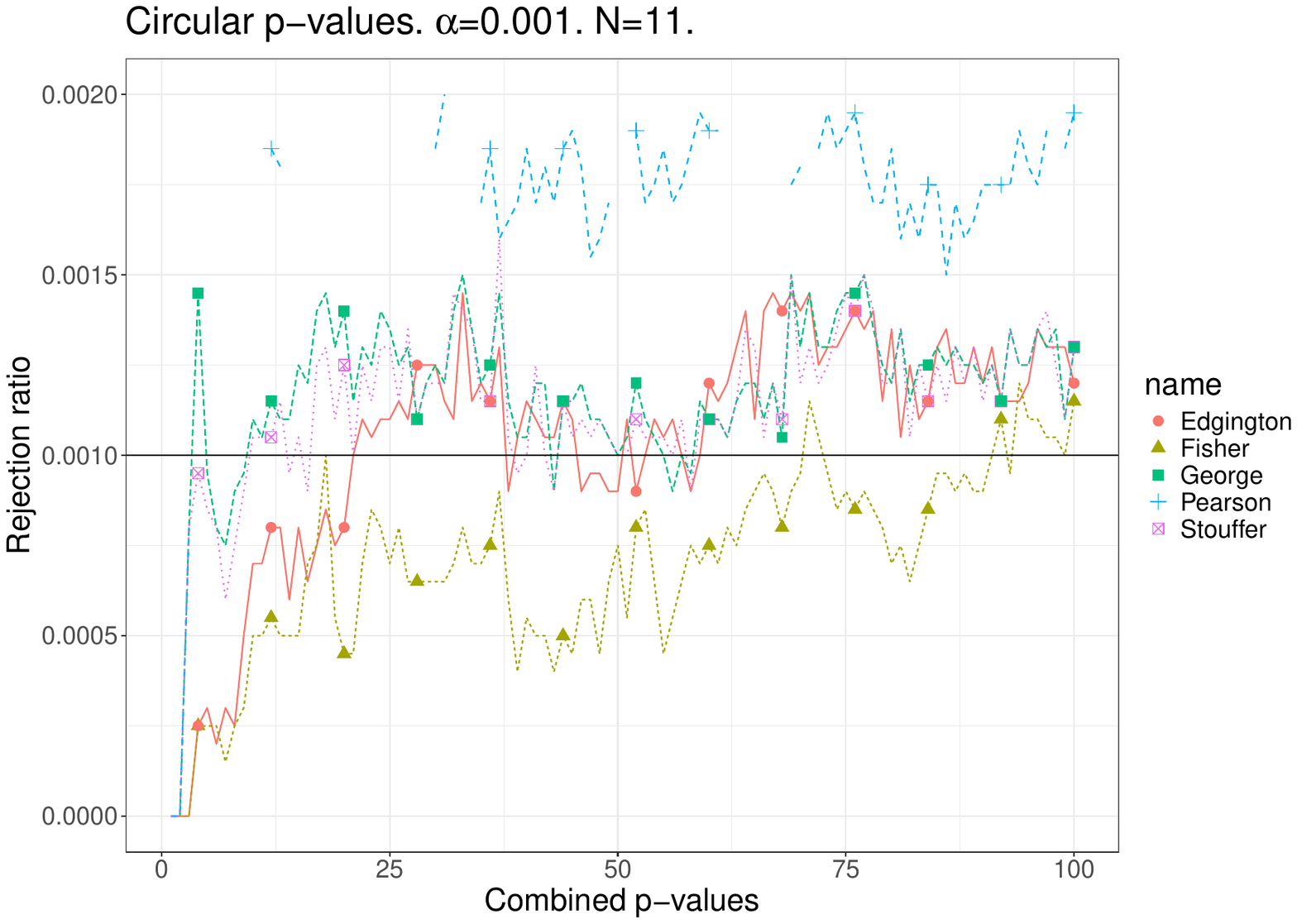} \\
		\includegraphics[width=0.5\textwidth]{pvcontcirculara5n199} 
		\includegraphics[width=0.5\textwidth]{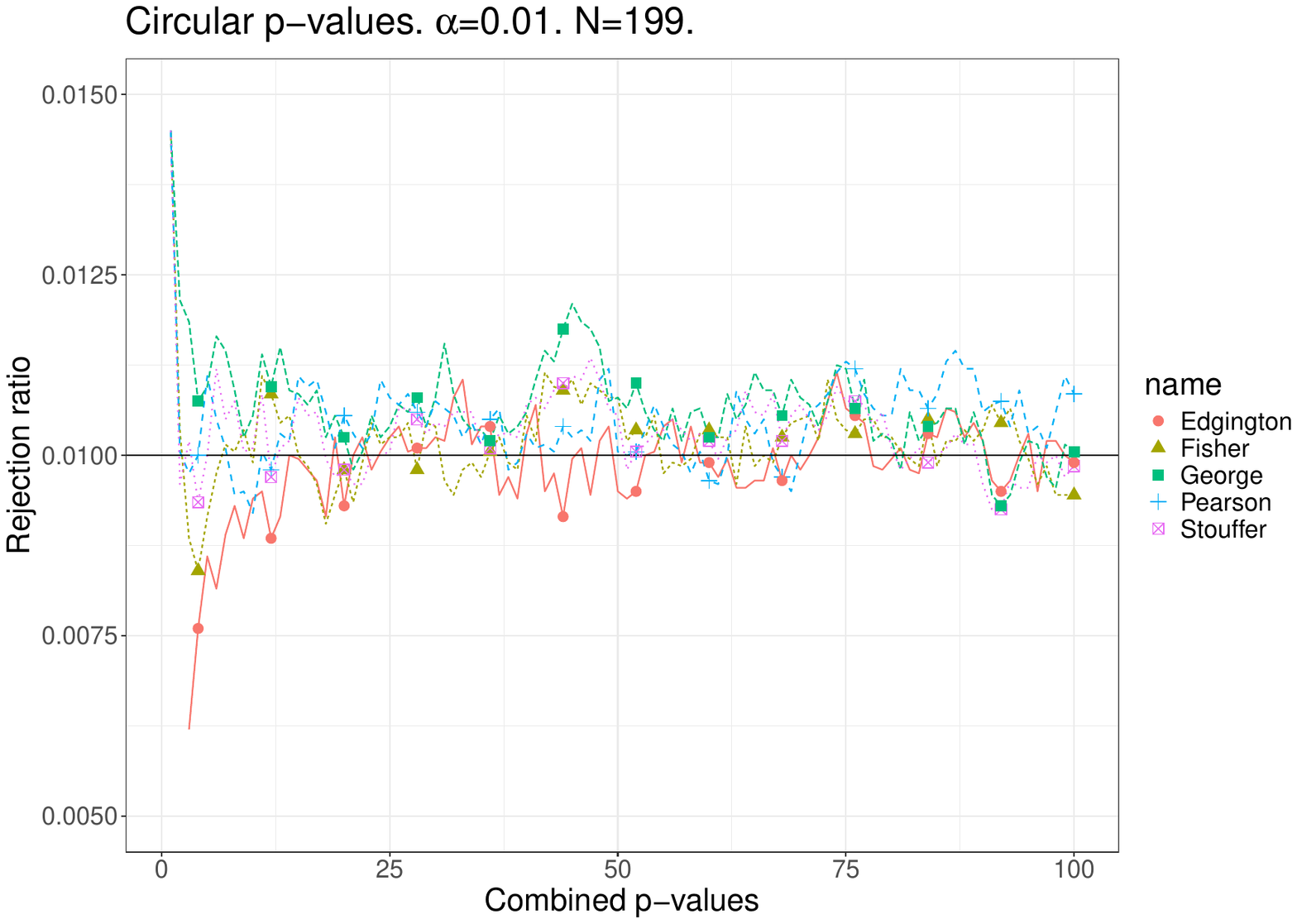} 
		\includegraphics[width=0.5\textwidth]{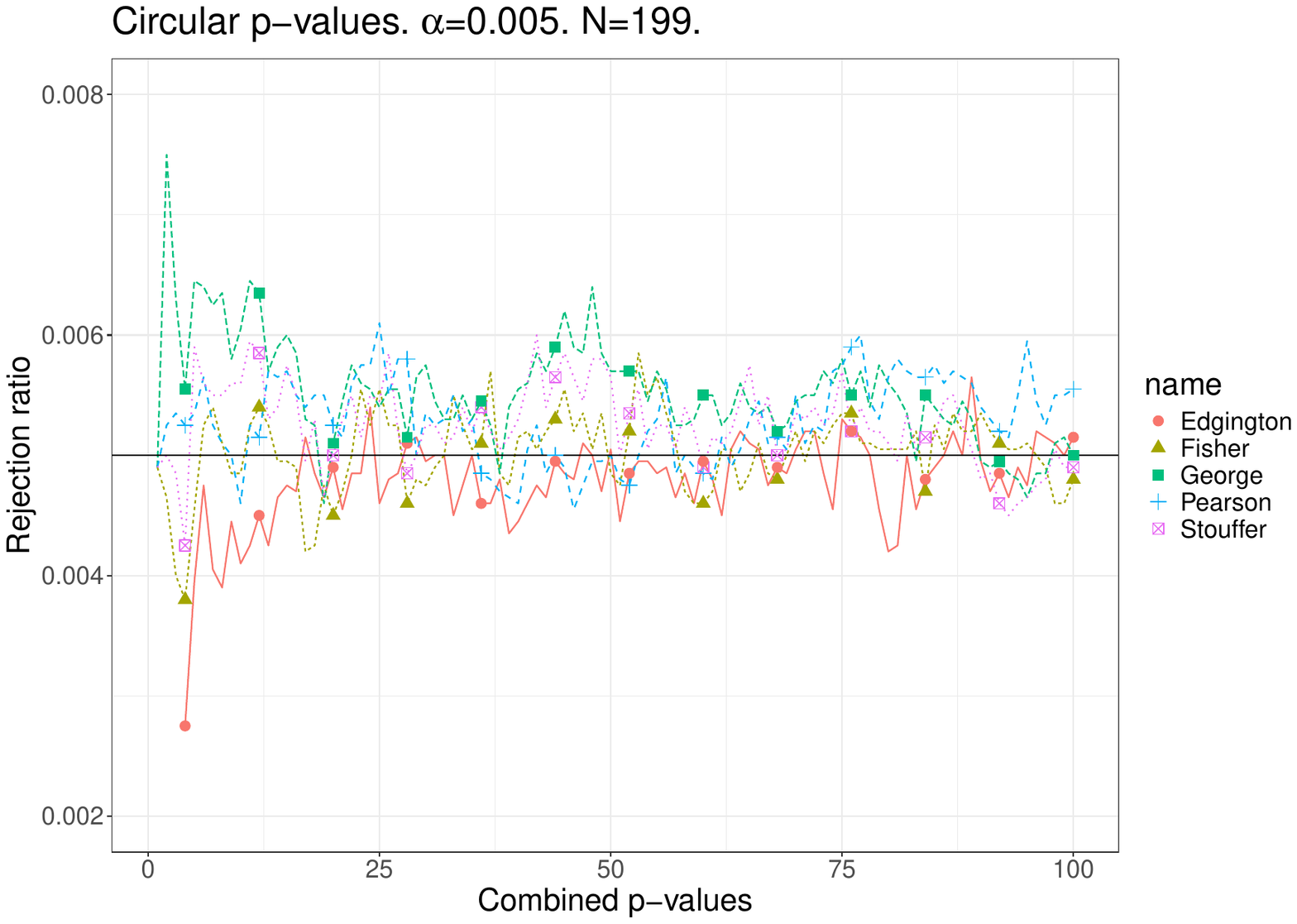} 
		\includegraphics[width=0.5\textwidth]{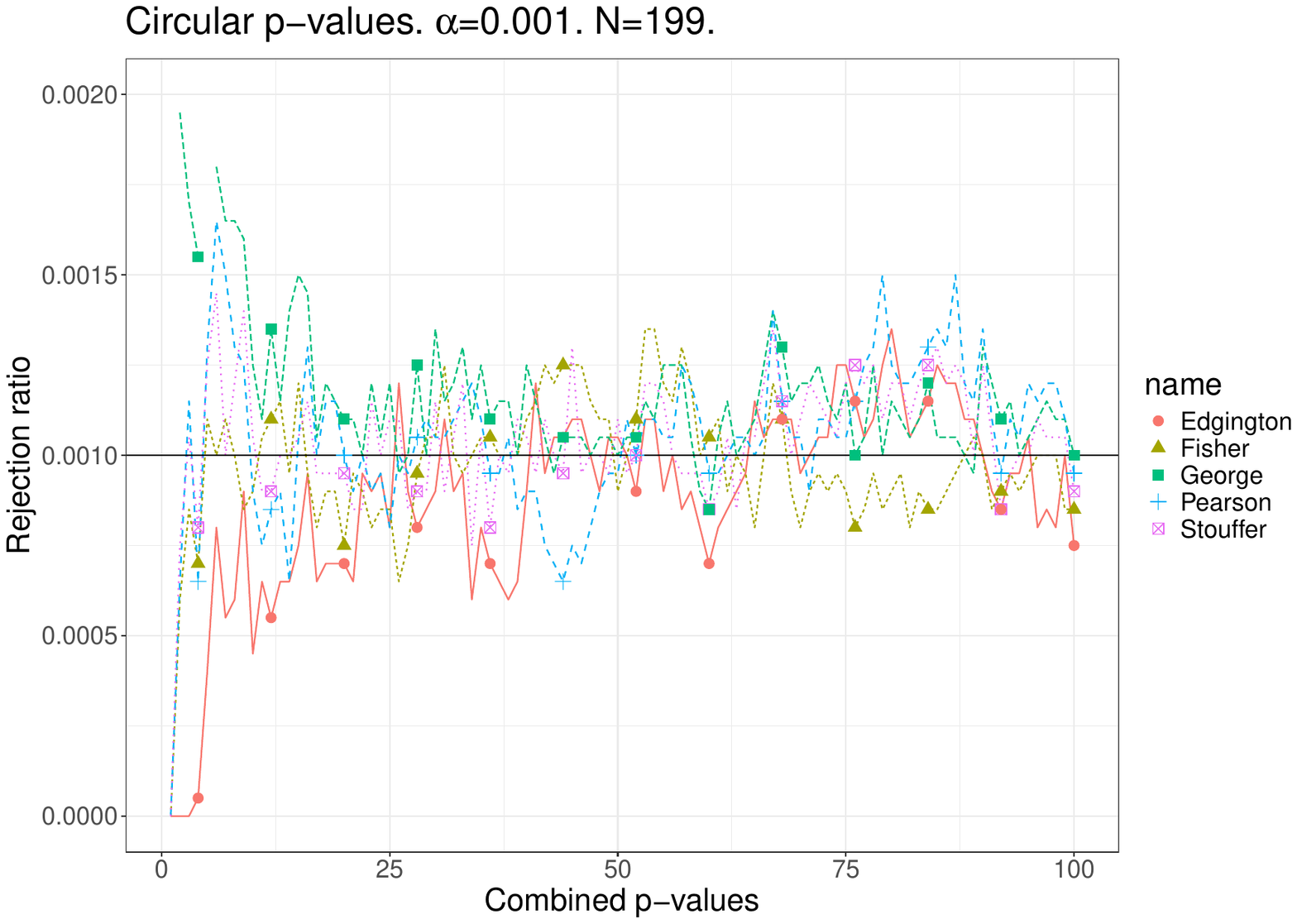} \\
		\caption{Empirical Type I error rates (proportion of rejections) as a function of the number of combined $p$-values $n$, for circular data with $N=11$ (top two rows) and $N=199$ points. Each panel corresponds to a different nominal Type I error rate $\alpha \in \{0.05, 0.01, 0.005, 0.001\}$. Results are based on 20,000 simulation replicates. The horizontal line indicates the nominal $\alpha$ level.}
		\label{fig:pvcontcirc}
	\end{figure}
\end{center}

\begin{center}
	\begin{figure}[H] 
		\includegraphics[width=0.49\textwidth]{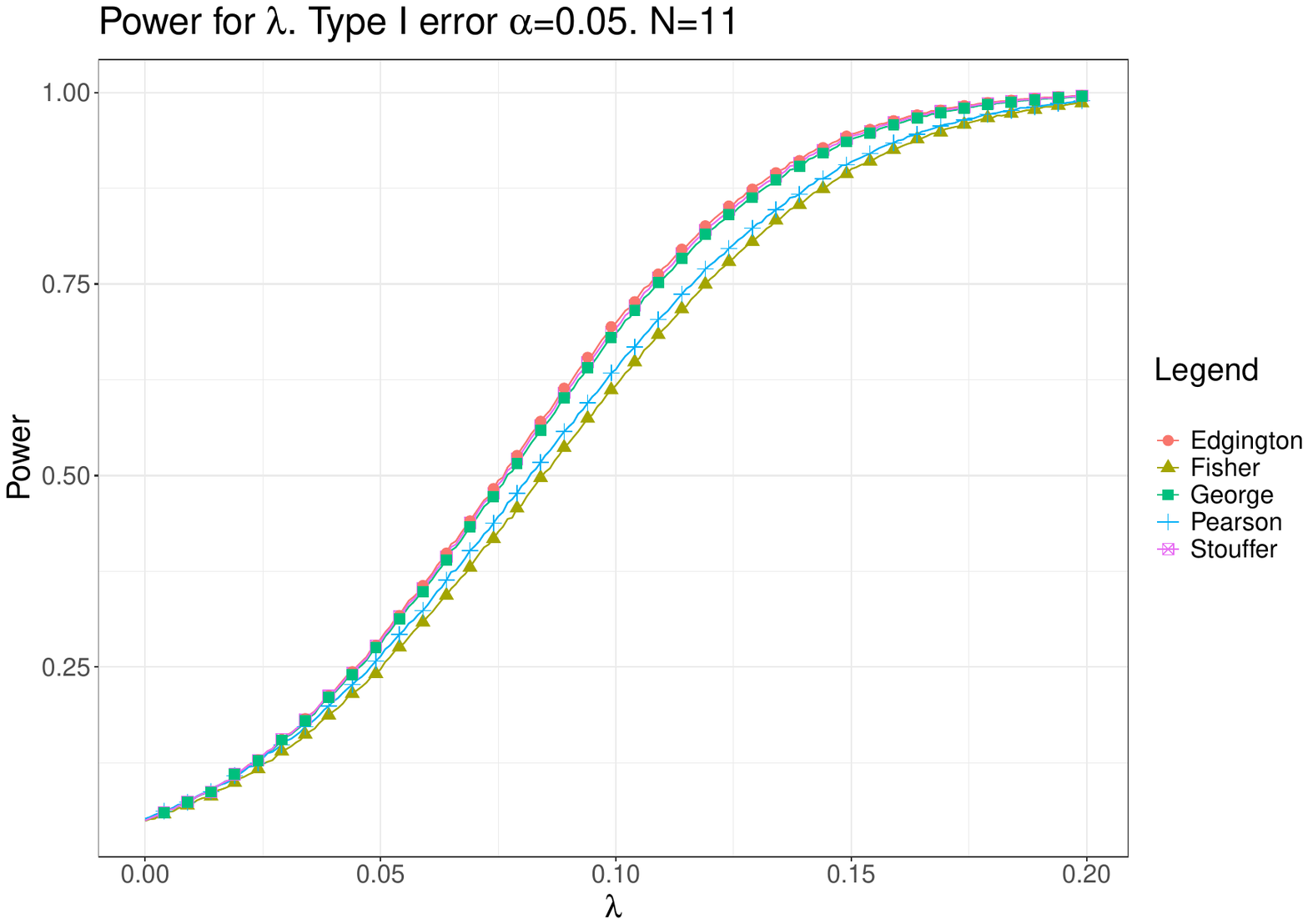} 
		\includegraphics[width=0.49\textwidth]{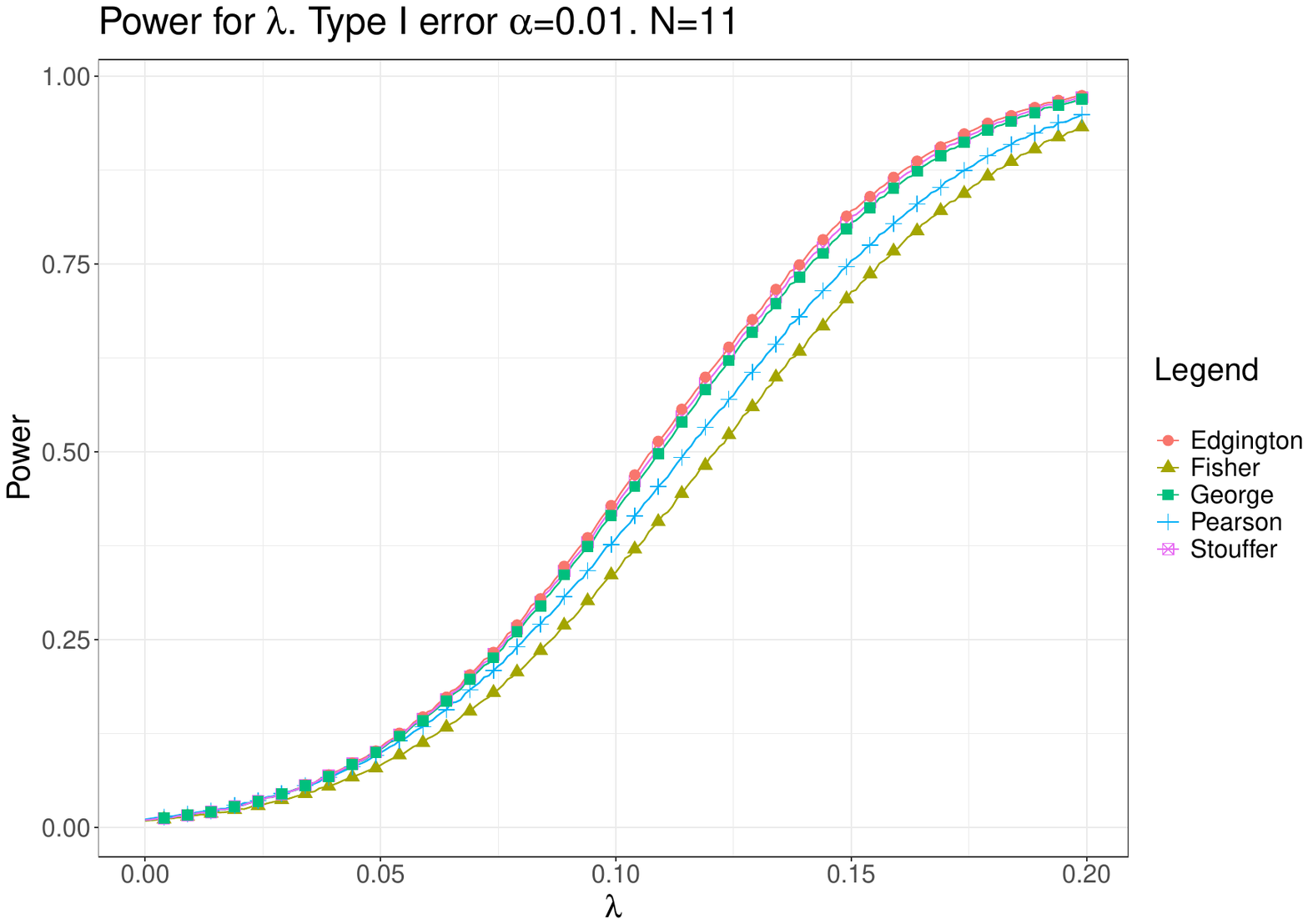} \\
		\includegraphics[width=0.49\textwidth]{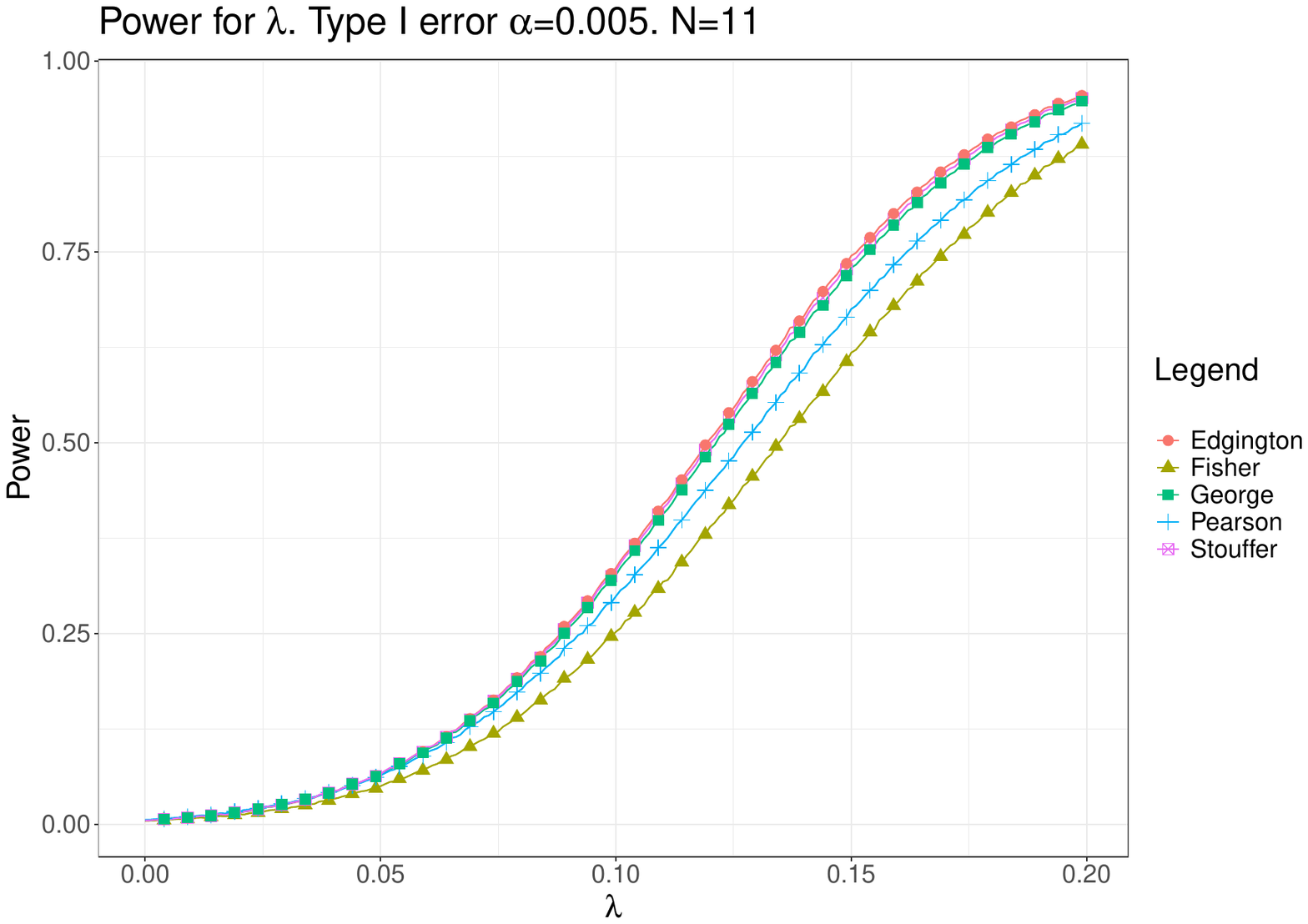} 
		\includegraphics[width=0.49\textwidth]{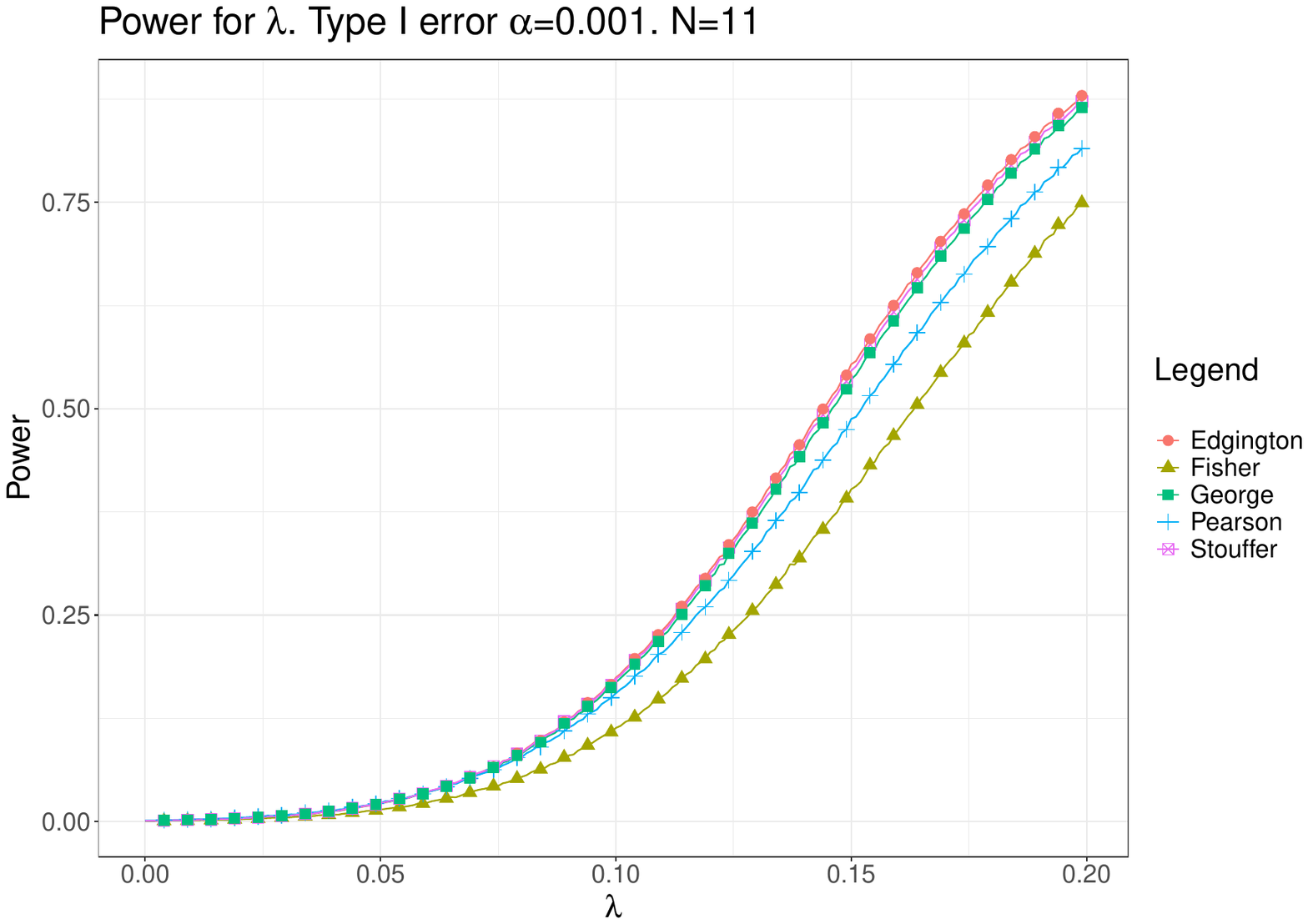} \\
		\caption{Empirical power curves for combining $n=100$ $p$-values from circular data with $N=11$ evenly spaced points on the unit circle, as a function of the alternative parameter $\lambda \in (0, 0.2)$. Each curve shows the proportion of rejections (power) over 100{,}000 simulated datasets, for four different nominal Type I error rates.}
		\label{fig:powercirc}
	\end{figure}
\end{center}

	\begin{center}
	\begin{figure}[H] 
		\includegraphics[width=0.49\textwidth]{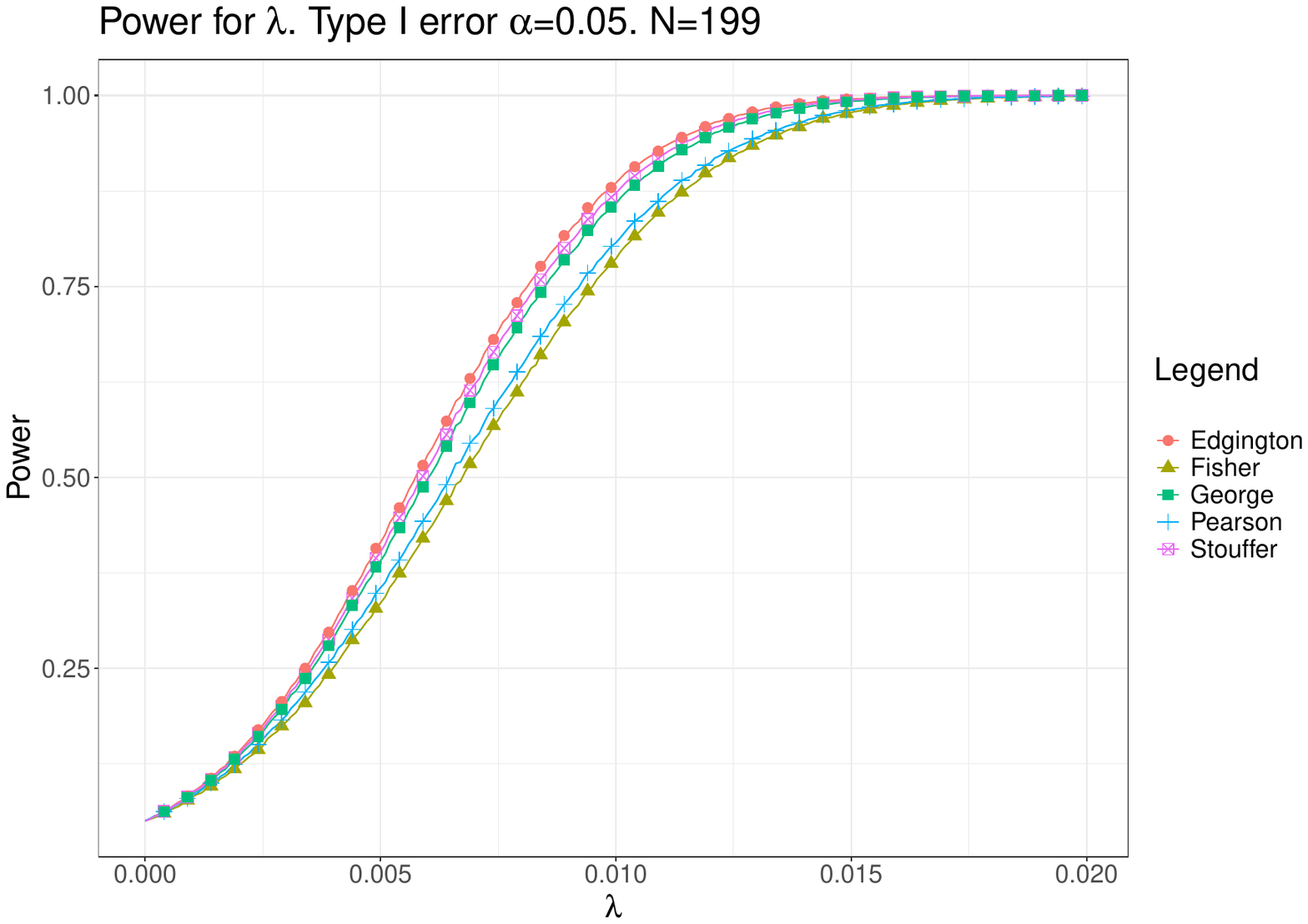} 
		\includegraphics[width=0.49\textwidth]{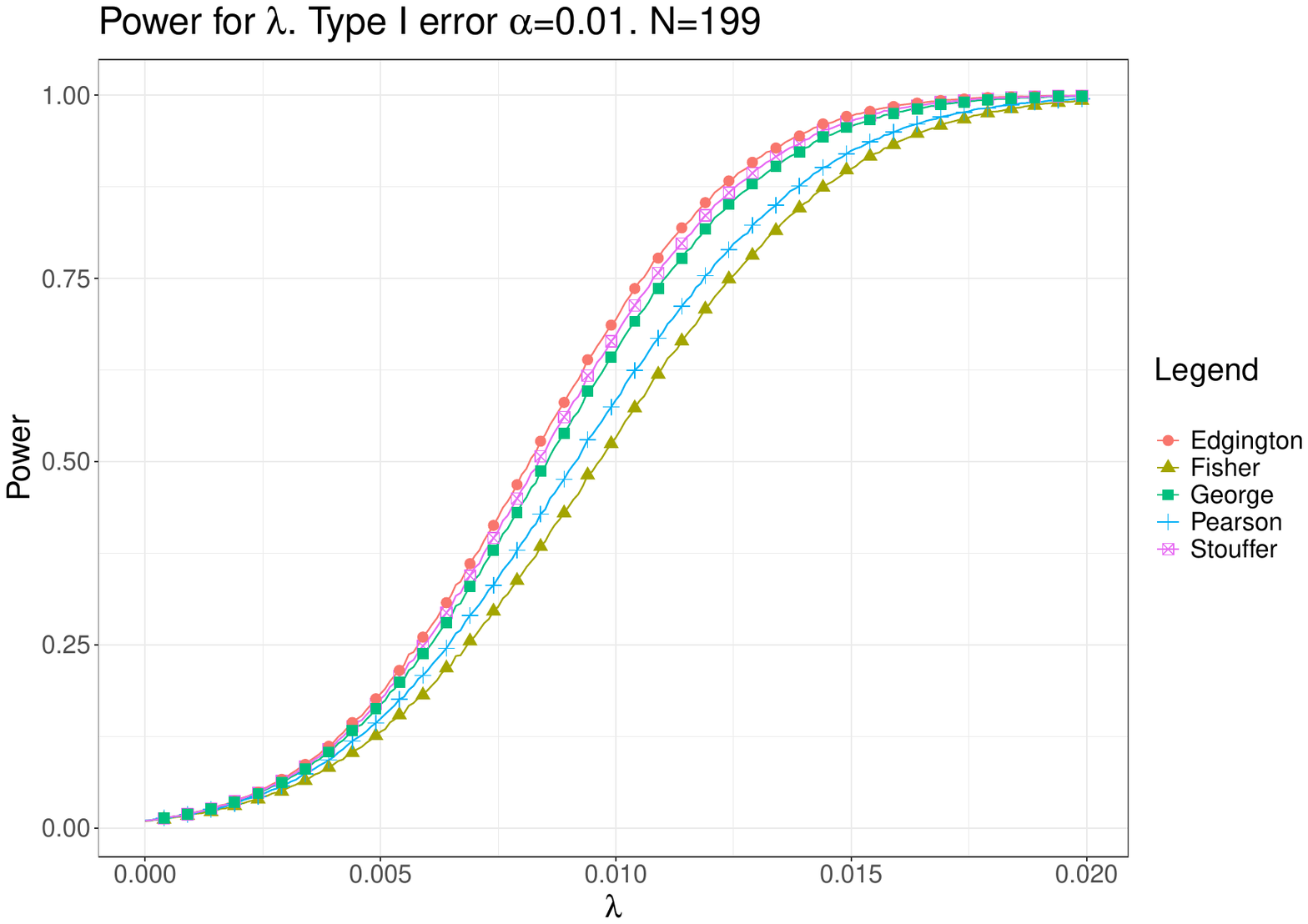} \\
		\includegraphics[width=0.49\textwidth]{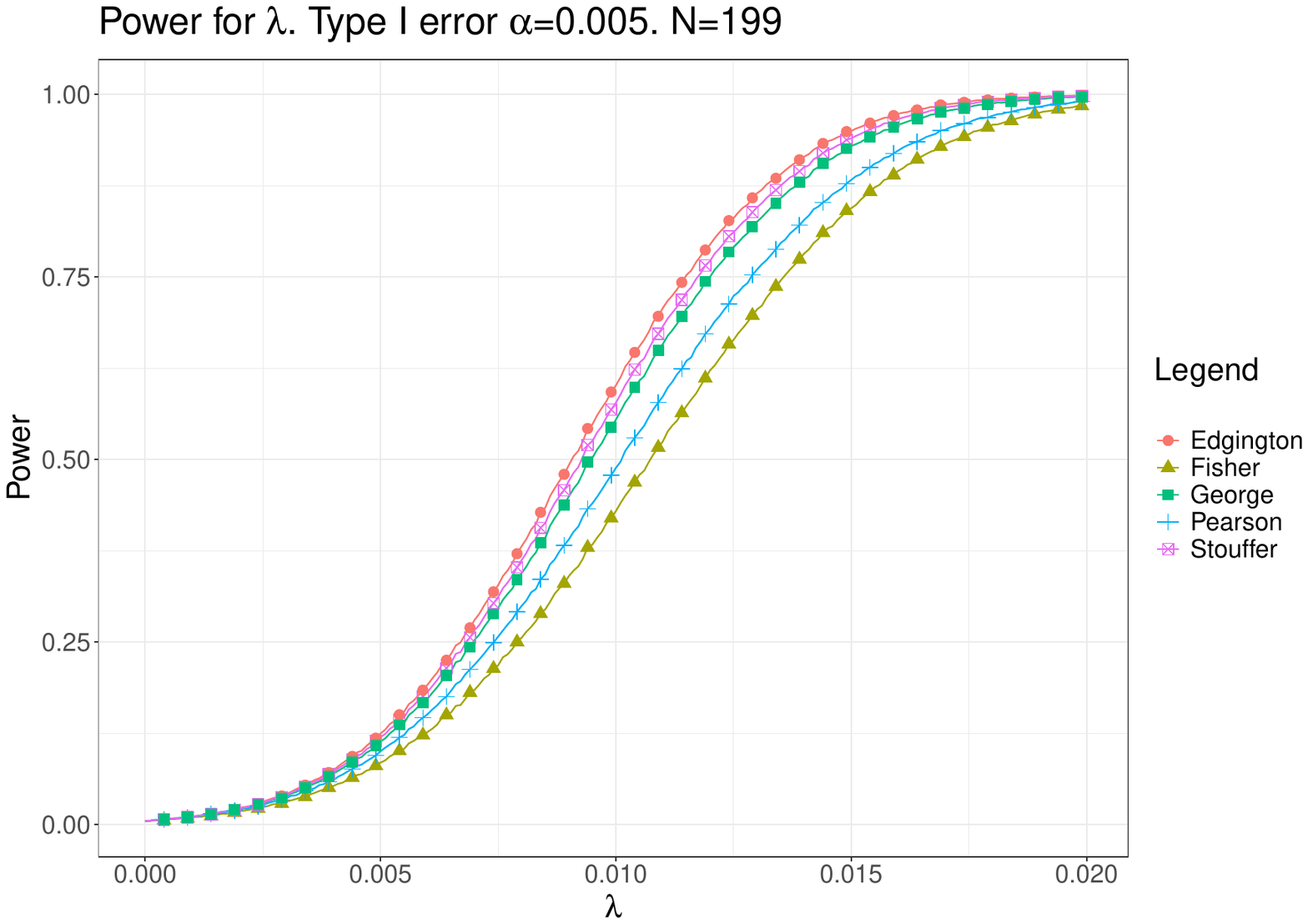} 
		\includegraphics[width=0.49\textwidth]{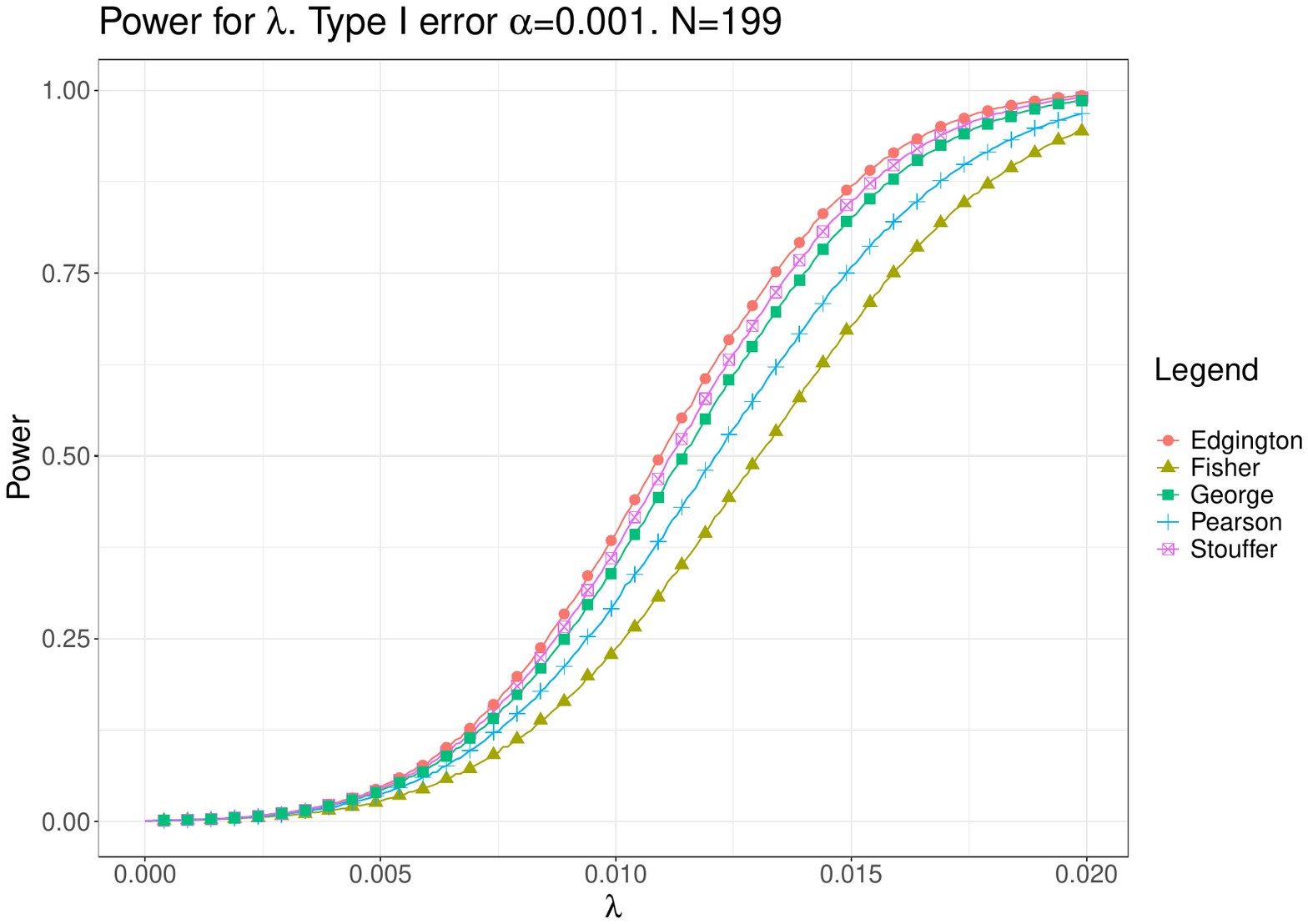} \\
		\caption{Empirical power curves for combining $n=100$ $p$-values from circular data with $N=199$ evenly spaced points on the unit circle, as a function of the alternative parameter $\lambda \in (0, 0.02)$. Each curve shows the proportion of rejections (power) over 100{,}000 simulated datasets, for four different nominal Type I error rates.}
		\label{fig:powercirc2}
	\end{figure}
\end{center}

\section{Supplementary Material for Example in Section \ref{sec:example}}\label{sec:casecontrol}

\begin{table}[H]
	\caption{Summary of mutation counts per SNP.}
	\centering
	\begin{tabular}{l c c c c c c c c c c c c c c c}
		\toprule
		SNP & 1  & 2  & 3  & 4  & 5  & 6  & 7  & 8  & 9  & 10 & 11 & 12 & 13 & 14 & 15 \\ 
		\midrule
		Mutations in total & 19 & 16 & 16 & 10 & 13 & 12 & 10 & 12 & 11 & 16 & 19 & 9  & 14 & 8  & 7  \\ 
		Mutations in cases & 13 & 11 & 11 & 7  & 9  & 8  & 7  & 8  & 8  & 11 & 10 & 3  & 6  & 5  & 4  \\ 
		\bottomrule
	\end{tabular}
	\label{table:casecontrol}
\end{table}

\section{Additional Example: Binomial Data Analysis}\label{sec:bin}
Through a simulation study, we illustrate the workflow of model comparison and global testing using the proposed tools for data analysis. We consider i.i.d. discrete test statistics $X_j \sim \text{Binomial}(5,\theta)$, for $j=1,\ldots,n$ and $\theta \in (0.001,0.999)$, and their corresponding $p$-values derived from the global hypothesis: 
\begin{equation}
	\label{eq:H0A}
	H_0:\theta=\theta_0 \text{ vs. } H_A:\theta < \theta_0, 
\end{equation}
where we assume a left-sided alternative. Note that for $X \sim \text{Binomial}(n, \theta)$, if $Y \sim \text{Binomial}(n, 1-\theta)$ and $k \in \{0, \ldots, n\}$, then $\P(X \leq k) = \P(Y \geq n - k)$. Thus, the results of this numerical study for left-sided $p$-values can be directly extended to right-sided $p$-values by symmetry. For a fixed value of $\theta$, we have $F_0 = 0$ and $F_i = \sum_{k=0}^{i-1} \binom{5}{k} \theta^k (1-\theta)^{5-k}$ for $i \in \{1,2,3,4,5,6\}$. The largest value of $F_i - F_{i-1}$ occurs when $\theta \approx i/5$, meaning the $p$-value distribution has a large mass on the left for small $\theta$ and on the right for large $\theta$.

\begin{center}
	\begin{figure}[h!] 
		\includegraphics[width=1\textwidth]{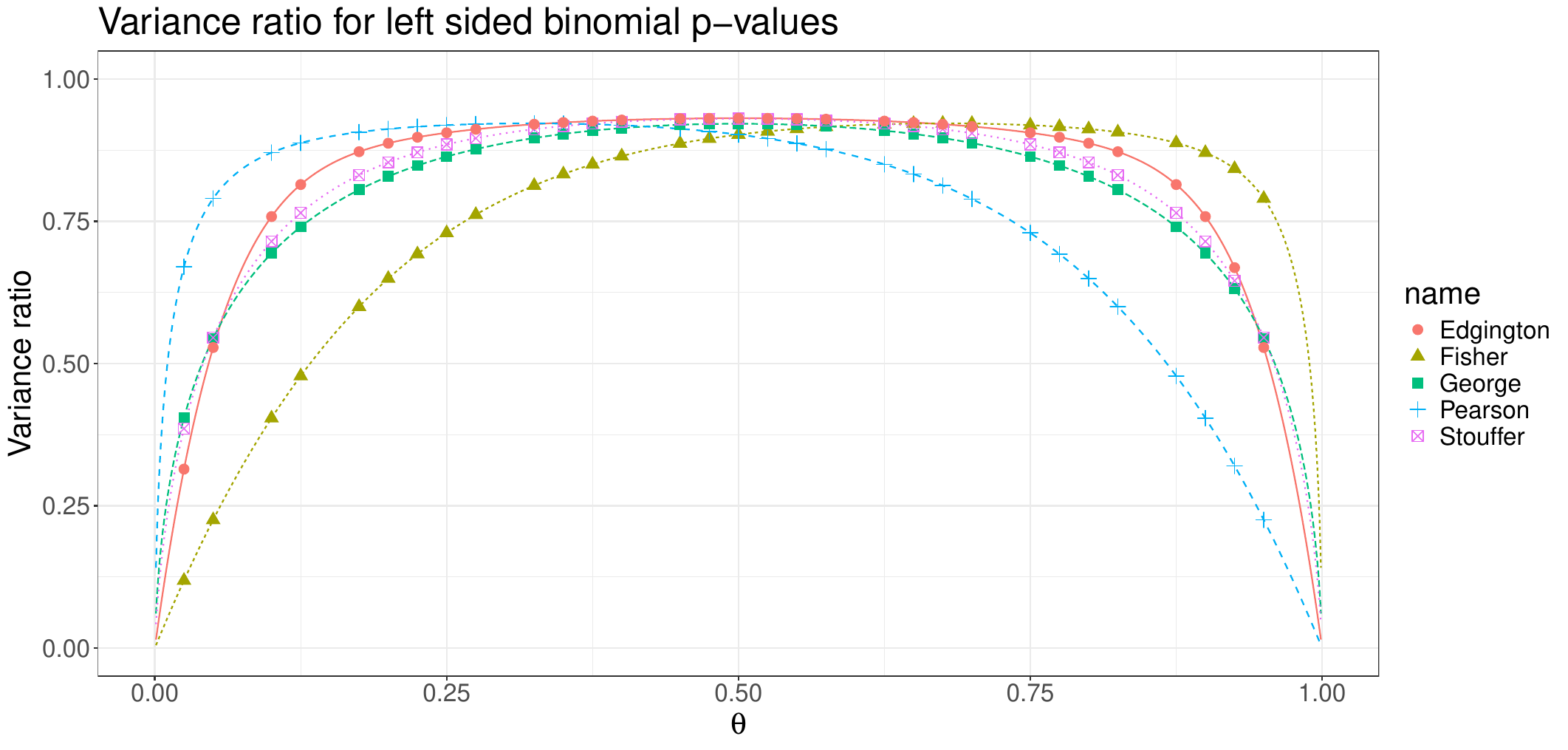} 
		\caption{Variance ratios $\Var(Z)/\Var(Y)$ for all proposed combination statistics as a function of the binomial probability parameter $\theta \in (0.001, 0.999)$.}
		\label{fig:varratios}
	\end{figure}
\end{center}

We begin by computing the variance ratios ($\Var(Z)/\Var(Y)$) introduced in Section~\ref{sec:t1e} for all proposed combination statistics in \eqref{eq:combinations} as a function of $\theta$. As shown in Figure~\ref{fig:varratios}, Pearson's statistic \eqref{eq:sumP} achieves the highest variance ratio for small $\theta$, while Fisher's statistic \eqref{eq:sumF} achieves the highest ratio for large $\theta$. For intermediate values of $\theta$ (e.g., $\theta \in (0.4, 0.6)$), the other statistics (Stouffer, Edgington, and George) have higher variance ratios. This demonstrates that the choice of combination method, according to the variance ratio criterion, depends on the underlying value of $\theta$.

We focus on three representative null parameter values to analyze Type I error control:
\begin{itemize}
	\item $\theta_0=0.1$, for which the CDF of the null $p$-value is $F_0=0, F_1=0.59049, F_2=0.91854, F_3=0.99144, F_4=0.99954, F_5=0.99999, F_6=1$, with a large mass on the left of its distribution. As shown in Figure \ref{fig:varratios}, Pearson's statistic \eqref{eq:sumP} achieves the largest variance ratio and Fisher's statistic \eqref{eq:sumF} achieves the smallest, whereas the other three statistics achieve similar ratios among themselves, with numerical values $\nu_P/4=0.871 > \nu_E/(1/12)=0.758 > \nu_S=0.715 > \nu_G/(\pi^2/3)=0.694 > \nu_F/4=0.404$. Given both the shape of the CDF and the variance ratios, our recommendation is to use Pearson's statistic in this case.
		
	\item $\theta_0=0.5$, for which the CDF of the null $p$-value is $F_0=0, F_1=0.03125, F_2=0.18750, F_3=0.5, F_4=0.81250, F_5=0.96875, F_6=1$, with the largest masses at the center of its distribution. As shown in Figure \ref{fig:varratios}, all five statistics achieve high variance ratios, with very similar numerical values: $\nu_S=0.932 > \nu_E/(1/12)=0.931 > \nu_G/(\pi^2/3)=0.921 > \nu_P/4=\nu_F/4=0.902$. Due to their very similar variance ratios, our recommendation would be to use either Stouffer's, Edgington's, or George's statistic in this case. 
		
	\item $\theta_0=0.9$, for which the CDF of the null $p$-value is $F_0=0, F_1=0.00001, F_2=0.00046, F_3=0.00856, F_4=0.08146, F_5=0.40951, F_6=1$, with a very large mass on the right of its distribution. As shown in Figure \ref{fig:varratios}, in contrast to the case $\theta_0=0.1$, Pearson's statistic \eqref{eq:sumP} achieves the smallest variance ratio and Fisher's statistic \eqref{eq:sumF} achieves the largest, whereas the other three statistics achieve similar ratios among themselves, with numerical values $\nu_F/4=0.871 > \nu_E/(1/12)=0.758 > \nu_S=0.715 > \nu_G/(\pi^2/3)=0.694 > \nu_P/4=0.404$. Given both the shape of the CDF and the variance ratios, our recommendation is to use Fisher's statistic in this case.
\end{itemize}

We evaluate the performance of the recommended combination statistics through simulation studies, focusing on both Type I error control and power.

\begin{center}
	\begin{figure}[H] 
		\includegraphics[width=0.5\textwidth]{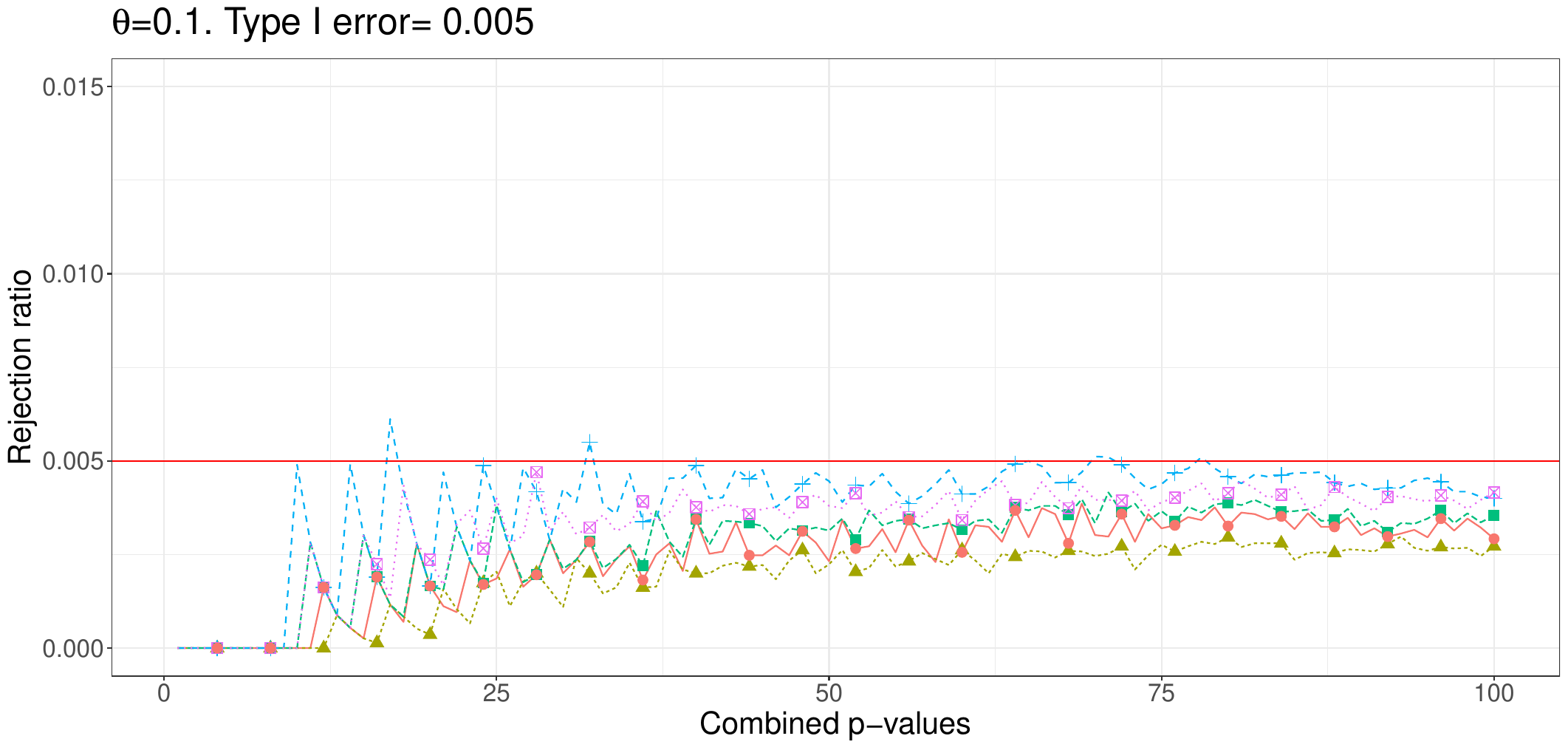}
		\includegraphics[width=0.5\textwidth]{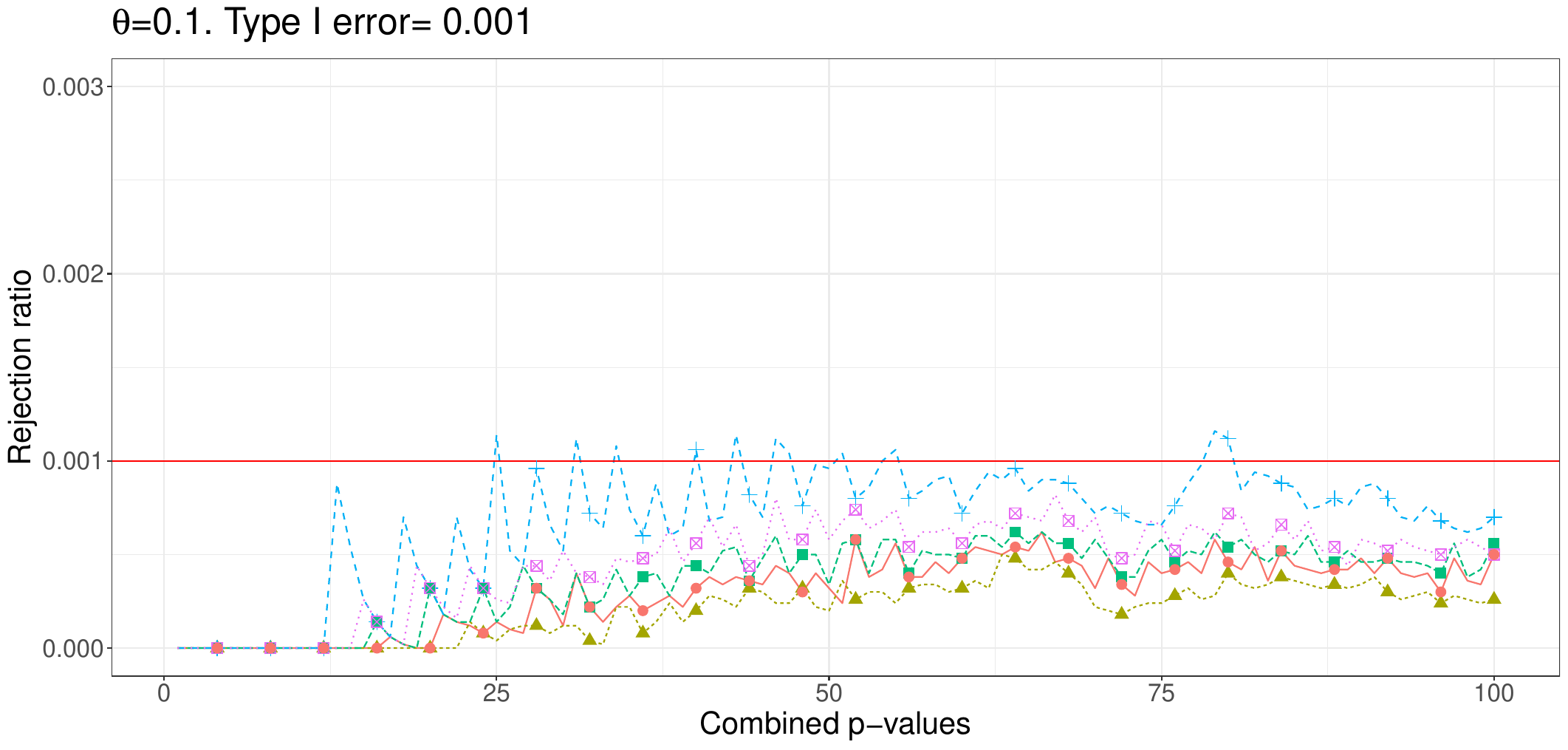}\\				
		\includegraphics[width=0.5\textwidth]{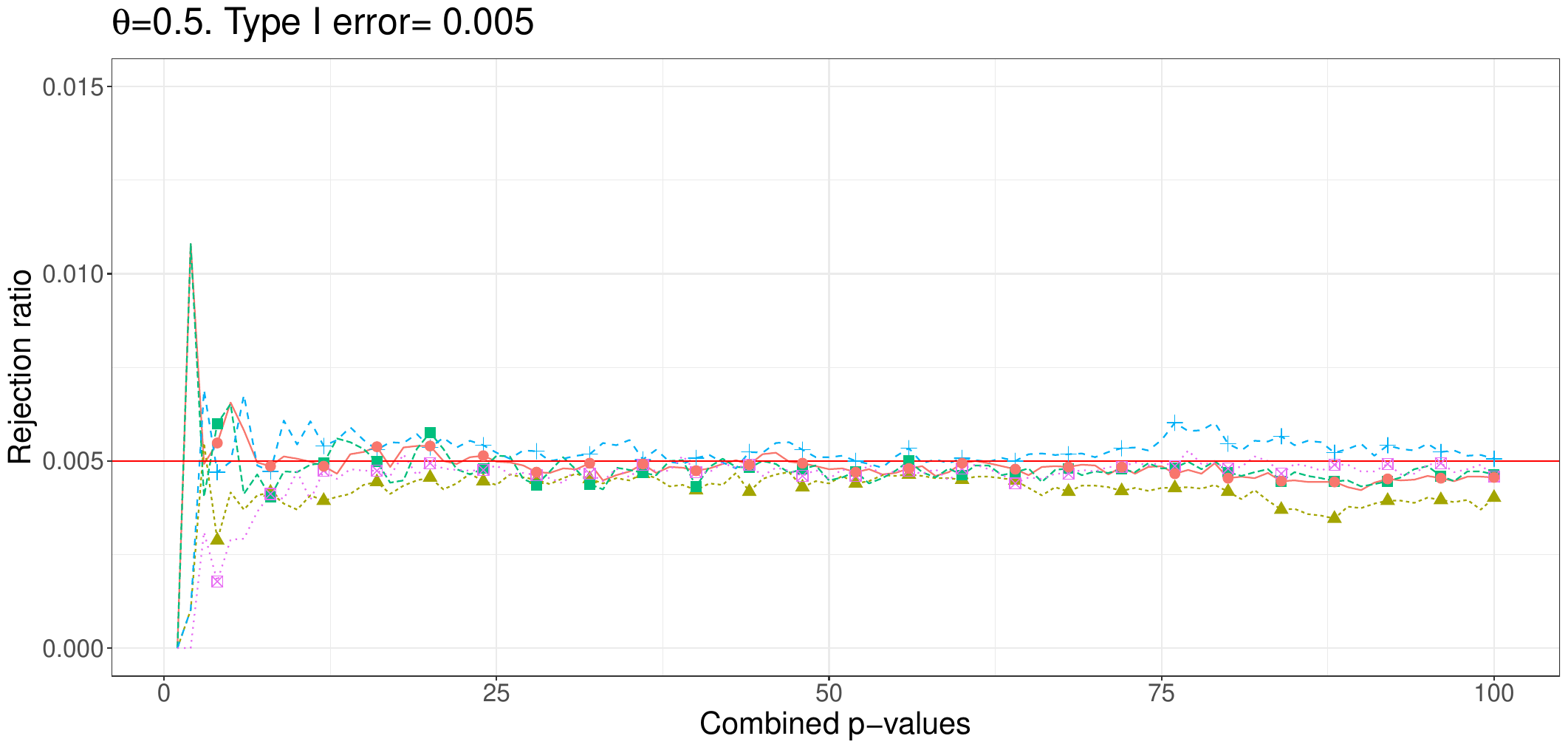}
		\includegraphics[width=0.5\textwidth]{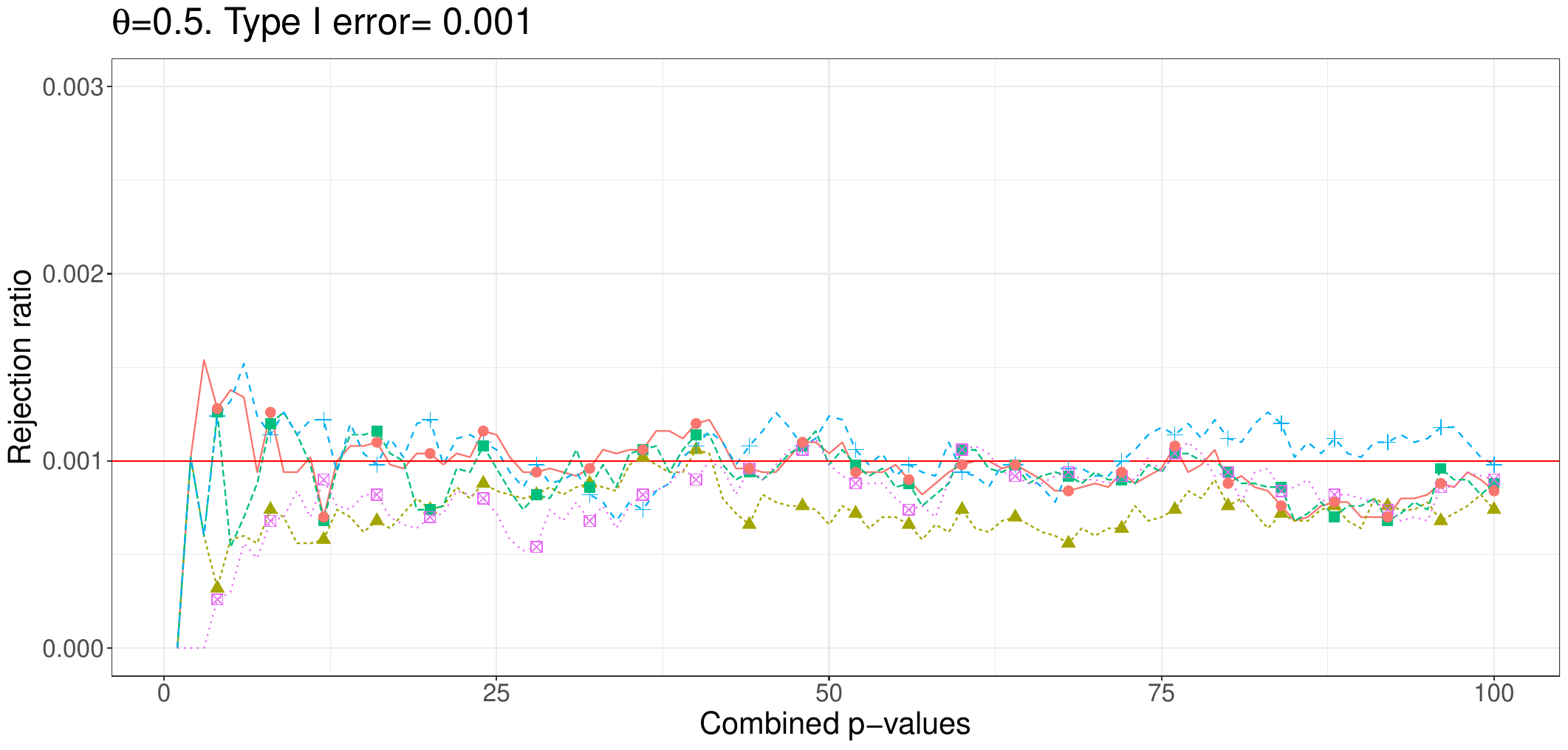}\\
		\includegraphics[width=0.5\textwidth]{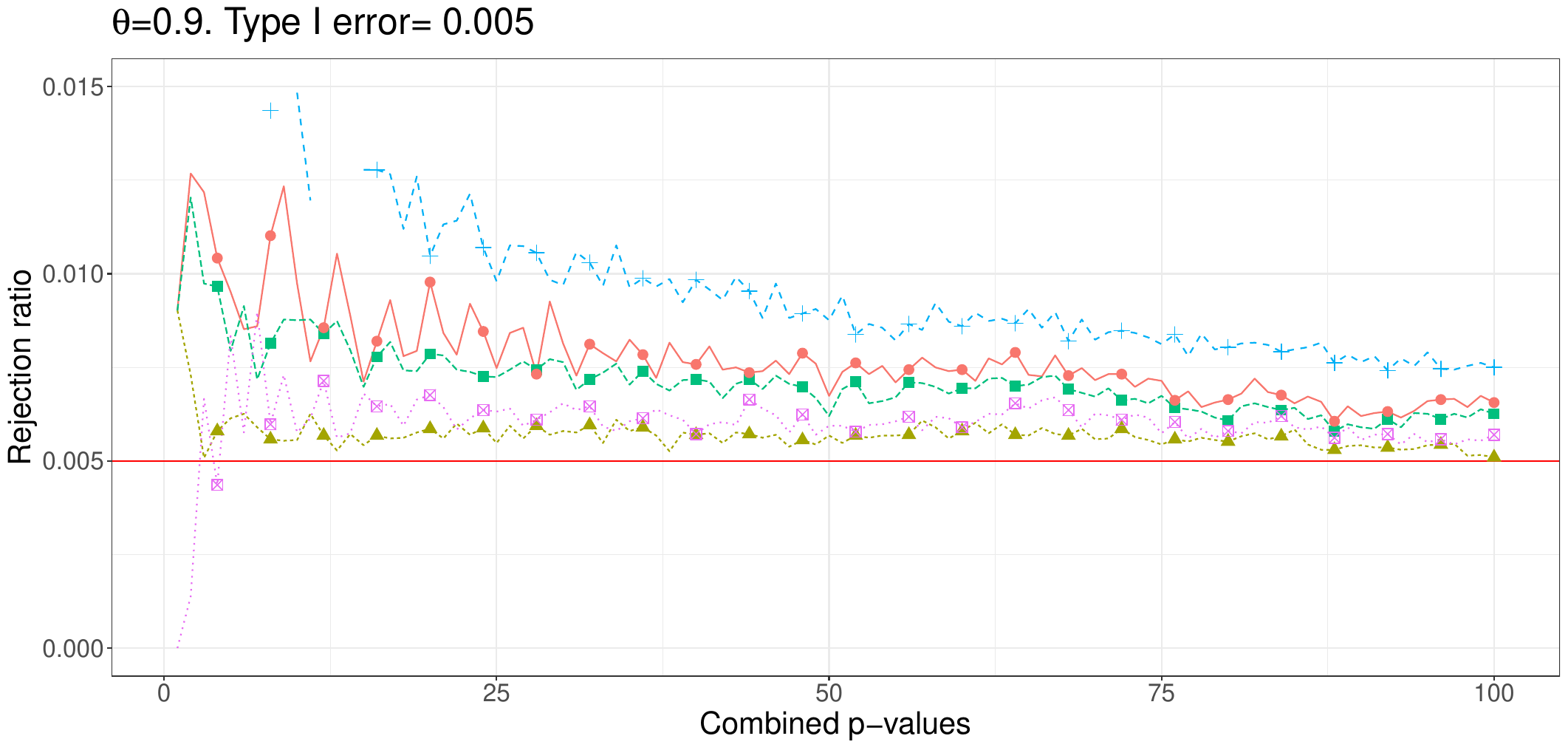}
		\includegraphics[width=0.5\textwidth]{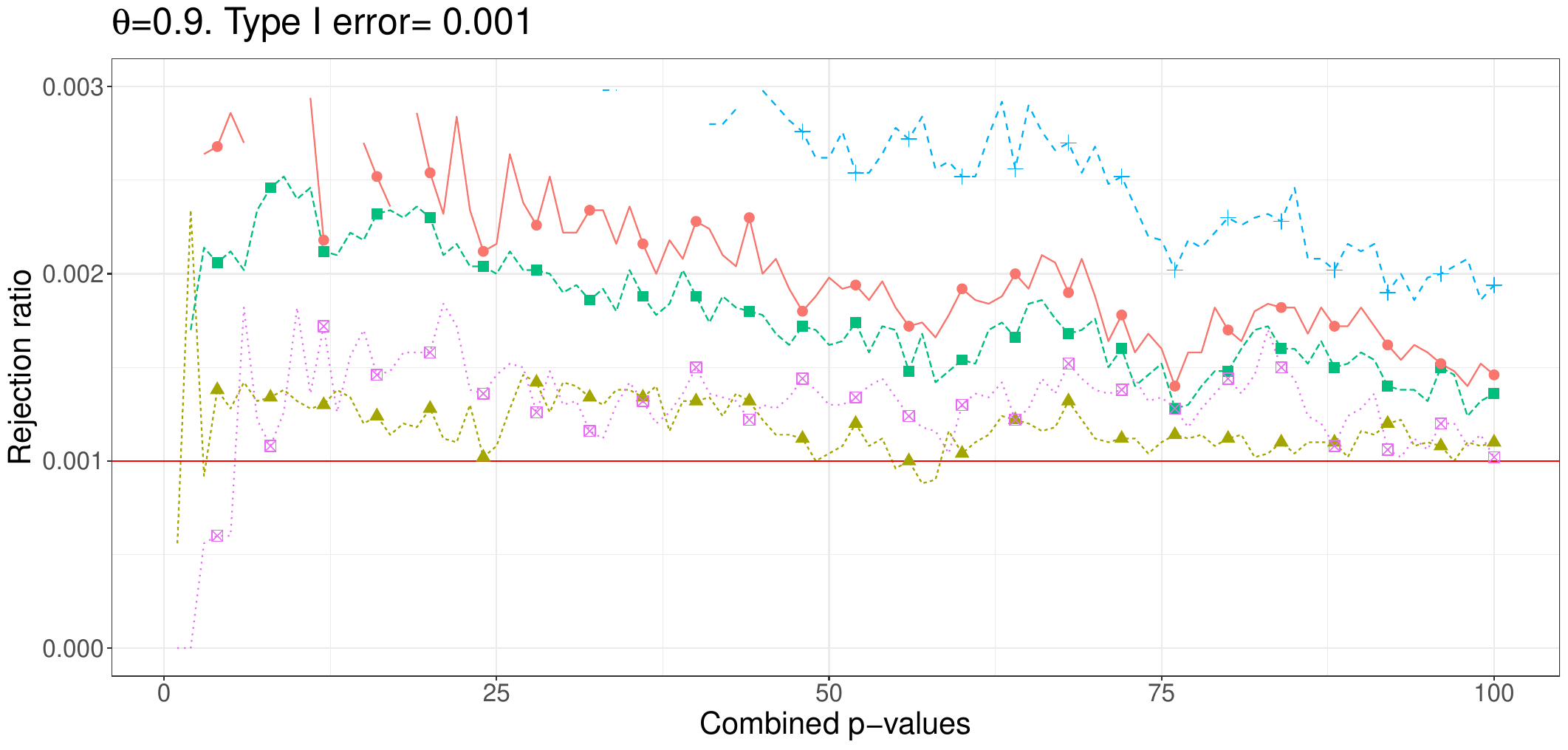} 
		\caption{Empirical Type I error rates at nominal $\alpha=0.005$ (left column) and $\alpha=0.001$ (right column) for binomial parameters $\theta_0=0.1$ (top row), $\theta_0=0.5$ (middle row), and $\theta_0=0.9$ (bottom row). Methods: Fisher (brown triangles), Pearson (blue crosses), Stouffer (purple squares with cross), Edgington (red circles), George (green solid squares). The red horizontal line indicates the nominal Type I error rate.}
		\label{fig:typeIerrornum}
	\end{figure}
\end{center}

For Type I error control, we combine up to $n=100$ Binomial($5,\theta_0$) $p$-values using the statistics in \eqref{eq:combinations}. Rejection is based on the quantiles of their surrogate distributions at significance levels $\alpha=0.005$ and $\alpha=0.001$. Each scenario is repeated 50,000 times, and the empirical Type I error is estimated as the proportion of rejections. As shown in Figure~\ref{fig:typeIerrornum}, the statistics recommended by the variance ratio criterion achieve empirical Type I error rates closest to the nominal level for $\theta_0=0.1$ and $\theta_0=0.9$. For $\theta_0=0.5$, where all variance ratios are above $0.9$, all methods provide accurate Type I error control.

\begin{center}
	\begin{figure}[H]
		\includegraphics[width=0.93\textwidth]{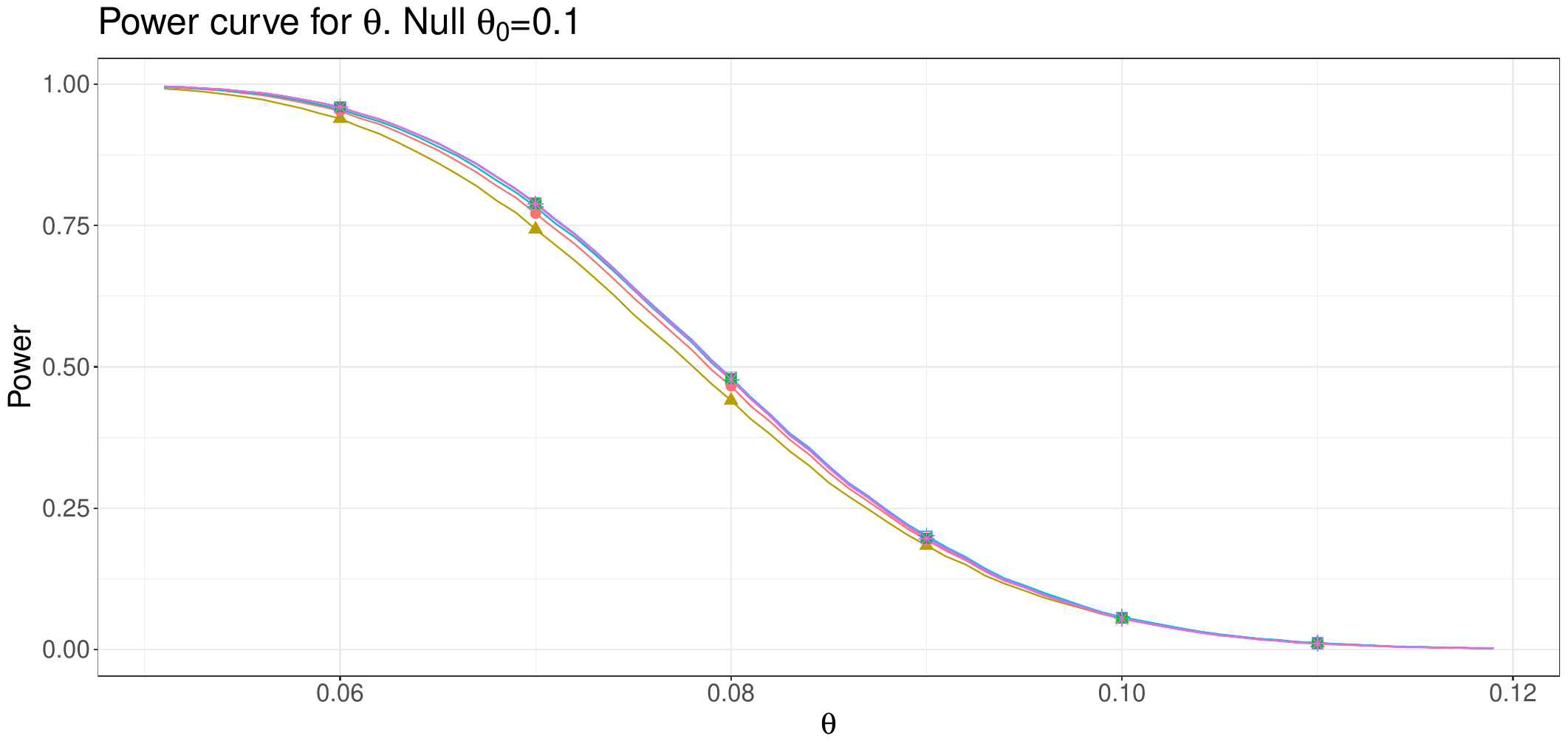}\\
		\includegraphics[width=0.93\textwidth]{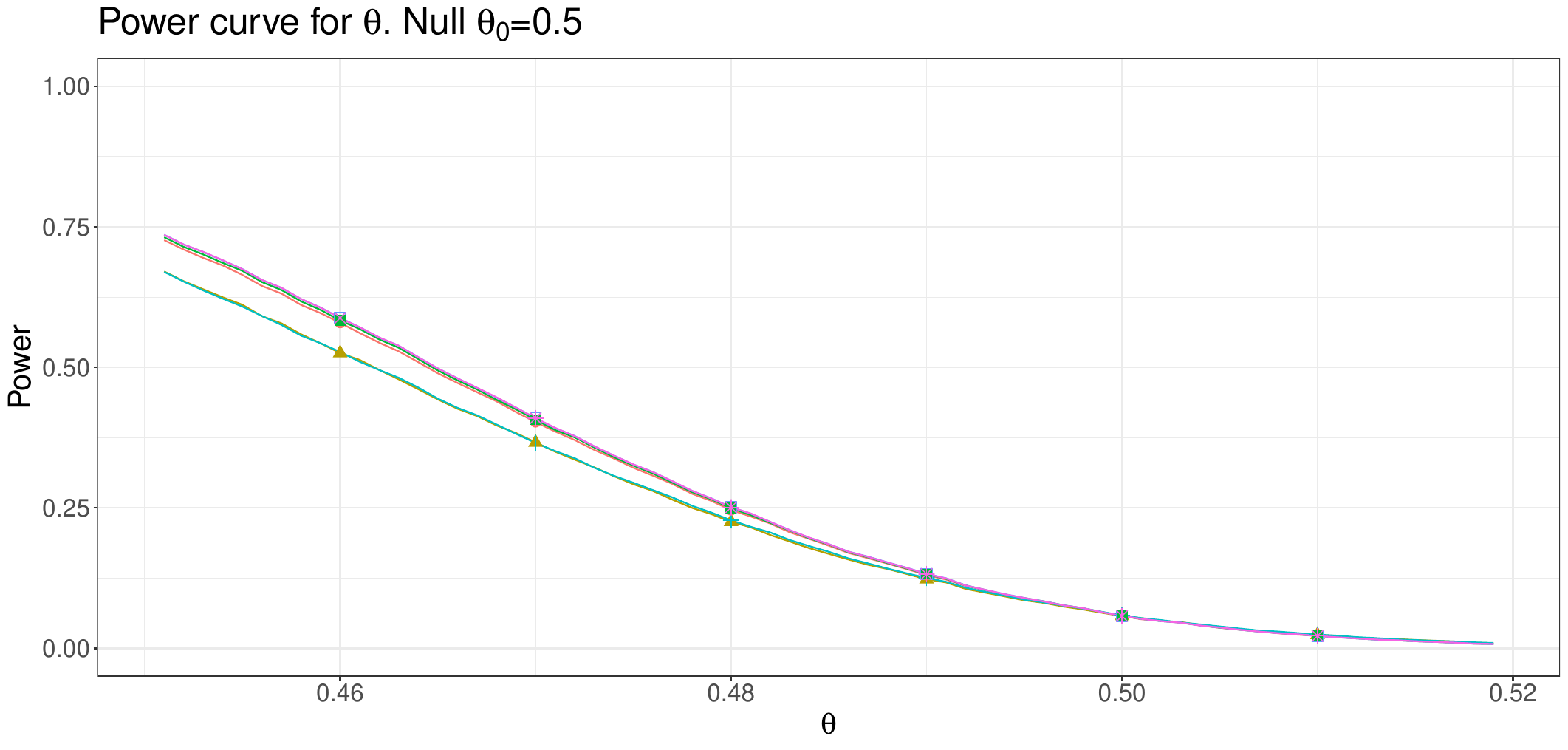}\\
		\includegraphics[width=1.05\textwidth]{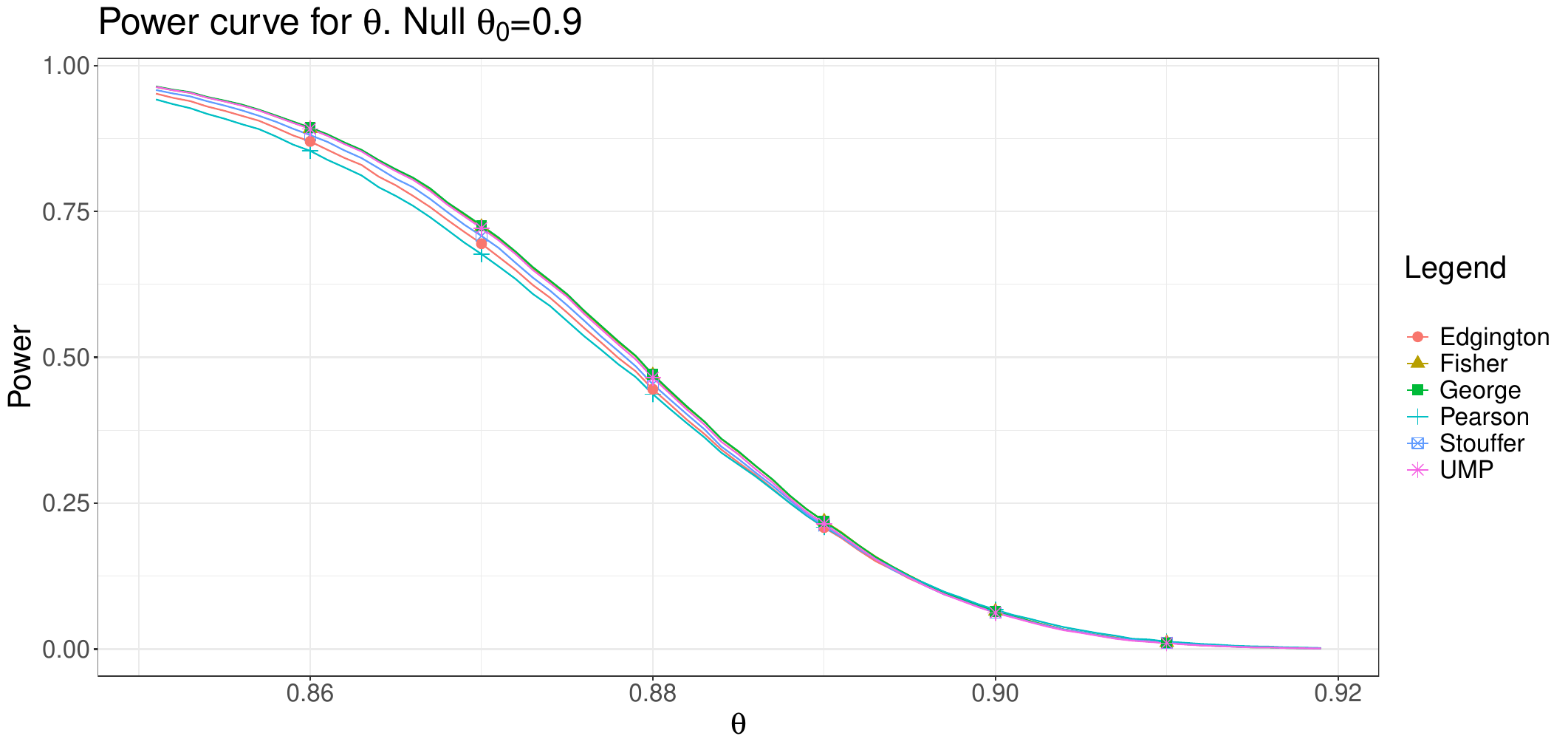}
		\caption{Empirical power curves for combining $n=100$ binomial $p$-values under three different null parameters: $\theta_0=0.1$ (top), $\theta_0=0.5$ (middle), and $\theta_0=0.9$ (bottom). Each curve shows the proportion of rejections (power) over $10^5$ simulated datasets, as a function of the true binomial parameter $\theta$. The UMP test is compared with the five combination methods: Fisher, Pearson, Stouffer, Edgington, and George.}
	\label{fig:powernum}
	\end{figure}
\end{center}

To assess power, we simulate binomial variables with true parameter $\theta \in (\theta_0-0.05,\,\theta_0+0.02)$. For each set of $100$ binomial variables, we compute the corresponding $p$-values under the null and combine them using the five proposed methods. We also compute the $p$-value for the UMP likelihood ratio test, which is the probability that a binomial random variable with parameters $500$ and $\theta_0$ is less than or equal to the sum of the $100$ simulated values. Using a significance threshold of $\alpha=0.05$, we repeat this process $10^5$ times to estimate the empirical power at each $\theta$ as the proportion of rejections for each method. Figure~\ref{fig:powernum} shows that the methods with the highest variance ratios achieve power curves nearly identical to the UMP test: Pearson's method for $\theta_0=0.1$, Fisher's for $\theta_0=0.9$, and all three recommended methods for $\theta_0=0.5$. These results demonstrate that the proposed testing procedures can preserve statistical power of the UMP test.

%


\end{appendices}
\newpage
\bibliographystyle{plain}
\bibliography{main}

\end{document}